\documentclass[a4paper,11pt]{article}
\pdfoutput=1

\usepackage{jcappub}
\usepackage{amsmath}
\usepackage{amsfonts}
\usepackage{amssymb}
\usepackage{mathtools}
\usepackage{graphics}
\usepackage{tabularx}
\usepackage{adjustbox}
\usepackage{siunitx}
\usepackage{booktabs}
\usepackage{multirow}
\usepackage{citesort}
\usepackage{graphicx}
\usepackage{url}
\usepackage{soul}
\usepackage{bm}
\usepackage{xcolor}
\usepackage{multicol}
\usepackage[utf8]{inputenc}
\usepackage[normalem]{ulem}
\usepackage{hyperref}

\usepackage{tikz,xcolor,hyperref}

\definecolor{lime}{HTML}{A6CE39}
\DeclareRobustCommand{\orcidicon}{\hspace{-1mm}
	\begin{tikzpicture}
	\draw[lime, fill=lime] (0,0) 
	circle [radius=0.16] 
	node[white] {{\fontfamily{qag}\selectfont \tiny \,ID}};
	\draw[white, fill=white] (-0.0525,0.095) 
	circle [radius=0.007];
	\end{tikzpicture}
	\hspace{-3mm}
}

\foreach \x in {A, ..., Z}{\expandafter\xdef\csname orcid\x\endcsname{\noexpand\href{https://orcid.org/\csname orcidauthor\x\endcsname}
			{\noexpand\orcidicon}}
}

\title{High energy neutrino production in gamma-ray bursts: dependence of the neutrino signal on the jet composition}

\author[a,b]{Valentin De Lia\orcidA{}}
\author[a]{and Irene Tamborra\orcidB{}}

\affiliation[a]{Niels Bohr International Academy and DARK, Niels Bohr Institute, University of Copenhagen, Blegdamsvej 17, 2100 Copenhagen, Denmark}
\affiliation[b]{{\'E}cole Normale Sup{\'e}rieure Paris-Saclay, Universit{\'e} Paris-Saclay, 4 Avenue des Sciences, 91190 Gif-Sur-Yvette, France}

\emailAdd{valentin.de\_lia@ens-paris-saclay.fr}
\emailAdd{tamborra@nbi.ku.dk}

\abstract{Heavy nuclei can be synthetized or entrained in gamma-ray bursts (GRBs)  with implications on the high-energy neutrino emission.  
By means of a Monte-Carlo algorithm, we model  nuclear cascades and investigate their impact on the neutrino production considering  kinetic dominated jets (in the internal shock model, including a dissipative photosphere) as well as  Poynting flux dominated jets (for a jet model invoking internal-collision-induced magnetic reconnection and turbulence, ICMART). 
We find that the ICMART model allows for efficient nuclear cascades leading to  an overall larger neutrino fluence than in the other two  jet models. 
The survival of nuclei and inefficient nuclear cascades lead to an overall reduction of the neutrino fluence up to one order of magnitude. However,  if nuclei are disintegrated, the neutrino fluence may be comparable to the one emitted from a jet loaded  with protons.
Exploring the   parameter space of jet properties, we  conclude that  the composition and  the bulk Lorentz factor have  significant impact on the  efficiency of nuclear cascades as well as the spectral shape of the expected neutrino fluence. On the other hand, the neutrino spectral distribution is less sensitive to  the  power-law index of the  accelerated population of protons or heavier nuclei. 
For what concerns the diffuse emission of neutrinos from GRBs, we find that the uncertainty due to the jet composition can be at most  comparable to the one related to the GRB cosmological rate. 
}

\begin{document}
\maketitle

\section{Introduction}
Gamma-Ray Bursts (GRBs) are among the most energetic transient astrophysical phenomena discovered up to date,  releasing isotropic energies up to $E_\text{iso} = 1.2 \times 10^{55}$ erg within a few seconds~\cite{Frederiks:2023bxg}. 
The observation of GRB 170817A in connection with the gravitational wave event GW 170817 has provided the first confirmation that binary neutron star mergers can harbor ultra-relativistic jets~\cite{Eichler:1989ve,Paczynski:1991aq,Nakar:2007yr,LIGOScientific:2017zic,LIGOScientific:2017ync,Mooley:2017enz,Goldstein:2017mmi}, known as short GRBs. Supernovae have instead been observed in connection with long duration gamma-ray bursts~\cite{Woosley:2006fn,Modjaz:2015cca}. However, recent observations challenge the  conjectured different origin of long and short GRBs~\cite{Yang:2023mqt,Troja:2022yya,Gillanders:2023zys}. 
Observed long duration GRBs can be divided in high-luminosity (HL, with isotropic luminosity $\mathcal{O}(10^{51}$--$10^{53})$~erg/s) and low-luminosity (LL, with isotropic luminosity smaller than $\mathcal{O}(10^{49})$~erg/s) GRBs~\cite{Sun:2015bda,Virgili:2008gp}. It is believed that HL and LL GRBs share similar properties, with LL GRBs having smaller Lorentz boost factor or perhaps corresponding to HL GRBs observed off-axis. The dynamical evolution of the GRB jet is linked to the properties of the central engine powering the jet; the jet could be magnetically driven, if the central engine harbors a strong magnetic field, or it would be better described by a fireball when the magnetic activity subsides~\cite{Paczynski:1986px,Drenkhahn:2002ug}.

Gamma-ray bursts are observed as irregular pulses of gamma-rays, with a non-thermal photon spectral energy distribution. Such non-thermal features suggest  efficient particle acceleration  in the jet~\cite{Rees:1994nw,Spitkovsky:2008fi}. Nevertheless, the physics driving the observed signals is still subject of active debate~\cite{Beloborodov:2017use,Peer:2016mqn,Drenkhahn:2002ug,Spruit:2000zm,Rudolph:2023auv}. 

Gamma-ray bursts are expected to be sources of ultra-high energy cosmic rays and high energy neutrinos~\cite{Waxman:2003vh,Meszaros:2017fcs,Piran:2004ba,Moore:2023sgo,Guarini:2023rnd,Boncioli:2018lrv}, produced also thanks to heavier nuclei synthetized or entrained in the jet~\cite{Beloborodov:2002af,Horiuchi:2012by} and eventually interacting with the photon background. In the aftermath of particle acceleration, the inelastic interaction of non-thermal 
photons with ultra-relativistic protons or nuclei could lead to the production of high-energy neutrinos, see e.g.~Refs.~\cite{Biehl:2017zlw,Pitik:2021xhb,Waxman:2003vh,Heinze:2020zqb,Rudolph:2022ppp,Murase:2008sp}. 

It is expected that the GRB jet may entrain  nuclei, e.g.~loaded at the base of the jet,  captured from the  stellar material surrounding the jet and entrained during propagation, or synthetized in situ~\cite{Beloborodov:2002af,Horiuchi:2012by,Wang:2007xj}. Such nuclei could be accelerated in the GRB jet and therefore survive photodisintegration against the intense photon field, only if the dissipation radius and the bulk Lorentz factor are large enough. Assuming that such conditions are fulfilled, the resulting neutrino production would be strongly affected.

Previous work~\cite{Biehl:2017zlw,Murase:2010gj}  investigated  the impact of the jet composition  on the high energy neutrino flux, focusing on the internal shock model~\cite{Rees:1994nw}. In the light of the strong dependence of the neutrino signal on the GRB emission mechanism~\cite{Pitik:2021xhb}, we extend such exploration to the photospheric~\cite{Toma:2010xw}  and the internal-collision-induced magnetic reconnection and turbulence (ICMART)~\cite{Zhang:2010jt} models, and explore the neutrino production for a range of jet properties. 

This work is organized as follows. Section~\ref{sec:Main_models} provides an overview of the internal shock model, the photospheric one, and the ICMART model. Section~\ref{sec:cooling_times} offers a summary of the acceleration and cooling timescales adopted to compute the maximum energies of protons and nuclei. A description of how  neutrinos are produced through   
 photohadronic interactions of protons and neutrons or nuclei is presented in Sec.~\ref{sec:neutrino_from_pn}. Section~\ref{sec:heavy_nuclei} focuses on the formalism adopted to compute the neutrino emission from jets loaded with heavier nuclei,  including the modeling of the nuclear cascades using a Monte-Carlo method. 
 Our  findings on the impact of the jet composition on the neutrino emission for our three jet models are presented in Sec.~\ref{sec:results}, together with an exploration of the jet parameter space. We also discuss the consequences that the jet composition should have on the diffuse  emission of high-energy neutrinos in Sec.~\ref{sec:diffuse_flux}. Finally, we critically review our findings and conclude in  Sec.~\ref{sec:discussion}. Details on the Monte Carlo simulation of nuclear cascades are provided in Appendix~\ref{App:MCNC_params}, while the nuclear photohadronic model is outlined in Appendix~\ref{App:photomeson}.  In addition, Appendix~\ref{App:Comparison_Biehl} provides a comparison between our setup and the one adopted in Ref.~\cite{Biehl:2017zlw}.

\section{Gamma-ray burst models} \label{sec:Main_models}
This section highlights the main features of the kinetic dominated and Poynting flux dominated jets considered in this work as well as the adopted photon spectral energy distributions. We refer the interested reader to  Ref.~\cite{Pitik:2021xhb} for a broader overview. 

In the following, we distinguish among three reference frames: the comoving jet frame, the  central engine frame, and the observer frame; any jet quantity $X$ is denoted  as $X^\prime$, $\Tilde{X}$, or $X$ respectively according to the adopted reference frame. For example, energy or length transform as $X = {\Tilde{X}}/{(1+z)}  = {\Gamma \; X^\prime}/{(1+z)}$, while time transforms as $X = (1+z)\;\Tilde{X} = {(1+z)\;X^\prime}/{\Gamma}$.
We consider a relativistic jet propagating with bulk Lorentz factor $\Gamma$ with respect to the central engine. During the prompt phase, it is expected that $\theta_{\rm op} \gg \Gamma^{-1}$, with $\theta_{\rm op}$ being the half opening angle~\cite{Bromberg:2011fg}. This justifies the employment of the isotropic equivalent energy ($\Tilde{E}_\text{iso}$) representing the energy content of the outflow. Additionally, GRBs are characterized by two timescales. The first one is the total duration of the event $t_{\text{dur}}$ and the second one is the average variability timescale of the central engine $t_{\text{var}}$. We thus define the total isotropic luminosity $\Tilde{L}_\text{iso} = \Tilde{E}_\text{iso} / \Tilde{t}_{\text{dur}}$. Any burst is also characterized by its redshift $z$.

\subsection{Internal shock model}
The internal shock (IS) model has been widely investigated  for the GRB prompt emission~\cite{Rees:1994nw,Kobayashi:1997jk,Daigne:1998xc} because it allows to easily explain the variability of the GRB lightcurves. Due to the erratic activity of the  the central engine, the outflow results in several shells, each with different  Lorentz factor $\Gamma$. When a faster shell hits a slower shell, dissipation of kinetic energy occurs. In the following, we focus on a one-zone collision model for the sake of simplicity, assuming that all collisions are identical. However, multi-zone collision models are expected to lead to less efficient neutrino emission~\cite{Guetta:2000ye,Bustamante:2016wpu,Rudolph:2019ccl,Globus:2014fka,Rudolph:2022ppp}.
We assume that, for a GRB of duration $\Tilde{t}_{\text{dur}}$ and variability time scale $\Tilde{t}_{\text{var}}$, $N_\text{shock} = \Tilde{t}_{\text{dur}} / \Tilde{t}_{\text{var}}$ identical shocks occur at the radius $R_\gamma = R_{\text{IS}} = 2\, \Gamma^2 c\,  \Tilde{t}_{\text{var}}$, within a volume $V'_s = 4\pi \, R_{\text{IS}}^2 \, c \, t'_{\text{var}}$. Unless otherwise stated, the benchmark jet properties adopted for this model are summarized in Table~\ref{tab:initial_conditions}.
While the assessment of the jet conditions leading to  a specific jet composition is beyond the scope of this work, we generalize our findings in the second part of this paper, exploring variations of some of the jet characteristic properties.

The fraction of energy dissipated in particle acceleration, and the fractions of  energy that  effectively accelerates seed particles (protons, nuclei), electrons, or contributes to amplify the magnetic field are  $\varepsilon_{\text{d}}  = 0.2$, $\varepsilon_{A} = 0.1$ (where $A$ indicates protons or nuclei), $\varepsilon_{e} = 0.01$, and $ \varepsilon_{B}=0.1$~\cite{Sironi:2010rb,Crumley:2018kvf}. 
 We stress that these microphysical parameters, and in particular the ratio between $\varepsilon_{e}$ and $\varepsilon_{A}$, are highly sensitive to the characteristic properties of the jet,  such as the initial magnetization and the geometry of the shock, see e.g. Ref.~\cite{Sironi:2010rb};  we have chosen to use the lower bounds of the microphysical parameters predicted by  particle-in-cell simulations of relativistic shocks~\cite{Sironi:2010rb,Crumley:2018kvf}.
The magnetic field is 
\begin{equation}
B' = \sqrt{8 \pi \frac{\varepsilon_B \,\varepsilon_{d}\; E'_\text{iso,s}}{V'_{s}}}\ , 
\end{equation}
with $E'_\text{iso,s} = E'_\text{iso} / N_\text{shock}$.

Shocks accelerate a certain fraction of the charged particles into a power-law distribution~\cite{Lipari:2007su}:
\begin{equation} \label{eq:acc_power_law}
    n'_A(E'_A) = C {E'_A}^{-k} \exp\left[ - \left(\frac{E'_A}{E'_{A,\max}} \right)^2 \right] \Theta (E'_A - E'_{A,\min})\ ,
\end{equation}
with  $k\simeq 2.2$; imposing that $U_A = \int_{E'_{A,\min}}^{E'_{A,\max}}  n'_A(E'_A) \;E'_A \;dE'_A$, we compute the normalization constant 
 $C$, where  $ U_A = {\varepsilon_{d}\; \varepsilon_A\; E'_{\text{iso}, s}}/{V'_{s}}$ is the energy density of accelerated seed particles. 
\begin{table}
    \caption{Summary of the characteristic parameters assumed for our  benchmark jet for the internal shock (IS), photospheric (PHOTO) and ICMART models, respectively.} 
    \label{tab:initial_conditions}
    \centering
    \begin{tabular}{c||c|c|c|}
    Parameter & IS & PHOTO & ICMART\\\midrule
    $\Tilde{E}_\text{iso}$  & \multicolumn{3}{c|}{$4.5\times10^{54}$ erg~\cite{Wang:2015vpa}}\\ \hline
    $z$ &\multicolumn{3}{c|}{$2$~\cite{IceCube:2017amx}}  \\ \hline
    $\Gamma$  &\multicolumn{3}{c|}{$300$~\cite{Ghirlanda:2017opl}} \\ \hline
    $t_{\text{dur}}$  &\multicolumn{3}{c|}{$30$~s~\cite{Zitouni:2018wre}}  \\ \hline
    $t_{\text{var}}$  &\multicolumn{3}{c|}{$0.5$~s~\cite{ANTARES:2020vzs}}  \\ \hline
    $\varepsilon_\text{d}$ & $0.2$~\cite{Kobayashi:1997jk, Guetta:2000ye} & $0.2$~\cite{Kobayashi:1997jk, Guetta:2000ye} & $0.35$~\cite{Deng:2015xea} \\ \hline  
    $\varepsilon_e$  & $0.01$~\cite{Sironi:2010rb, Crumley:2018kvf} & $0.01$~\cite{Sironi:2010rb, Crumley:2018kvf} & $0.5$~\cite{Sironi:2015eoa} \\ \hline
    $\varepsilon_X$ & $0.1$~\cite{Crumley:2018kvf} & $0.1$~\cite{Crumley:2018kvf}& $0.5$~\cite{Sironi:2015eoa} \\ \hline
    $\varepsilon_B$  & $0.1$~\cite{Crumley:2018kvf} & $0.1$~\cite{Crumley:2018kvf} & n/a \\ \hline
    $k$ &$2.2$~\cite{Groselj:2024dnv, Sironi:2013ri} &  $2.2$~\cite{Groselj:2024dnv, Sironi:2013ri} & $2.0$~\cite{Zhang:2023lvw, Sironi:2014jfa}\\ \hline   
    $\sigma$& n/a &  n/a & $45$~\cite{Pitik:2021xhb} \\ \hline
    \end{tabular}
\end{table} 

The photon spectrum is modeled through a Band function~\cite{Band:1993eg}:
\begin{equation} \label{eq:band_spectrum}
    n'_\gamma(E'_\gamma) = C 
    \begin{cases}
         \displaystyle \left(\frac{E'_\gamma}{E'_0} \right)^\alpha \exp \left(- \frac{(\alpha+2) E'_\gamma}{E'_{\gamma,p}} \right) & \text{if } E'_{\gamma,\text{min}} \leq E'_\gamma \leq E'_{\gamma,c}\\
         \displaystyle \left(\frac{E'_\gamma}{E'_0} \right)^\beta \exp (\beta - \alpha ) \left(\frac{E'_{\gamma, c}}{E'_0} \right)^{\alpha- \beta}& \text{if } E'_{\gamma,c} \leq E'_{\gamma}\leq E'_{\gamma,\text{max}}\\
    \end{cases}
\end{equation}
with $E'_{\gamma,p}$, $E'_{\gamma,c}$, $E'_{\gamma,\text{min}}$, $E'_{\gamma,\text{max}}$  being the peak, cooling, minimum and maximum energies of the spectrum,  $E'_{\gamma,0} =  [{(1+z)}/{\Gamma}] \times 100$~keV, respectively. Relying on  GRB observations from Fermi, we choose $\alpha \simeq -1.1$ and $\beta \simeq -2.2$~\cite{Gruber:2014iza}. The normalization constant is given by  
$U_\gamma =  \int_{E'_{\gamma, \min}}^{E'_{\gamma, \max}}  n'_\gamma(E'_\gamma) \;E'_\gamma \;dE'_\gamma$
with $U_\gamma = {\varepsilon_{d}\; \varepsilon_e\; E'_{\text{iso}, s}}/{V'_s}$. The peak energy is computed following the Amati relation~\cite{Amati:2006ky}:
    $\Tilde{E}_{\gamma,p} = 80 \; \left({\Tilde{E}_{\gamma,\text{iso}}}/{10^{52}\; \text{erg}} \right)^{0.57} \text{keV}$
with $\Tilde{E}_{\gamma,\text{iso}} = \varepsilon_d \varepsilon_e \Tilde{E}_{\text{iso}} $, and the cooling energy is  
$E'_{\gamma,c} = ({\alpha -\beta})/({\alpha +2}) E'_{\gamma,p}$. Note that our benchmark $\tilde{E}_{\rm{iso}}$ is such that, in the internal shock model, $\tilde{L}_{\rm{iso}, \gamma} \simeq \varepsilon_d \varepsilon_e \tilde{E}_{\rm{iso}}/t_{\rm{dur}}$ is slightly lower than what  often considered in the literature. Yet, this value of  $\tilde{E}_{\rm{iso}}$~\cite{Wang:2015vpa} allows to recover the observed $\tilde{L}_{\rm{iso}, \gamma}$ for the photosperic and ICMART models;  we choose to rely on the same jet benchmark properties to allow for a fair comparison across jet models.

Note that we do not consider neutrino production below the photosphere~\cite{Bahcall:2000sa,Murase:2008sp,Murase:2013hh,Kashiyama:2013ata,Wang:2008zm,Guarini:2022hry,Rudolph:2023auv}, as neutrinos would be produced with GeV--TeV energies, while we are interested in comparing different models for neutrino production in the optically thin region with overall larger energies.

\subsection{Photospheric  model}
The photospheric model (PHOTO) is a variation of the internal shock model that assumes that dissipative processes shape the radiation bulk produced in the optically thick region below the photosphere~\cite{Beloborodov:2017use}. We assume all characteristic jet parameters as being  the same  as the ones of the  internal shock model, except for the ones related to the photon spectrum. Protons or nuclei interact with a photon spectrum described by three components~\cite{Toma:2010xw,Pitik:2021xhb}: a non-thermal photospheric component defined by a Band spectrum and undergoing dissipation at the internal shock radius,  a  photospheric Compton up-scattered spectrum, and  a synchrotron component due to electrons cooling in the shock region. 
The accelerated particle spectrum is given by  Eq.~\eqref{eq:acc_power_law} with $k=2.2$; the photon spectrum is~\cite{Pitik:2021xhb}:
\begin{equation} \label{eq:photo_spectrum}
    n'_\gamma(E'_\gamma) = \left(\frac{R_{\text{PH}}}{R_{\text{IS}}}\right)^2 n'_{\gamma, \text{PH}}(E'_\gamma) + n'_{\gamma, \text{SYNC}}(E'_\gamma) + n'_{\gamma, \text{UP}}(E'_\gamma)\ ,
\end{equation}
with $R_\text{PH} =  {\sigma_T \; \Tilde{L}_\text{iso}}/({ 4 \pi \Gamma^3 m_p c^3})$ being the radius of the photosphere. Each spectral component  is normalized  at the emission radius independently, using the fractions $x_\text{PH}$, $x_{\text{SYNCH}}$ and $x_\text{UP}$ of  $E'_\text{iso}$  going into each part of the spectrum, respectively. We have $x_{\text{PH}} = 0.2$, $x_{\text{SYNCH}} = 1.6 \times 10^{-4}$, and $x_{\text{UP}} = 0.002$~\cite{Pitik:2021xhb}. The spectral component $n'_{\gamma, \text{PH}}(E'_\gamma)$ is given by  Eq.~\eqref{eq:band_spectrum}. The spectral components $n'_{\gamma, \text{SYNC}}(E'_\gamma)$ and $n'_{\gamma, \text{UP}}(E'_\gamma)$ are joint broken power laws, describing synchrotron self-Compton emission of charged particles in the fast-cooling regime~\cite{2018pgrb.book.....Z}:
\begin{equation} \label{eq:broken_p_low}
    n'_\gamma(E'_\gamma) = C 
    \begin{cases}
         \displaystyle \left(\frac{E'_\gamma}{E'_{\gamma,c}} \right)^{-2/3} & \text{if } E'_{\gamma, \min} \leq E'_\gamma \leq E'_{\gamma,c}\\
         \displaystyle \left(\frac{E'_\gamma}{E'_{\gamma,c}} \right)^{-3/2} & \text{if } E'_{\gamma,c} \leq E'_\gamma \leq E'_{\gamma,p}\ .\\
         \displaystyle \left(\frac{E'_{\gamma, p}}{E'_{\gamma,c}} \right)^{-3/2} \; \left(\frac{E'_\gamma}{E'_{\gamma,p}} \right)^{-(k+2)/2}& \text{if } E'_{\gamma,p} \leq E'_\gamma \leq E'_{\gamma, \max}\\
    \end{cases}  
\end{equation}
For the synchrotron component, 
    $E'_{\gamma,i}  = ({3}/{2}) ({\hbar e }/{ m_e c}) {\gamma'_i}^2 B'$ where $i=p, c$,
$\gamma'_p = ({m_p}/{m_e})\times ({k_e - 2})/({k_e - 1}) \; \varepsilon_\text{d} \,\varepsilon_e$ and $\gamma'_c = ({3 \; m_e \Gamma})/({4 m_p n'_b x_\text{PH} \sigma_T R_{\text{IS}} })$~\cite{2018pgrb.book.....Z}, $\sigma_T = 6.65 \times 10^{-25}$ cm$^2$ is the Thompson cross-section, and $n'_b  = {\Tilde{L}_\text{iso}}/({4 \pi {R_{\text{IS}}}^2 m_p c^3 \Gamma^2})$ is the total baryon density. For the up-scattered component, we use the same $E'_{\gamma,c}$ and $E'_{\gamma,p}$ as for the Band spectrum.  Unless otherwise stated, the benchmark jet properties adopted for this model are summarized in Table~\ref{tab:initial_conditions}.

\subsection{ICMART model}
The ICMART model belongs to the class of Poynting flux dominated  jets and assumes that energy is dissipated at very large distance from the central engine~\cite{Zhang:2010jt}.
In the ICMART model, it is assumed that the  jet is composed of magnetized shells  with  constant initial magnetization; in this work we adopt $\sigma = 45$ in order to facilitate a comparison between our findings and the ones of Ref.~\cite{Pitik:2021xhb}--however, the initial magnetization is very uncertain and this has implications on the neutrino fluence; we refer the interested reader to  Refs.~\cite{Gottlieb:2022old,Guarini:2022hry} for an investigation of the impact of $\sigma$ on the expected neutrino fluence.
Similarly to the internal shock model, these shells collide with each other. 
These internal shocks in the optically thick region alter the ordered magnetic field configuration, triggering  magnetic reconnection and the release of stored magnetic energy at large radii~\cite{Zhang:2010jt}.  
The radius  at which gamma-rays and neutrinos are emitted is independent of the jet properties ($R_\gamma = R_\text{ICMART} =  10^{15}$~cm).

For the ICMART model, the photon spectrum is also a Band function, see Eq.~\eqref{eq:band_spectrum}, and the accelerated particle spectrum is a cut-off power law (Eq.~\ref{eq:acc_power_law}). However, magnetic reconnection is more efficient than relativistic shocks at accelerating particles, meaning that a smaller power law index ($k \simeq 2.0$) and larger  microphysical parameters than the ones characteristic of the internal shock model are considered: $\varepsilon_{\text{d}} = 0.35$, $ \varepsilon_{A}= 0.5$ and $ \varepsilon_{e} = 0.5$ \cite{Deng:2015xea, Sironi:2015eoa, Werner:2016fxe, Guo:2015ydj}.  Table~\ref{tab:initial_conditions}  summarizes the benchmark jet properties adopted for this model. Moreover, the magnetic field is~\cite{Zhang:2010jt} 
\begin{equation}
    B' = \sqrt{\frac{2\Tilde{L}_\text{iso}}{c \,\Gamma^2 \, R_\text{ICMART}^2}  \,\frac{\sigma}{\sigma +1 }}\ .
\end{equation}

\section{Acceleration and cooling timescales} \label{sec:cooling_times}
We present in this section the acceleration and cooling timescales used  to compute the maximum energy ($E'_{A, \max}$) up to which a nucleus of mass number $A$ and charge $Z$ inside the jet can be accelerated (see Eq.~\ref{eq:acc_power_law}). We highlight that any equation presented in the following section is also valid for protons and neutrons, which correspond to  $(A=1,~Z=1)$ and $(A=1,~Z=0)$ respectively. 

The inverse of the acceleration  timescale is
\begin{equation}
\label{eq:tau_acc}
  \tau^{\prime -1}_{\text{acc}} = \frac{c\,Ze\,B'}{E'_A}\ ,
\end{equation}
with $E'_A$ being the energy of the nucleus $^A_ZX$. We need to compare $\tau^{\prime -1}_{\text{acc}}$ with the inverse of the total cooling timescale  given by 
\begin{equation}
\label{eq:sum_tau}
\tau^{\prime -1}_{\rm{cool}} = \sum_i \tau^{\prime -1}_i\ ;
\end{equation}
$\tau_i$ stands for the adiabatic cooling, synchrotron, inverse Compton, Bethe-Heitler pair creation,  
and collisions with thermal protons~\cite{Gao:2012ay,2009herb.book.....D}:
\begin{align} 
\label{eq:tau_ad}
    \tau^{\prime -1}_{\text{ad}} &= \frac{c \Gamma}{R}\ , \\
    \tau^{\prime -1}_{\text{synch}} &= \frac{4 Z^4\, \sigma_T m_e^2 E'_A B'^2}{8 \pi \times 3 m_A^4 c^3 }\ , \\
    \tau^{\prime -1}_{\text{IC}} &= \frac{3 (m_e c^2)^2\, Z^2 \sigma_T c}{16 {\gamma'_A}^2( {\gamma'_A} -1)\beta'_A } \int_{E'_{\gamma,\min}}^{E'_{\gamma,\max}} F(E'_\gamma , \gamma'_A ) \;n'_{\gamma}(E'_\gamma) \frac{dE'_\gamma }{{E'_\gamma}^2}\ ,  \\
    \tau^{\prime -1}_{\text{BH}} &= \frac{\alpha r_e^2 c Z^2 (m_e c^2)^2}{E'_A } \int_{2}^{\frac{2\gamma'_A E'_{\gamma, \max}}{m_e c^2} } n'_\gamma\left(\frac{\epsilon'_\gamma}{2\gamma'_A}\right) \frac{\varphi(\epsilon'_\gamma)}{{\epsilon'_\gamma}^2}  d \epsilon'_\gamma\ , \\
 \label{eq:tau_Ap} \tau^{\prime -1}_{Ap} &= c n'_p \;\sigma_{Ap}(E'_A)\ , 
\end{align}
and $\tau^{\prime -1}_{A\gamma} \equiv R^{\prime}_{A\gamma}$ is the photohadronic interaction rate: 
\begin{equation}
    R^{\prime}_{A\gamma} = \frac{ c}{2{\gamma'_A}^2 }\int_{\frac{E^\mathrm{th}_{\gamma,r}}{2 \gamma'_A}}^{+\infty}
 \frac{n'_{\gamma}(E'_\gamma) }{{E'_\gamma}^2} \int_{E^\mathrm{th}_{\gamma,r}}^{2 \gamma'_A E'_\gamma} E_{\gamma,r}\; \sigma_{A\gamma}(E_{\gamma,r}) \;d E_{\gamma,r} \; dE'_\gamma\ .  \label{eq:tau_Agamma}\ 
\end{equation}

In the equations above $\epsilon'_\gamma = E'_\gamma / (m_e c^2)$, $E_{\gamma,r}$ is the energy of the photon in the reference frame of the nucleus, $\gamma'_A = E'_A / ( m_A \,c^2)$, $ r_e =e^2/ (m_e \,c^2) \approx 2.82\times10^{-13} $ cm is the classical radius of the electron, $\alpha \approx 1/137$ is the fine structure constant, and $n'_p = \Tilde{L}_\mathrm{iso} / ( 4 \pi  R_\mathrm{IS}^2 m_p c^3 \Gamma^2)$ is the density of thermal protons in the jet. The functions $\varphi(\epsilon'_\gamma)$ and $F(E'_\gamma , \gamma'_A)$ model the Bethe-Heitler and inverse Compton processes and are respectively defined in Eqs.~(9.36) and (9.37) of Ref.~\cite{2009herb.book.....D}  and Eqs.~(13)--(15) of Ref.~\cite{PhysRev.137.B1306}. The hadronic  cross-section is $\sigma_{Ap}$; we follow Ref.~\cite{Kafexhiu:2014cua} for  the proton-proton cross-section  and Ref.~\cite{Lebedev_1964} for the extension of such cross section  to any nucleus. The total inelastic photohadronic cross-section is $\sigma_{A\gamma}(E_r)$ and is modeled as described in  Appendix~\ref{App:MCNC_params}. Note that we include the contribution of meson production   in the total photonuclear cross section, even if this is not  a disintegration event; meson production leads to the cooling of  nuclei without disintegrating them. Since the branching ratio for pion production is small ($\sigma_{A\gamma \rightarrow \pi } / \sigma_{A\gamma, \mathrm{tot} } \approx 0.07 $--see Fig.~11 of Ref.~\cite{Morejon:2019pfu}), this correction has a negligible impact. We rely on the photohadronic interaction rate instead of the photohadronic cooling timescale since any isotope undergoing a photohadronic interaction is photodisintegrated, leaving behind a new isotope which is then consistently treated as another nuclear species in our  Monte-Carlo algorithm  (see Sec.~\ref{sec:heavy_nuclei}). This approach warrants the conservation of the nucleon number and energy. For the sake of simplicity,   the set of  cooling timescales and photohadronic interaction rate is dubbed  ``cooling timescale'' hereafter.

\begin{figure}
    \centering
    \includegraphics[width = \linewidth]{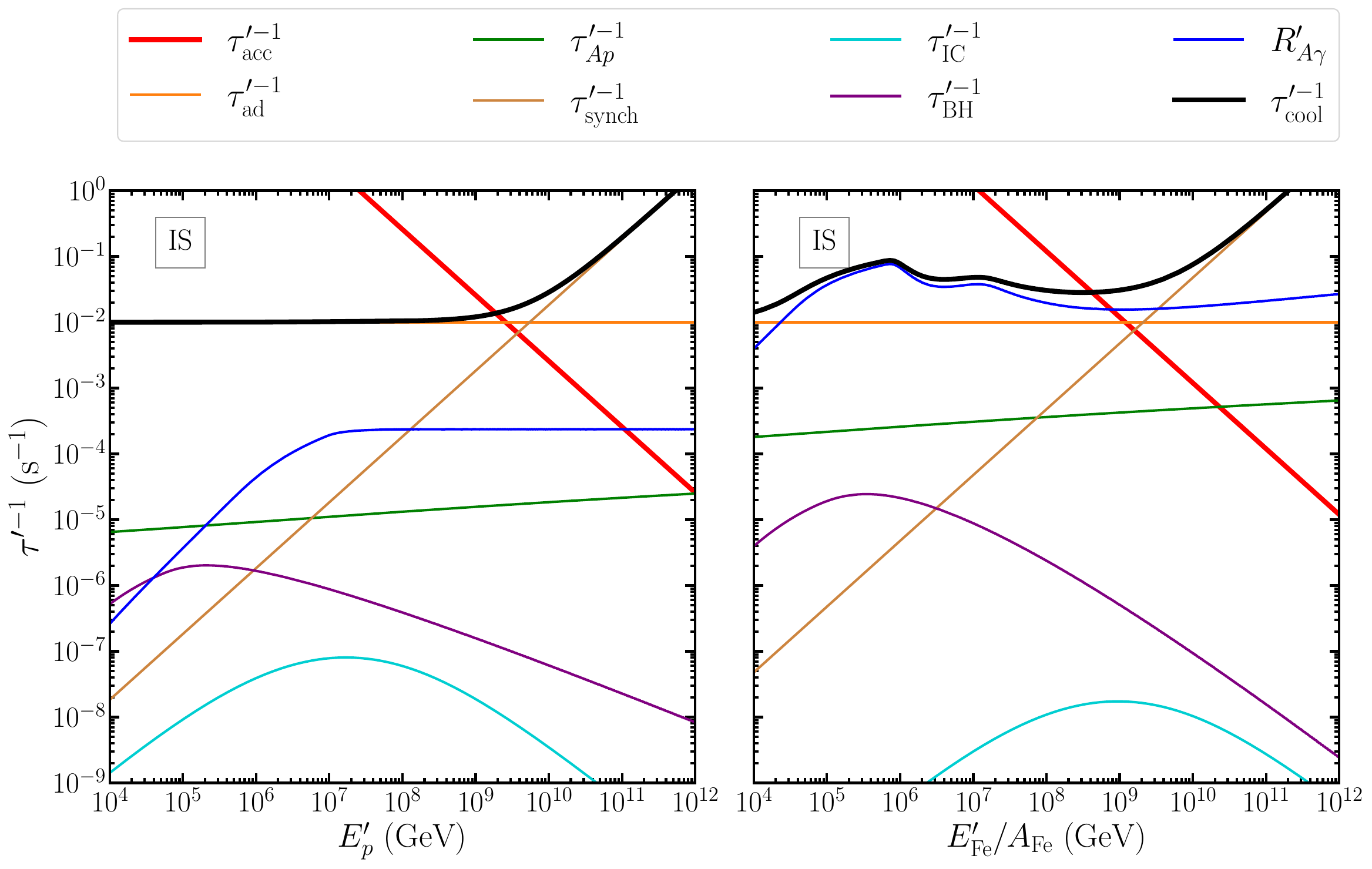}
    \caption{Acceleration (Eq.~\ref{eq:tau_acc}), inverse cooling (Eqs.~\ref{eq:sum_tau}--\ref{eq:tau_Ap})   timescales, and photohadronic interaction rate (Eq.~\ref{eq:tau_Agamma})  for protons (on the left) and iron (on the right) as functions of the comoving proton energy and normalized iron energy respectively. We consider a jet with characteristic properties as in Table~\ref{tab:initial_conditions} for the internal shock  model. The maximum comoving energy is obtained computing the energy for which $\tau^{\prime -1}_{\rm{cool}}=\tau_{\rm{acc}}^{\prime -1}$ (cf.~thick black and red curves). One can see that the normalized maximum  energy for iron is smaller than the proton one because of photonuclear interactions contributing to  the total cooling of iron and being irrelevant for protons.
     For iron (right panel), $R'_{A\gamma}$ displays a double peak, which is due to the giant dipole resonance and the photomeson resonance (see Fig.~\ref{fig:Fe_cross_section}).
}
    \label{fig:timescales}
\end{figure}

For illustrative purposes, Fig.~\ref{fig:timescales} shows a comparison among the  cooling and acceleration timescales obtained for protons and  iron, assuming a jet with characteristic properties summarized in Table~\ref{tab:initial_conditions} for the internal shock model. The maximum energy at which $p$ or $^{56}$Fe can be accelerated is given by  the crossing between the acceleration and total cooling timescales. 
The iron energy  useful to compute the neutrino emission is  the energy per nucleon. In this example, $E'_{p, \max} = 1.9 \times 10^9$ GeV and $E'_{\text{Fe}, \max} / A_{\text{Fe}} = 4.2 \times 10^8$ GeV, i.e.~the energy per nucleon is lower for iron than for protons.
This result is due to the fact that photonuclear interactions add an important contribution to the radiative processes associated to iron, while this is not the case for protons.

\section{Neutrino production from proton and neutron photohadronic interactions} \label{sec:neutrino_from_pn}
Neutrinos can be produced thanks to photohadronic interactions of protons and neutrons or nuclei. In this section, we focus on $p\gamma$ interactions, following Refs.~\cite{Hummer:2010vx,Lipari:2007su}. We adopt the same approach for what concerns  $n\gamma$ interactions after applying charge symmetry on the mesons. We also illustrate how to compute the neutrino fluence expected at Earth.

\subsection{Photohadronic interactions}

\begin{figure}
    \centering
    \includegraphics[width=\linewidth]{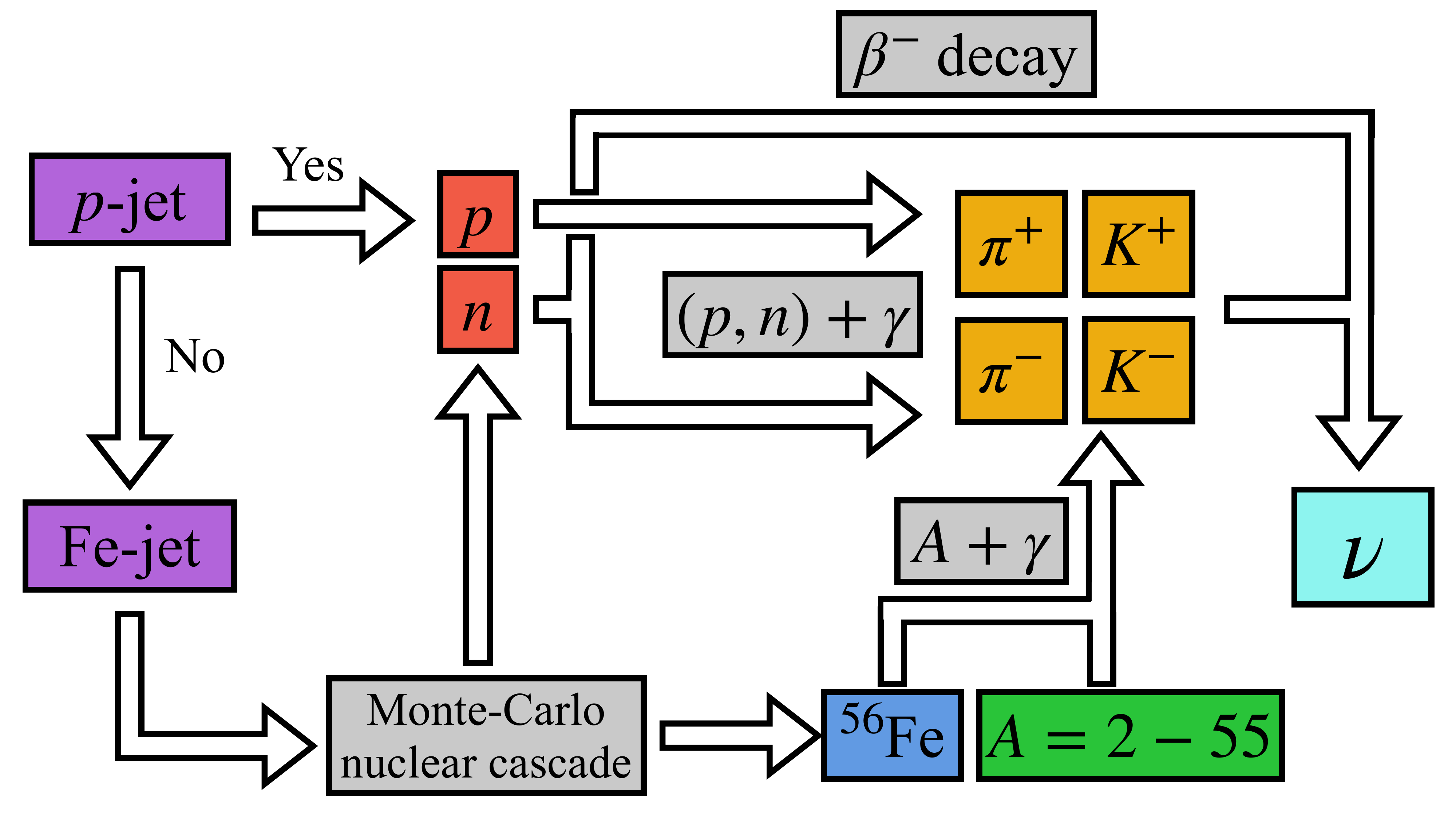}
    \caption{Flow chart illustrating the approach adopted to compute  neutrino production in jets loaded  with protons only ($p$-jet) and jets loaded with iron (Fe-jet). }
    \label{fig:flow_chart}
\end{figure}
The physical processes leading to the production of neutrinos,  discussed in this section, are illustrated in  Fig.~\ref{fig:flow_chart}. Photohadronic interactions can lead to the production of mesons, if the energy of the photon in the reference frame of the proton is larger than $140$~MeV~\cite{Morejon:2019pfu}. The production channels can be classified as follows~\cite{Hummer:2010vx}: 
\begin{align}
\text{Resonances: } &\; p+\gamma \rightarrow (\Delta, N, \rho) \rightarrow n + \pi^+ \label{eq:resonances} \\
\text{Direct production: } &\;p+\gamma \rightarrow n + \pi^+ \label{eq:direct_prod}\\
\text{Multi-pion production: } &\;p+\gamma \rightarrow n + \pi_1 + \pi_2 + ... \label{eq:multi_prod}
\end{align}
The production  rate for the particle $\mathcal{M}$ (with $\mathcal{M} = \pi^+$, $\pi^-$, $K^+$ or $K^-$) is defined as the number of particles created per volume, per unit of time and per energy interval (in units of cm$^{-3}$ s$^{-1}$ GeV$^{-1}$) and it is given by:
\begin{equation}
\label{eq:QM}
  Q'_\mathcal{M}(E'_\mathcal{M}) = \sum_{\alpha} \int n'_p(E'_p) \;\Gamma^{\alpha}_{p\gamma \rightarrow \mathcal{M}}\; \frac{d {n^{\prime \alpha}_{p\gamma\rightarrow \mathcal{M}}}}{d E'_p} (E'_p, E'_\mathcal{M})\; d E'_p\ .  
\end{equation}
The sum runs over all  possible interactions $\alpha$, see Eqs.~\eqref{eq:resonances}--\eqref{eq:multi_prod}. The seed proton density per unit energy is $n'_p(E'_p)$ (in units of  cm$^{-3}$ GeV$^{-1}$), while $\Gamma^{\alpha}_{p\gamma \rightarrow \mathcal{M}}(E'_p)$ is the production rate for a given interaction (in units of s$^{-1}$), and ${d {n^{\prime \alpha}_{p\gamma\rightarrow \mathcal{M}}}}/{d E'_p}$ represents the probability density for production of a meson with energy $E'_\mathcal{M}$ given that the incident proton has an energy $E'_p$.

In order to simplify the numerical solution of the integral above,
we  assume that all pions are produced with a certain  energy that does not depend on the parent proton energy: 
\begin{equation}
   \frac{d {n^{\prime \alpha}_{p\gamma\rightarrow \mathcal{M}}}}{d E'_p} \simeq M_{\mathcal{M}}^{\alpha} \times \delta(E'_\mathcal{M} - \chi_\mathcal{M}^{\alpha} \;E'_p)\ , 
\end{equation}
where $M_{\mathcal{M}}^{\alpha}$ is the multiplicity of the interaction and $\chi_\mathcal{M}^{\alpha}$ describes the amount of energy  transferred to the produced meson. 
We also assume an isotropic photon distribution. Hence, Eq.~\eqref{eq:QM} becomes:
\begin{equation} \label{eq:Q_def}
   Q'_\mathcal{M}(E'_\mathcal{M}) = \sum_{\alpha}  n'_p\left(\frac{E'_\mathcal{M}}{\chi_\mathcal{M}^{\alpha}}\right) \; \frac{m_p c^2}{E'_\mathcal{M}} M_{\mathcal{M}}^{\alpha}\; \int_{E_{\gamma,r}^{\mathrm{th}, \alpha} /2 }^{+\infty} f^{\alpha}(y) \; n'_\gamma \left(y \frac{ \chi_\mathcal{M}^{\alpha} m_p c^2}{E'_\mathcal{M}} \right) dy\ ,  
\end{equation}
with 
\begin{equation} \label{eq:f_def}
    f^{\alpha}(y) = \frac{1}{2y^2}\; \int_{E_{\gamma,r}^{\mathrm{th}, \alpha} }^{2y} \;E_{\gamma,r} \;\sigma^{\alpha}(E_{\gamma,r}) \; d E_{\gamma,r}\ .
\end{equation}
The parameters characterizing each interaction $\alpha$ (namely $\chi_\mathcal{M}^{\alpha}$, $M_{\mathcal{M}}^{\alpha}$, $f^{\alpha}(y)$ and $E_{\gamma,r}^{\mathrm{th}, \alpha}$) can be found in Ref.~\cite{Hummer:2010vx}.

\subsection{Neutrino production} \label{sec:neutrino_prod}
The mesons ($\pi^+$, $\pi^-$, $K^+$ and $K^-$) produced in the reactions illustrated above  can decay into neutrinos through the following channels:
\begin{eqnarray}
\pi^+ &\rightarrow& \mu^+ + \nu_\mu \rightarrow e^+ + \nu_e +\bar{\nu}_\mu + \nu_\mu\\
K^+ &\rightarrow& \mu^+ + \nu_\mu \rightarrow e^+ + \nu_e +\bar{\nu}_\mu + \nu_\mu
\end{eqnarray}
and similar reactions hold for the antiparticle conjugates. Other channels of kaon decay are neglected since their branching ratio is smaller and they would generate lower energy neutrinos.

The neutrino production for the flavor $\nu_\beta$  from the decay of $\mathcal{M}$ is given by~\cite{Hummer:2010vx}:
\begin{equation}
    Q'_{\nu_\beta}(E'_{\nu_\beta}) = \int_{E'_{{\nu_\beta}}}^{+\infty} Q'_{\mathcal{M}}(E'_{\mathcal{M}}) \left[ 1 - \exp{\left(- \frac{\tau'_\text{cool}}{\tau'_{l}} \right)}\right] \frac{1}{E'_{\mathcal{M}}} F_{\mathcal{M}\rightarrow \nu_\beta}\left(\frac{E'_{\nu_\beta}}{E'_{\mathcal{M}}}\right) d E'_{\mathcal{M}}\ .
\end{equation}
with $\nu_\beta$ representing one neutrino (or antineutrino, $\bar\nu_\beta$) species. The exponential term in the equation above takes into account the cooling of the decaying particle, with $\tau'_\text{cool}$ and $\tau'_{l}$ being the total cooling timescale and the mean lifetime of $\mathcal{M}$, respectively. The function $F_{\mathcal{M}\rightarrow \nu_\beta}\left({E'_{\nu_\beta}}/{E'_{\mathcal{M}}}\right)$ encapsulates the reaction kinematics and is defined as in Ref.~\cite{Lipari:2007su}. In particular, the muon decay into neutrinos depends on the muon helicity, as illustrated in Ref.~\cite{Lipari:2007su}. 

\subsection{Neutrino fluence at Earth} \label{sec:fluence}

En route  to the Earth, neutrinos change their flavor, with transition and survival probabilities defined as follows~\cite{Anchordoqui:2013dnh}: 
\begin{eqnarray}
    P_{\nu_e \rightarrow \nu_\mu} = P_{\nu_\mu \rightarrow \nu_e} &=& \frac{1}{4}\; \sin^2{2 \, \theta_{12}}\ ,\\
    P_{\nu_\mu \rightarrow \nu_\mu} &=&  \frac{1}{8} \;(4 - \sin^2{2\,\theta_{12}})\ ,\\
    P_{\nu_e \rightarrow \nu_e} &=&  1- \frac{1}{2} \; \sin^2{2\, \theta_{12}}\ ,
\end{eqnarray}
with $\theta_{12} = 33.5^\circ$~\cite{Esteban:2020cvm}. Transition and survival probabilities are the same assuming antineutrinos instead of neutrinos. Hence the neutrino fluence $F_{\nu_\beta}$ (in units of cm$^{-2}$ GeV$^{-1}$) is:
\begin{equation}
    F_{\nu_\beta}(E_{\nu_\beta}) = V'_{s}\; t_{\text{dur}}\; \frac{(1+z)^2}{4\pi d_L(z)^2} \sum_\alpha P_{\nu_\alpha\rightarrow \nu_\beta} \, Q'_{\nu_\alpha} \left(\frac{E_{\nu_\beta}(1+z)}{\Gamma} \right)\ .
\end{equation}
Within a flat $\Lambda$CDM cosmology, the luminosity distance is given by 
\begin{equation}
     d_L \;(z) = \frac{c \; (1+z)}{H_0} \int_0^z \frac{dz^\prime}{ \sqrt{\Omega_\Lambda + \Omega_M (1+z^\prime)^3}}\ , 
\end{equation}
where $H_0 = 67.4$~km s$^{-1}$ Mpc$^{-1}$, $\Omega_M \simeq  0.315$, and $\Omega_\Lambda \simeq 0.685$~\cite{Planck:2018vyg}.

\section{Neutrino production from nuclear photohadronic interactions} \label{sec:heavy_nuclei}

The outflows of GRB  jets may contain a certain fraction of nuclei~\cite{Beloborodov:2002af}. Nuclei can be synthetized along the jet, loaded at the base of the jet, or produced in the stellar ejecta and entrained in the jet. However, photodisintegration from high-energy photons and spallation from protons and neutrons may disintegrate such nuclei. In fact, Ref.~\cite{Horiuchi:2012by} concluded that  the survival of nuclei up to the  region of neutrino production is unlikely in HL-GRBs in the internal shock and photospheric models, but nuclei could easily survive in  magnetized jets or LL-GRBs due to the smaller  photon density. In this section, we introduce the method adopted to model jets containing nuclei through photohadronic interactions.

\subsection{Monte Carlo simulation of nuclear cascades}
In order to model the neutrino emission, interactions of 
nuclei in the jet should be taken into account. To this purpose, Refs.~\cite{Biehl:2017zlw, Boncioli:2018lrv} numerically solved a system of partial differential equations to compute the spectrum of any nucleus as a function of time and energy; while Ref.~\cite{Globus:2014fka} relied on  the SOPHIA event generator to compute energy loss and photodisintegration processes of accelerated nuclei. In the following, we first model nuclear disintegration and nuclear decay. Then, along the lines of Ref.~\cite{Globus:2014fka}, we outline a Monte-Carlo algorithm that generates nuclear cascades, the Nuclear Cascade Monte Carlo (NCMC). The steps leading to the production of neutrinos, presented in this section, are illustrated in Fig.~\ref{fig:flow_chart}. 

 The main difference between our approach and the one adopted in Ref.~\cite{Globus:2014fka} is that the latter focuses on  acceleration, propagation and emission of ultra-high energy cosmic rays using a Monte-Carlo algorithm  that simulates the  trajectories of nuclei in the acceleration region of the jet. Their algorithm relies on the photonuclear cross sections derived by Refs.~\cite{Khan:2004nd, Rachen:1996zeh}.
Conversely, the goal of this work is to use NCMC to take into account photodisintegration events and follow the energy distribution of the different nuclear species within the acceleration region.
 We rely on the  empirical cross-section model from Ref.~\cite{Morejon:2019pfu}; despite the different approach (NCMC vs.~solution of partial differential equations), the employment of  different cross-section models is  the main source of difference between our findings and the ones of  Ref.~\cite{Biehl:2017zlw}, which instead  assumed a single-particle model (cf.~Appendix~\ref{App:Comparison_Biehl} for  details).

\subsubsection{Nuclear photodisintegration} \label{sec:photodis_model}
Nuclei undergo  photon-nucleus and nucleus-nucleus collisions. However, since  the photon density in the jet is several orders of magnitude larger than the  density of nuclei, we  only focus on photon-nucleus interactions and neglect  any nucleus-nucleus collision.

Photonuclear interactions are fundamentally different than  $p\gamma$ and $n\gamma$ interactions. Indeed, nuclei are made of bonded nucleons.  Photons with enough energy can  break the nuclear bond, leading to the emission of one or several nucleons. We should distinguish between two ranges for the energy of the incident photon in the rest frame of the nucleus: The Giant Dipole Resonance (GDR) with $E_{\gamma,r} \in [E^\mathrm{th}_{\gamma, r}, 140 \; \text{MeV}]$ and the photomeson (PM) range for $E_{\gamma,r} > 140 \;\text{MeV}$ \cite{Morejon:2019pfu}. The GDR threshold energy $E^\mathrm{th}_{\gamma, r}$ varies as a function of $(Z,A)$, but it is approximately $10$~MeV~\cite{Stecker:1998ib}.  

Data from photonuclear reactions are relatively scarce compared to $(p,n)\gamma$ interactions.
However, a few models for the cross-section and branching ratios
for any nucleus with  mass number $A \leq 56$ exist. 
Major improvements on the estimation of these cross-sections have been made thanks to the Monte-Carlo event generation software Geant4~\cite{GEANT4:2002zbu}. We adopt fits from Refs.~\cite{2002EPJA...14..377K} and \cite{Morejon:2019pfu} for the GDR and PM ranges of the cross section. Regarding the branching ratios of these processes, we draw from  Refs.~\cite{Puget:1976nz} and  \cite{Morejon:2019pfu} for  the giant dipole resonance and the photomeson energy  ranges, as illustrated in Appendix~\ref{App:MCNC_params} (this appendix presents   an extensive description of the modeling of the cross section, the branching ratios and the adopted parameters).

\subsubsection{Nuclear decay} \label{sec:nuclear_decay}
Any stable nucleus that loses nucleons may become a radioactive isotope. We have refined the NCMC by taking into account both lifetimes and radioactive decay mechanisms of any isotope of mass number $A \leq 56$. To this purpose, we have retrieved nuclear decay data  from Ref.~\cite{IAEA} (cf.~Appendix~\ref{App:Radioactivity}). 

When a nucleus undergoes  photodisintegration, we compute its new isotope lifetime at rest $t_l$, randomly extracting    a time from a distribution following the exponential probability density:
\begin{equation} \label{eq:pdf_decay_lifetime}
    p(t_{l}) = \frac{1}{\bar{t_l}}\, e^{ - \frac{t_l}{\bar{t_l}}} \ , 
\end{equation} 
with  $\bar{t_l} $ being the mean decay time of the isotope at rest. The lifetime  in the reference frame of the jet is 
\begin{equation} \label{eq:lifetime}
    t'_{l} = \frac{E'_A}{m_A c^2} \; t_{l}\ , 
\end{equation} 
where $t'_l$ is the comoving lifetime of the isotope in the jet frame. Assuming that  the isotope is created at $t'=0$, if  photodisintegration does not  happen at $t'_{l}$, the isotope  decays as illustrated in  
Appendix~\ref{App:Radioactivity}.  

\begin{figure}
    \centering
    \includegraphics[width = \linewidth]{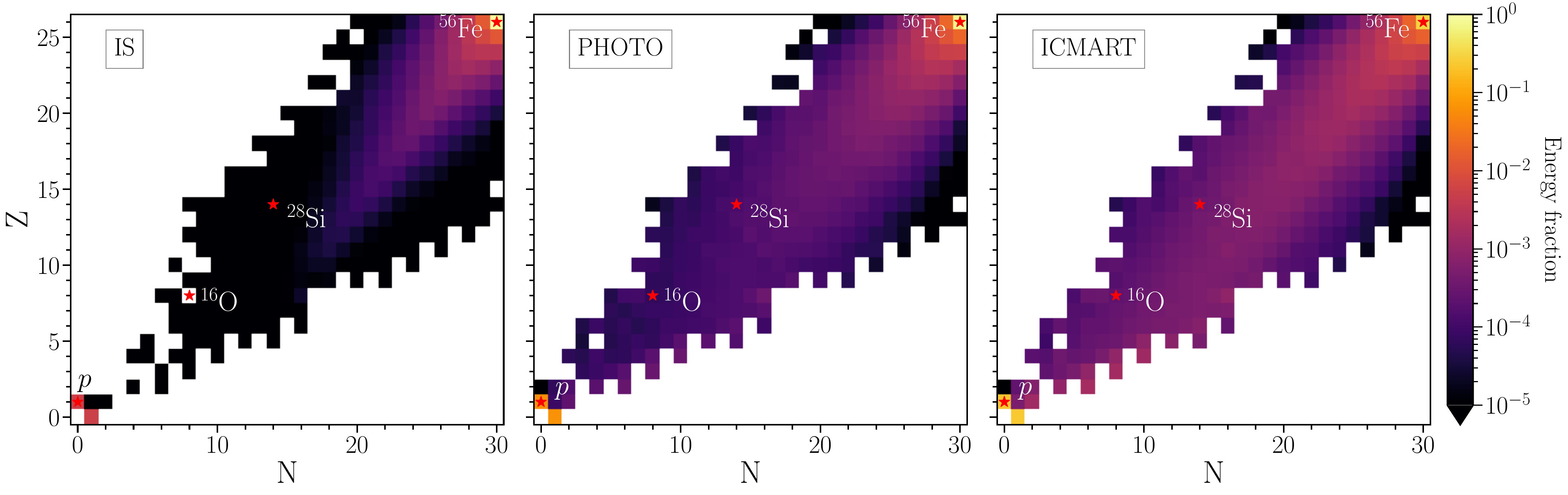}
\caption{Contour plot of the fraction of the total energy (initially injected in $^{56}$Fe) carried by the different isotopes in the plane spanned by $N$ and $Z$ for our benchmark GRB (cf.~Table~\ref{tab:initial_conditions} for Fe-jet). The results for the internal shock, photospheric and ICMART models are plotted from left to right, respectively. In order to guide the eye, $^{56}$Fe, $^{28}$Si, $^{16}$O, and $p$ are marked with red stars. The internal shock model does not favor efficient nuclear cascades with respect to the other two models. The ICMART model leads to the most efficient cascades, which in turn determine the largest neutrino emission as shown in Fig.~\ref{fig:fluence_comp}.}
    \label{fig:e_density_cascade}
\end{figure}

Following the computed branching ratios, we can see that it is unlikely for a nucleus to stray from the valley of stability, as shown in Fig.~\ref{fig:e_density_cascade}. Since we are interested in the high-energy tail of the spectral energy distribution of nuclei, the lifetime of radioactive nuclei in the comoving  frame is large (see Eq.~\ref{eq:lifetime}). This implies that the  nucleus is relatively stable and neutrino emission  from $\beta^+/ \beta^-$ processes from radioactive decays is negligible with respect to  photohadronic interactions~\cite{Biehl:2017zlw}. The only exception is the $\beta^-$ decay of free neutrons, which we take into account given that the related radioactive isotope  can  be abundant, see for example Fig.~\ref{fig:e_density_cascade}.

\subsubsection{Monte Carlo algorithm and particle spectra}

Accelerated nuclei can be injected  anytime during the interaction time interval, $t'_\text{shock} = t'_{\text{dur}} / N_\text{shock}$. This implies that we can compute  a random interaction time for each Monte-Carlo event  as 
$t'_\text{eff} = t'_{\text{shock}} \; U(0,1)$,
with $U(0,1)$ being the uniform probability density function between $0$ and $1$. For each Monte-Carlo event, $t'_{\text{eff}}$ is split into $N$ equal time steps 
    $\Delta t' = t'_{\text{eff}} / N$. 
Hence, the probability of a photonuclear interaction for a nucleus $^A _Z X$ of energy $E'_A$, during the time interval $\Delta t'$ in the comoving frame of the jet is:
\begin{equation}
\label{eq:pAgamma}
     p_{A\gamma}(E'_A, A, Z) = R'_{A\gamma} (E'_A, A, Z) \; \Delta t'\ ,
\end{equation}
with $ R'_{A\gamma}$ being the photonuclear interaction rate computed according to Eq.~\eqref{eq:tau_Agamma}. At each time step, nuclei can  interact or not with  photons according to Eq.~\eqref{eq:pAgamma}. If an interaction occurs,  one of the physical processes described in  Sec.~\ref{sec:photodis_model} is randomly selected according to the branching ratios. In parallel, nuclei lose energy at each time step, following the adiabatic cooling; excluding the photodisintegration cooling which is already taken into account through the NCMC, adiabatic processes always dominate the total cooling timescale, as shown in Fig.~\ref{fig:timescales}. 
The photonuclear cooling timescale due to meson production ($\tau'_{A\gamma \rightarrow \pi}$) should not be confused with the interaction rate $R'_{A\gamma}$ displayed  in Fig.~\ref{fig:timescales} ($\tau'_{A\gamma} \approx \chi_A R'_{A\gamma} \sigma_{A\gamma \rightarrow \pi } / \sigma_{A\gamma, \mathrm{tot}} $, where $\chi_A \approx 1/A$ is the average energy loss after meson production and $\sigma_{A\gamma \rightarrow \pi } / \sigma_{A\gamma, \mathrm{tot}} \approx 0.1$ is the branching ratio~\cite{Morejon:2019pfu}). The $A\gamma \rightarrow \pi $ cooling is about three orders of magnitude below the interaction rate, displayed  in Fig.~\ref{fig:timescales}; this explains why it is subleading  in all cases considered in this work.
If no interaction takes place, when the nucleus is unstable it radioactively decays, as described in  Sec.~\ref{sec:nuclear_decay}.

We  assume that, after a photonuclear interaction that generates $(N_p, N_n)$ protons and neutrons, only nucleons and no nuclei can be emitted. For instance, the loss of $2$ protons and $2$ neutrons  generates  $2$ free protons and $2$ free neutrons, and not $_2^4$He. 
The energy of the produced nucleon $\mathcal{N}=(p,n)$ is
    $E'_\mathcal{N} = {E'_A}/{A}$,
with $E'_A$ being the energy of the parent particle $^A _Z X$ in the comoving frame of the jet. 
 For the sake of simplicity, we do not  implement a probability distribution function taking into account  all  possible configurations of photodisintegration of nuclei.
Overall, this approximation results in an underestimation of the density of intermediate nuclei in the region of neutrino production when nuclear cascades are moderately efficient. In a more realistic modeling, a fraction of the nucleons produced in photodisintegration events should have been emitted as intermediate nuclei (see Fig.~\ref{fig:compar_biehl_density} and the related discussion in Appendix~\ref{App:Comparison_Biehl}).

For any accelerated nuclear species, the spectral energy distribution $n'_A(E'_A)$ (in units of cm$^{-3}$ GeV$^{-1}$) of the seed population is adopted to draw the injected nuclear species in the NCMC. 
For simplicity, we only inject  pure isotopes. 
In order to do this, we randomly select the initial energy of the seed nuclei following a uniform probability in logarithmic scale. Then, for any given energy, each seed nucleus is associated to a certain density computed following $n'_A(E'_A)$. The nuclear cascade is then generated and we retrieve the density and final energy of each isotope eventually created. 
We then add each contribution to the final spectra. For more details, we refer the reader to Appendix \ref{App:comp_spectra}.

\subsection{Nuclear photohadronic interactions} \label{sec:photomeson}
It is now possible to compute the neutrino production associated to each isotope, while for the production of mesons from protons and neutrons, we rely on Sec.~\ref{sec:neutrino_from_pn}.
For nuclei with mass number $A \leq 56$, 
the meson production $Q'_\mathcal{M}(E'_\mathcal{M})$ (in units of cm$^{-3}$ s$^{-1}$ GeV$^{-1}$) is given by~\cite{Hummer:2010vx}:
\begin{equation} \label{eq:Q_formula}
    Q'_\mathcal{M}(E'_\mathcal{M}) = \sum_{\Delta E}  n'_A\left(\frac{E'_\mathcal{M}}{\chi_\mathcal{M}^{\Delta E}}\right) \; \frac{m_A c^2}{E'_\mathcal{M}} M_{\mathcal{M}}^{\Delta E}\; \int_{\Delta E} f^{\Delta E}(y) \; n'_\gamma \left(y \frac{ \chi_\mathcal{M}^{\Delta E} m_A c^2}{E'_\mathcal{M}} \right) dy\ ,
\end{equation}
where $\mathcal{M}$ is any meson (i.e., pions and kaons) and the sum runs over  $\Delta E$, which is  an energy interval where the multiplicity and the fraction of the energy transferred to produced mesons are assumed to be constant (see Appendix~\ref{App:photomeson}).

Each interaction type is characterized by three parameters: $\chi_\mathcal{M}^{\Delta E}$ represents the amount of energy transferred from the seed nucleus to the newly created meson  ($ E'_\mathcal{M} = \chi_\mathcal{M}^{\Delta E} E'_A$); the multiplicity, $M_{\mathcal{M}}^{\Delta E}$, describes how many mesons are created on average; $f^{\Delta E}(y)$ is  computed as follows: 
\begin{equation}
    f^{\Delta E}(y) = \frac{1}{2y^2}\; \int_{E_{\gamma, r}^{\mathrm{th}, \Delta E} }^{2y} \;E_{\gamma,r} \;\sigma_{A\gamma\rightarrow \mathcal{M}}(E_{\gamma,r}) \; d E_{\gamma,r}\ , 
\end{equation}
with $E_{\gamma, r}^{\mathrm{th}, \Delta E}$ and $\sigma_{A\gamma \rightarrow \mathcal{M}}(E_{\gamma,r})$ being the threshold energy and the inelastic cross-section of photohadronic interactions for meson production, respectively. 

In order to compute the cross section, previous work (e.g.~Ref.~\cite{Biehl:2017zlw}) relied on  the Single Particle Model, according to which 
$\sigma_{A\gamma\rightarrow \mathcal{M}}(E_{\gamma,r}) \simeq A \; \sigma_{p\gamma\rightarrow \mathcal{M}}(E_{\gamma,r})$. 
However, since Ref.~\cite{Morejon:2019pfu} showed that the Single Particle Model can be inaccurate  in the photomeson regime, 
we adopt  the  empirical model proposed in Ref.~\cite{Morejon:2019pfu}, which provides more accurate results for  nuclei up to $^{56}$Fe. 
The function $f^{\Delta E}(y)$ is then computed using the empirical model total cross-section in the photomeson range (see Appendix~\ref{App:photomeson} for details). 
The branching ratio for the photohadronic production is extracted from  estimations of the cross-sections from Ref.~\cite{Morejon:2019pfu}.
Due to the lack of data, we model $M_{\mathcal{M}}^{\Delta E}$ and $\chi_\mathcal{M}^{\Delta E}$ adapting the method employed to compute $p\gamma$ interactions in  Ref.~\cite{Hummer:2010vx}. The average values of $M_{\mathcal{M}, p\gamma}^{\Delta E}$ and $\chi_{\mathcal{M}, p\gamma}^{\Delta E}$ are taken into account for each $\Delta E$, 
and we  define $\chi_\mathcal{M}^{\Delta E} = \chi_{\mathcal{M}, p\gamma}^{\Delta E} \; / A $, $M_{\mathcal{M}}^{\Delta E} = M_{\mathcal{M}, p\gamma}^{\Delta E}$.

For the sake of simplicity, we assume that each nucleus contains an equal number of  protons and  neutrons. This approximation is reasonable given that it is unlikely for a nucleus to stray away from the valley of stability (which approximately corresponds  to $N = Z = A/2$ for $A\leq 56$). This symmetry enables us to get comparable production of $\mathcal{M}$ and $\bar{\mathcal{M}}$, reducing the computational time. For details on the inputs adopted for the different parameters, we refer the interested reader to Appendix~\ref{App:photomeson}. Finally, the decay of each meson into neutrinos is computed as in Sec.~\ref{sec:neutrino_prod} and the neutrino fluence expected at Earth follows Sec.~\ref{sec:fluence}, respectively.

\section{Impact of the jet composition on the neutrino emission} \label{sec:results}
In this section, we present our findings on the dependence of the neutrino signal on the jet composition. We investigate how the jet composition impacts the neutrino emission for the internal shock, photospheric and ICMART jet models. We then explore how the jet composition affects the neutrino emission across the allowed jet parameter space. 

\subsection{The role of the jet composition in kinetic and Poynting flux dominated jets}
\label{sec:impact_composition}
In order to facilitate a comparison, we consider the   neutrino fluence obtained for a jet  loaded with protons ($p$-jet) and a jet loaded with  $^{56}$Fe (Fe-jet) for the internal shock, photospheric and ICMART jet models. Jets with mixed composition or loaded with intermediate isotopes would have a neutrino fluence falling in between these two extreme cases. We also note that it is highly unlikely to have a jet mostly composed by iron, yet we investigate this case in order to explore the largest variability of the expected neutrino signal according to the jet composition. We otherwise assume  identical jet properties (cf.~Table~\ref{tab:initial_conditions});  in particular, the  total energy fraction going into accelerated protons is the same as the one going into accelerated iron. This  implies that the same power law index is chosen for both accelerated protons and iron nuclei (see Sec.~\ref{sec:param_space} for an extensive discussion about this hypothesis).

We note that the survival of nuclei up to the  region of neutrino production may be unlikely for our benchmark model, especially for kinetic dominated jets~\cite{Horiuchi:2012by}. Yet, we  rely on this benchmark jet to  assess the impact of the jet composition for illustrative purposes.

For the internal shock model, in order to test our NCMC approach, we  compare our results with the  three benchmark cases presented in Ref.~\cite{Biehl:2017zlw}. In all cases, we find excellent agreement (results not shown here).  We also find comparable  neutrino fluences using the Single Particle Model, yet we prefer to employ the more accurate photohadronic model in the rest of this paper (see Sec.~\ref{sec:photomeson}).

\begin{figure}
\centering
     \vspace{-1cm}
     \includegraphics[width = 0.35\linewidth]{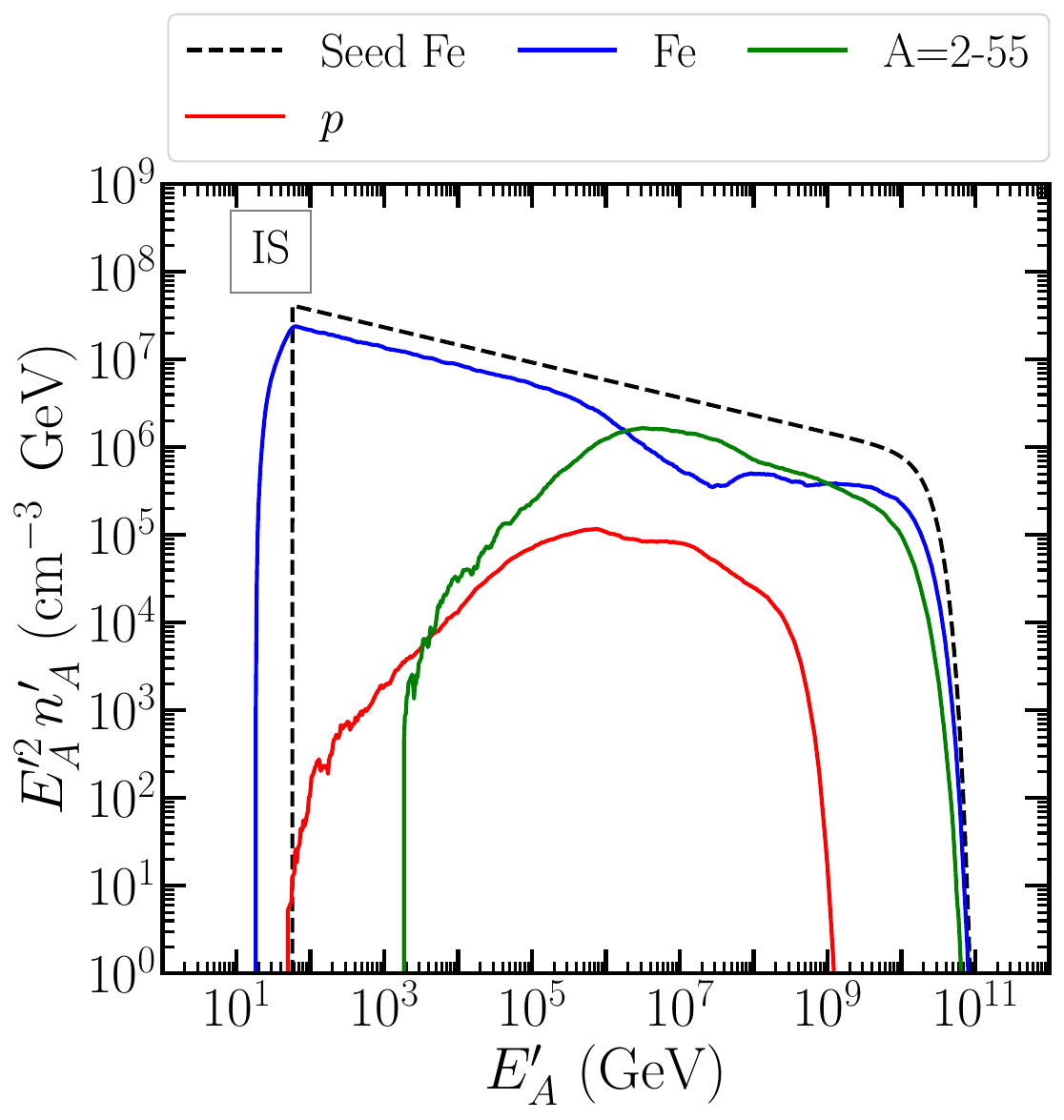}
     \includegraphics[width = 0.37\linewidth]{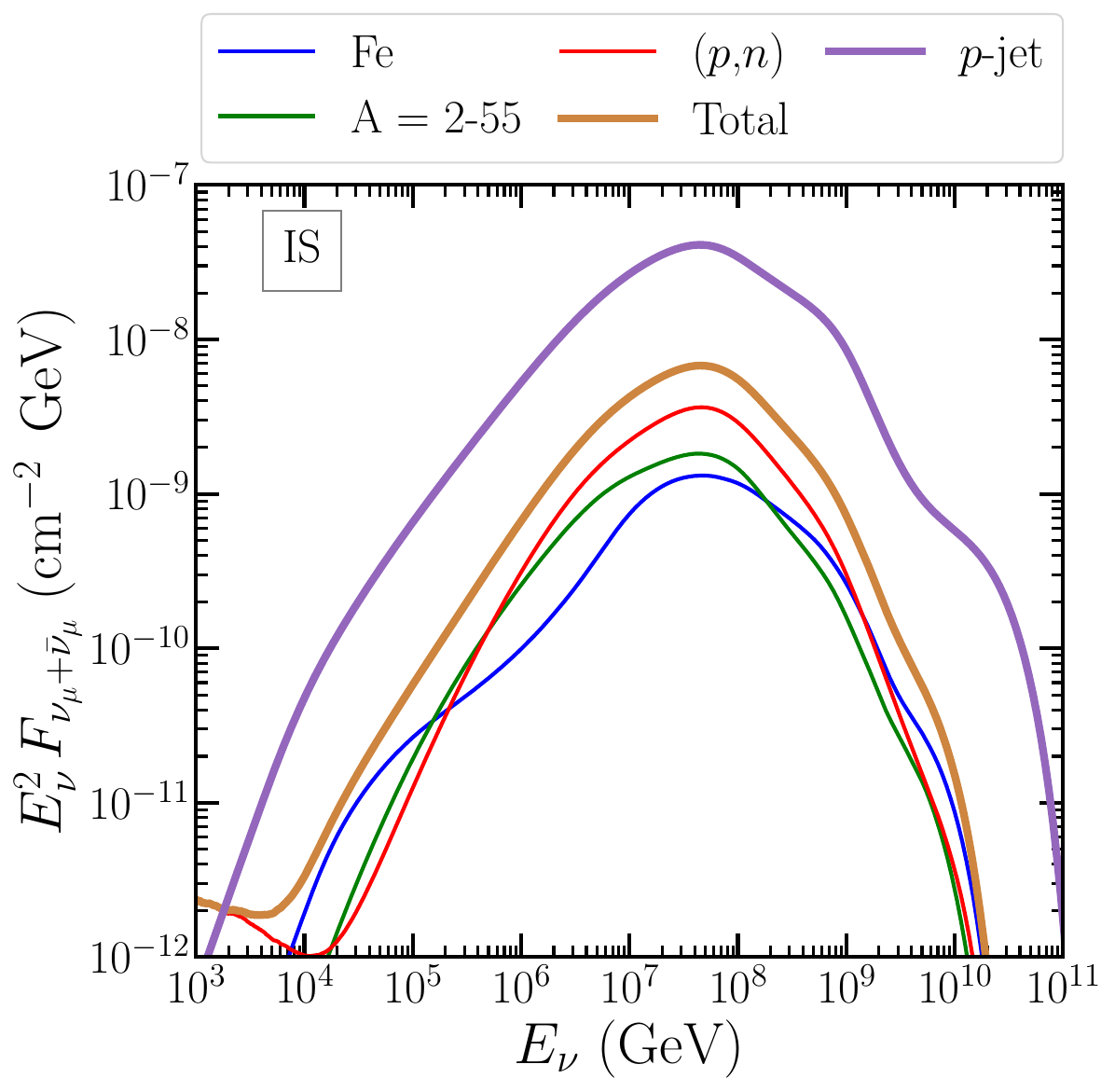} \\
     \includegraphics[width = 0.35\linewidth]{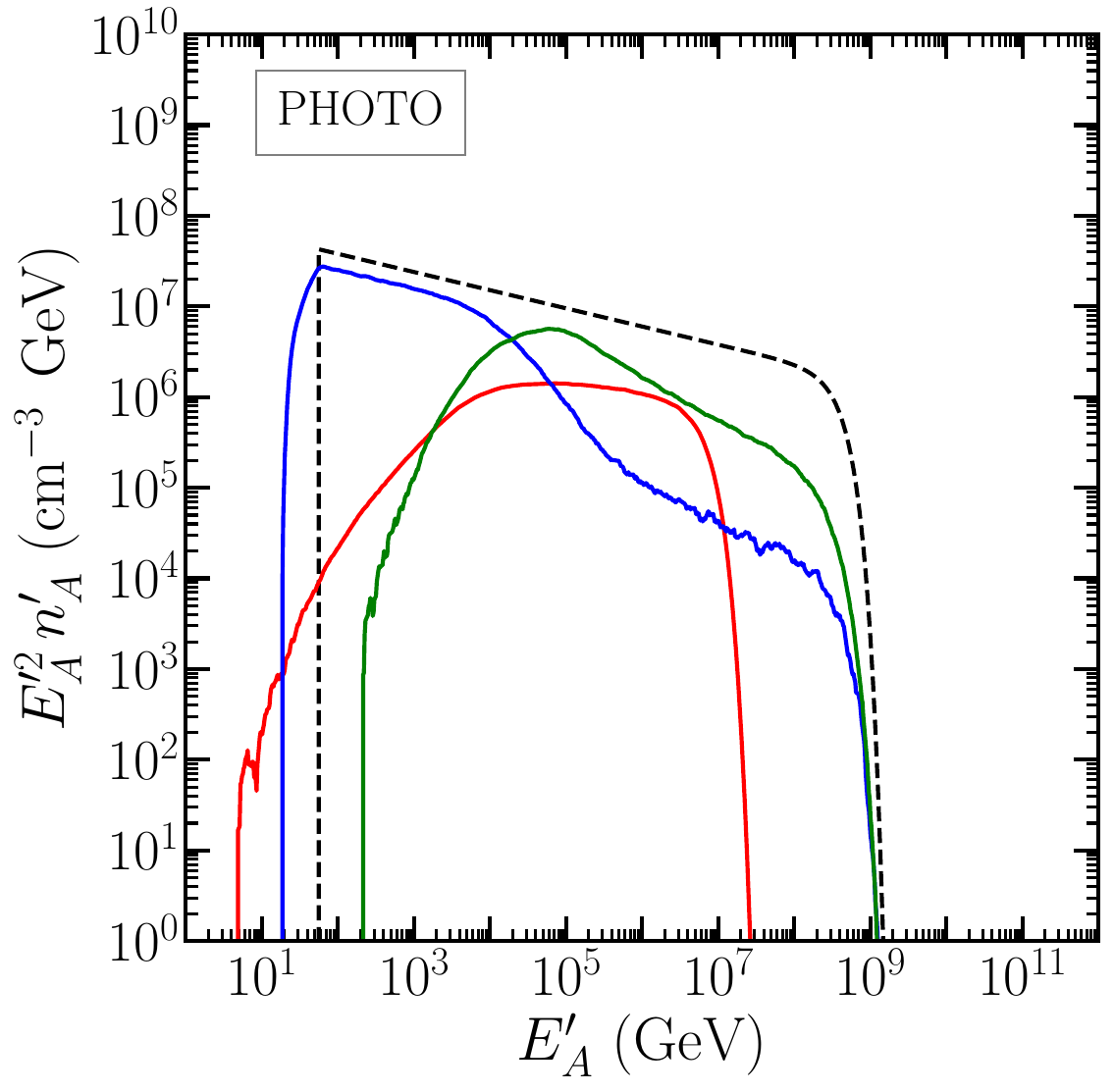}
     \includegraphics[width = 0.37\linewidth]{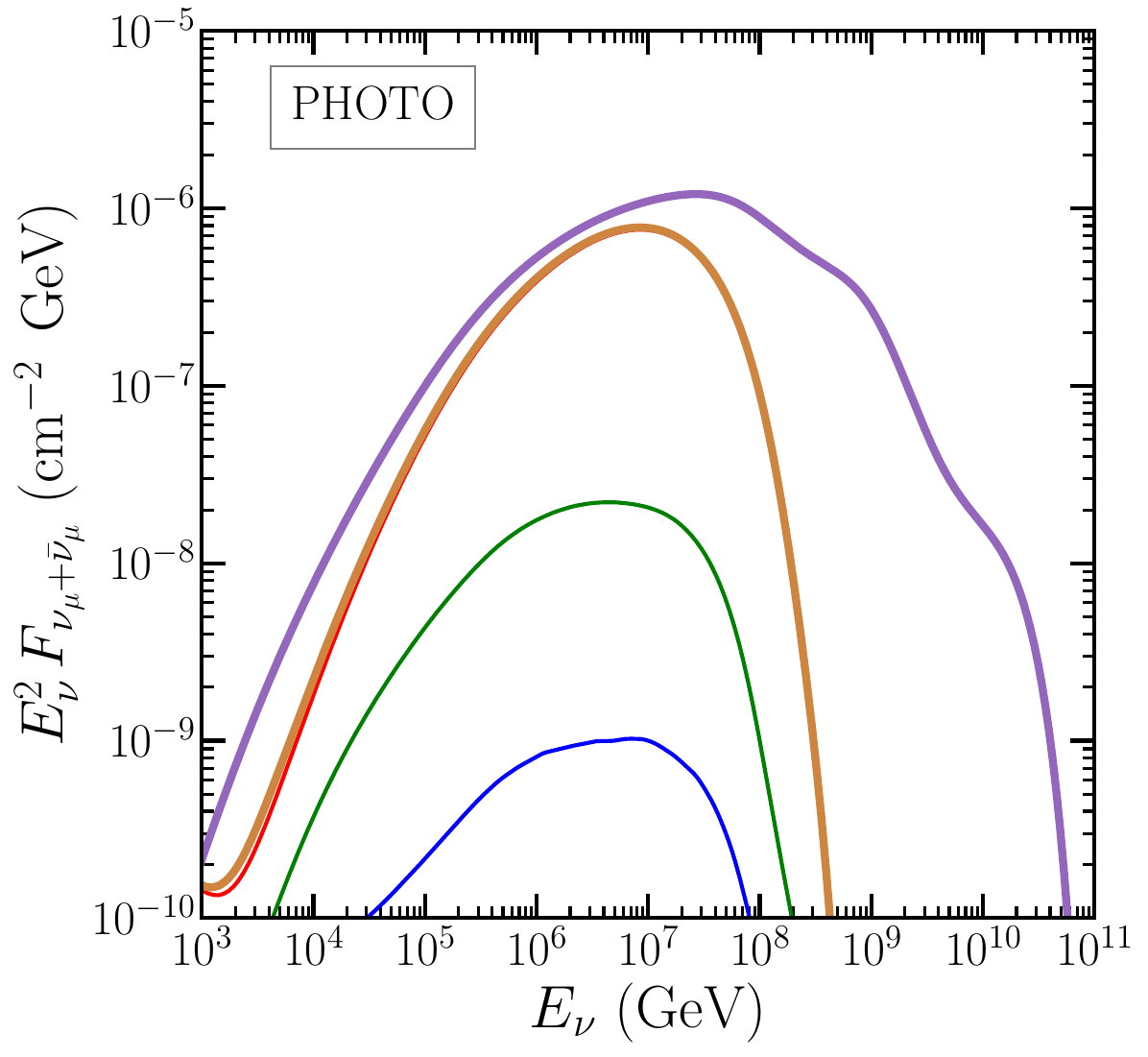} \\
    \includegraphics[width = 0.35\linewidth]{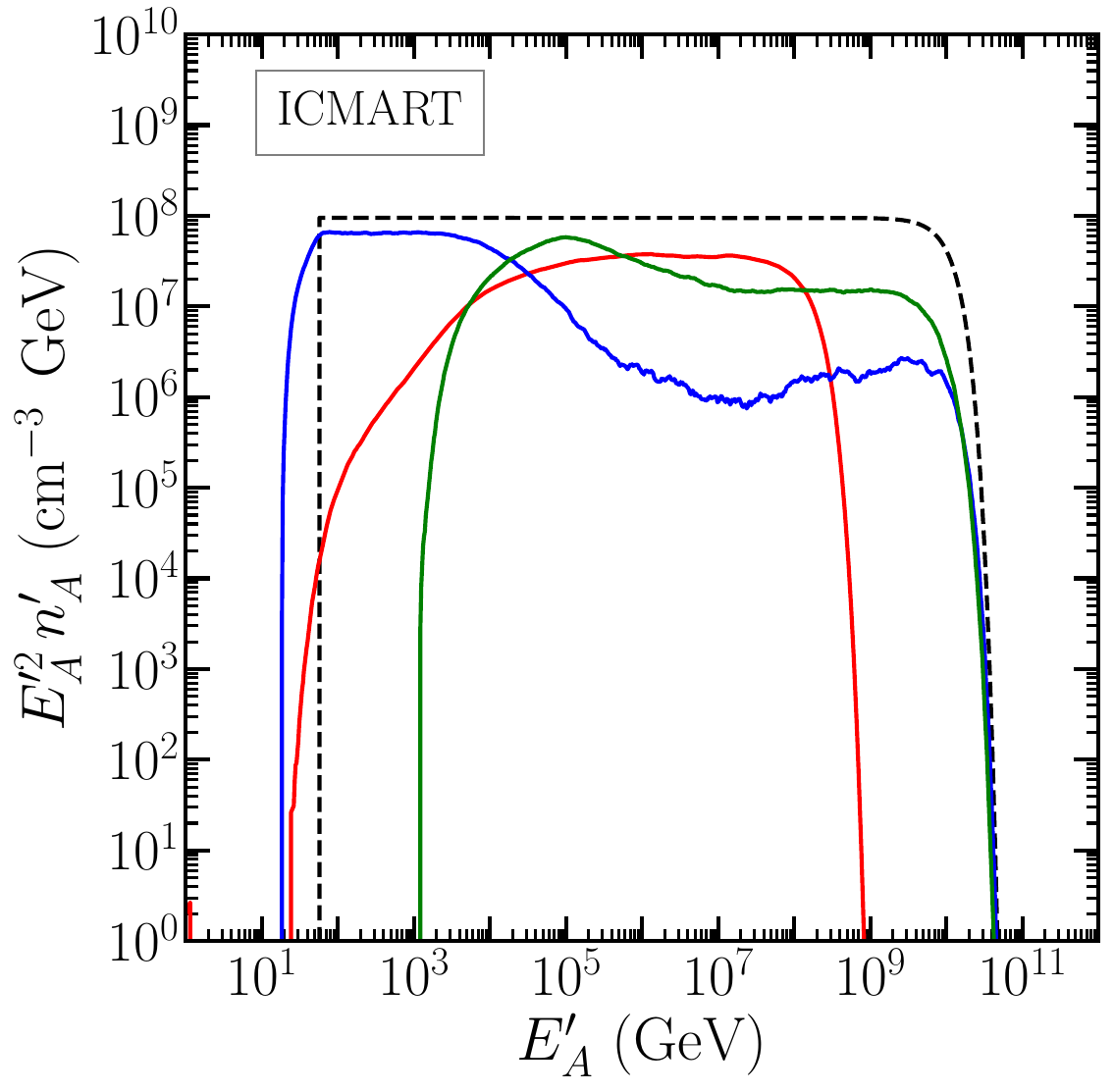}
     \includegraphics[width = 0.37\linewidth]{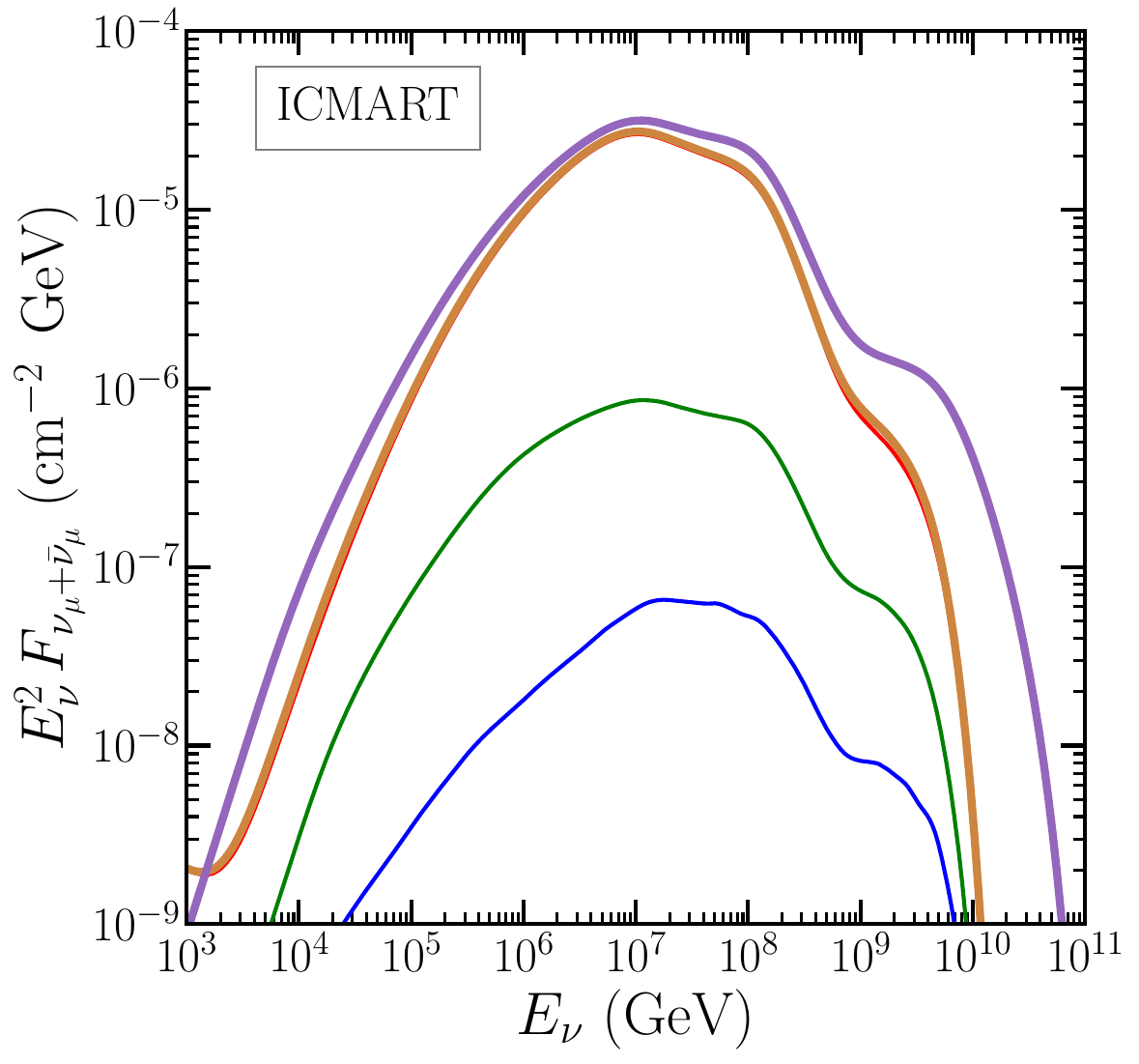}
    \caption{\textit{Left:} Spectral energy density of nuclei as a  function of the particle energy for our benchmark GRB (cf.~Table~\ref{tab:initial_conditions}) in the case of  Fe-jet, for the internal shock, photospheric, and ICMART models, from top to bottom respectively.   
    \textit{Right:} $\nu_\mu + \bar{\nu}_\mu$ neutrino fluence as a function of the neutrino energy for our benchmark GRB, in the case of a $p$-jet (indigo curve) and Fe-jet (the ochre curve represents the total neutrino emission). The relative contributions to the neutrino fluence from iron, all other lighter nuclei and nucleons are plotted in blue, green and red respectively (cf.~spectra of the corresponding parent particles in the left panels). If the density of iron nuclei is larger than the one of all other nuclei, neutrinos are produced mostly through iron photohadronic interactions.  Since  meson production from nuclei is less efficient than the one from protons and neutrons, the neutrino fluence for Fe-jet is smaller than for $p$-jet, independent of the jet model. Nuclear cascades are more efficient for the photospheric and  ICMART models than for the internal shock one; this leads to larger neutrino fluence than in the internal shock model for  Fe-jet.}
    \label{fig:fluence_comp}
\end{figure}
Figure~\ref{fig:fluence_comp} shows the spectral energy density  of nuclei (left panels) and the neutrino fluence (right panels) for the internal shock model, the photospheric model and the ICMART model, from top to bottom respectively. Comparing Fe-jet (ochre line) with the  $p$-jet (indigo line), we can see that the neutrino fluence is lower for  Fe-jet than for the $p$-jet up to one order of magnitude, independent of the jet model. 
For the internal shock case, the density of nuclei is  larger than the one of nucleons (cf.~blue/green vs.~red curves in the left panels), but neutrino production from nuclei is comparable to the one driven by  protons/neutrons. Indeed, since the meson production from nuclei is not so efficient as the one from protons or neutrons, the neutrino fluence is smaller than what could be expected given the associated nuclei densities. 
The shape of the neutrino fluence is overall the same for jets, independent of the composition. In particular, the maximum of $E^2_\nu F_{\nu_\mu+\bar\nu_\mu}$ is achieved at about the same energy $E^M_\nu \simeq 5\times10^7$~GeV.

For  the photospheric model, cascades are more efficient than in the internal shock  model due to the larger photon luminosity, 
leading to larger proton and neutron densities  (cf.~Fig.~\ref{fig:e_density_cascade}  which displays the fraction of the total energy carried by lighter isotopes and nucleons for  Fe-jet). The ICMART model  shows efficient nuclear cascades, as shown in Fig.~\ref{fig:e_density_cascade}, which lead to the largest neutrino fluence across all three jet models. However, compared to the photospheric model, the maximum fluence is achieved at the same energy for both $p$- and Fe-jets, and no sharp cutoff at high energies is observed. This is explained because  the crossing between the acceleration and cooling timescales happens at larger energies in the ICMART model since the magnetic field is stronger than in the photospheric model.  Hence, the acceleration processes are more efficient and the maximum energy per nucleon inside an iron nucleus is similar to the maximum energy found for protons.

Our findings are in general agreement with the ones presented in  Fig.~7 of Ref.~\cite{Biehl:2017zlw}, our internal shock case would correspond to their  ``empty cascade'' scenario, while our photospheric and ICMART models would correspond to their ``populated cascade'' scenario. 
The different  microphysical parameters among the internal shock, the photospheric, and the ICMART models (see Tab.~\ref{tab:initial_conditions}) do not allow for  comparable photon densities across models. A lower photon density results in a reduced efficiency of nuclear cascades and neutrino production,  which is the case for the internal shock model. Additionally, the radius $R_\gamma$ (where  high-energy neutrinos and gamma-rays are produced) is a function of $\Gamma$ and $\Tilde{t}_\mathrm{var}$ for the internal shock and photospheric models, while it is  constant for the ICMART model. The location of $R_\gamma$  affects the neutrino production since the density of nuclei and photons available for photohadronic interactions strongly depends on such radius, namely lower densities are found at higher radii (as we will show in Fig.~\ref{fig:fluence_gamma_var}).
However, our results differ for the ``inefficient cascade'' scenario of Ref.~\cite{Biehl:2017zlw} because of the empirical cross-section model adopted in this work (vs.~the Single Particle Model adopted in Ref.~\cite{Biehl:2017zlw}).
Following Ref.~\cite{Morejon:2019pfu}, the total photonuclear cross-section is roughly proportional to $A$ up to high photon energies. However, the  cross-section for meson production, taking into account the branching ratio and pion multiplicity, is proportional to $A^{2/3}$. This implies  a difference between the empirical and the single-particle model cross-sections used when computing the neutrino fluence. 
We find a difference of about one order of magnitude in the related neutrino fluence between the two approaches, see Appendices~\ref{App:photomeson} and \ref{App:Comparison_Biehl} for more details. This difference in the resulting neutrino fluence obtained  adopting  the empirical model or the single-particle model is consistent with the findings of   Ref.~\cite{Morejon:2019pfu}.

\begin{figure}
    \centering
    \includegraphics[width = 0.49\linewidth]{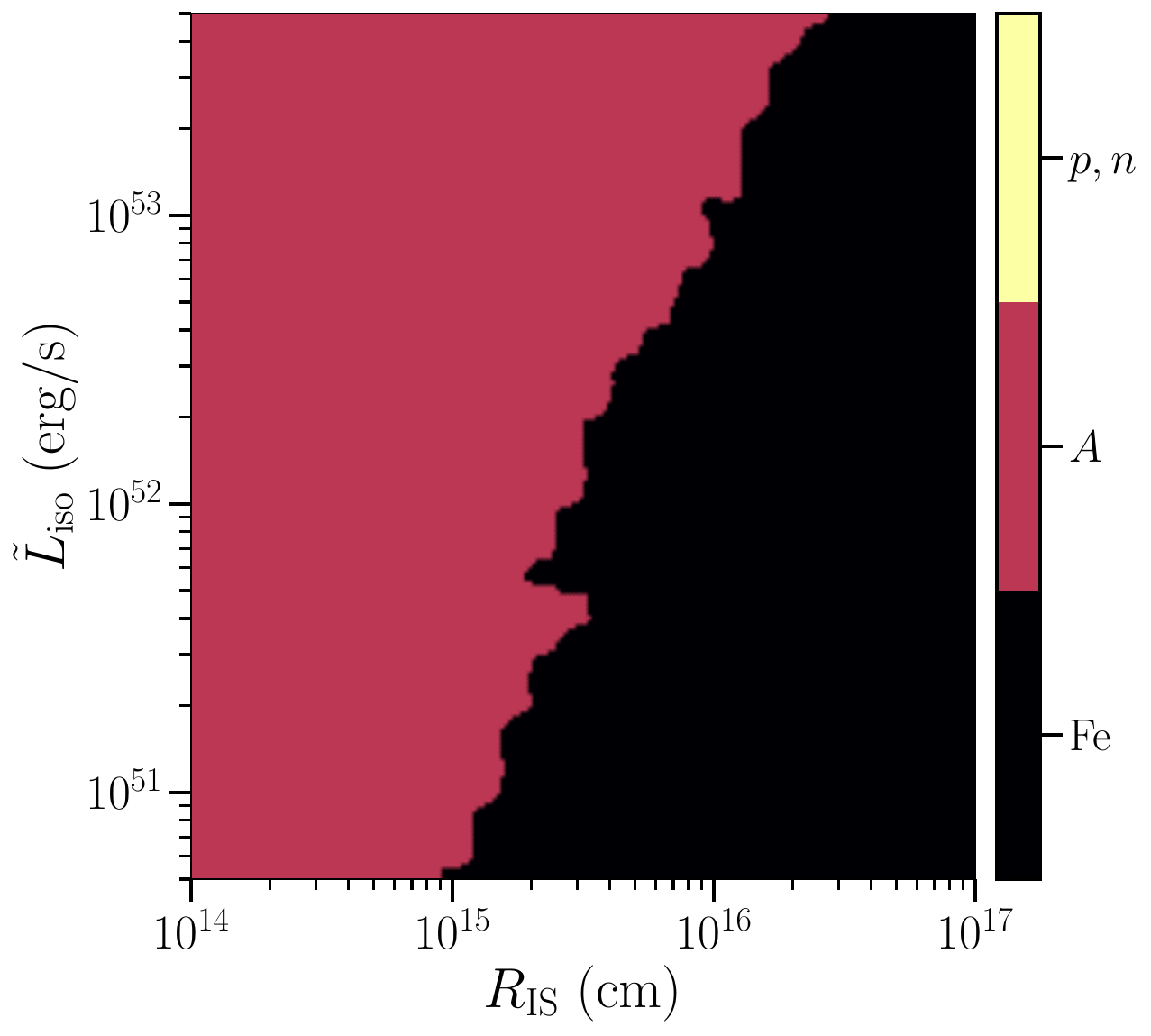}
    \includegraphics[width = 0.5\linewidth]{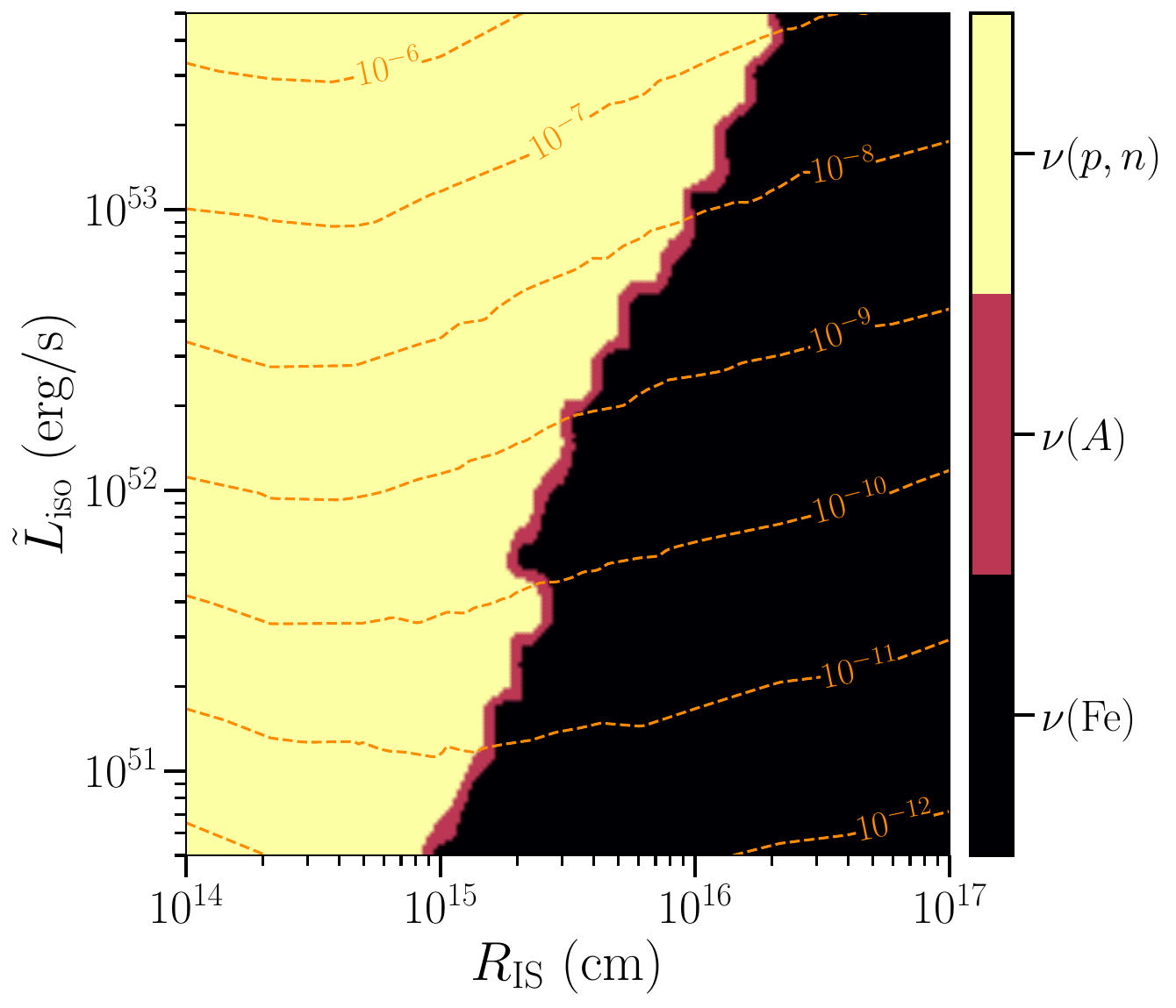}
    \caption{Nuclear cascade efficiency in the plane spanned by  the radius ($R_\mathrm{IS}$) of the interaction region and the isotropic luminosity ($\Tilde{L}_{\rm{iso}}$) for the photospheric model. \textit{Left:} The contours represent the region where the related isotopes (iron in black, any other lighter nucleus in magenta, and protons and neutrons in yellow) contribute the most to the total density. \textit{Right:} Same as the left panel, but it represents the dominant contribution to the total neutrino fluence. Dashed orange lines indicate  isocontours of $E_\nu^2 F_{\nu_\mu + \bar{\nu}_\mu}$ (in units of GeV cm$^{-2}$). For low luminosity and/or large radius, nuclear cascades are not efficient and iron dominates both the energy density of nuclei and the neutrino fluence. As the luminosity increases and/or the radius decreases, nuclear cascades are more efficient, and  intermediate isotopes can dominate the nuclear energy density with a resulting larger neutrino fluence. However,  protons and neutrons are responsible for most of the total neutrino fluence despite their lower density.}
    \label{fig:L_R_variation}
\end{figure}
Figure~\ref{fig:L_R_variation} highlights the  variation of nuclear cascade efficiency in the plane spanned by the  radius of interaction and the GRB luminosity. For illustrative purposes, the photospheric model is adopted as a proxy since the overall structure would be the same for the internal shock model as well, while the ICMART model would correspond to a slice in this plot drawn for constant radius. For low luminosity and/or large radius, nuclear cascades are not efficient and iron dominates both the particle density and the neutrino fluence. As discussed earlier (see Fig.~\ref{fig:fluence_comp}), we expect that the total neutrino fluence in this region is approximately reduced by one order of magnitude compared to a $p$-jet. As the luminosity increases and/or the radius decreases, nuclear cascades become more efficient, and we reach a region where intermediate isotopes dominate the nuclear energy density. However, except for a thin region (cf.~the right panel of Fig.~\ref{fig:L_R_variation}, magenta region), protons and neutrons are responsible for most of the total neutrino fluence despite their lower density. This is because nucleons are more efficient at producing neutrinos from photohadronic interactions than  nuclei.

\subsection{Dependence of the neutrino fluence on the injection spectral index of the parent nuclei and the jet Lorentz  factor} 
\label{sec:param_space}
We now extend our investigation of the dependence of the neutrino emission on the jet composition beyond our benchmark GRB jet, considering the variation of the neutrino fluence as a function of the bulk Lorentz factor ($\Gamma$) and the injection spectral index of the seed accelerated particle spectra ($k$). 
We choose  to compute the neutrino fluence for  $\Gamma = 10$--$1000$. Regarding the power law index, for relativistic shocks (internal shock and photospheric models), theoretical work predicts 
$k \simeq 2.2$~\cite{Achterberg:2001rx} and particle-in-cell  simulations find $k\simeq 2.0$--$2.5$~\cite{Groselj:2024dnv,Sironi:2013ri}; for magnetized jets, particle-in-cell  simulations suggest $k\simeq 2.0$~\cite{Zhang:2023lvw,Sironi:2014jfa}. 
It is crucial to highlight  that most of the existing work does not consider  populations of heavier nuclei because particle-in-cell simulations with heavier nuclei are  computationally intensive. 
In Ref.~\cite{Sironi:2013ri}, it has been found that the power law index does not depend strongly on the mass of the ion for ion masses ranging from $m_i = 25 \;m_e$ to $m_i = 1600 \;m_e \simeq m_p$. 
Thus, we extrapolate from this result, and consider that the power law index for an iron population would be similar to the one of a proton population and explore the efficiency of neutrino production  for  $k = 2.0$--$3.0$  for  $p$- and Fe-jets.

\begin{figure}
        \includegraphics[width = \linewidth]{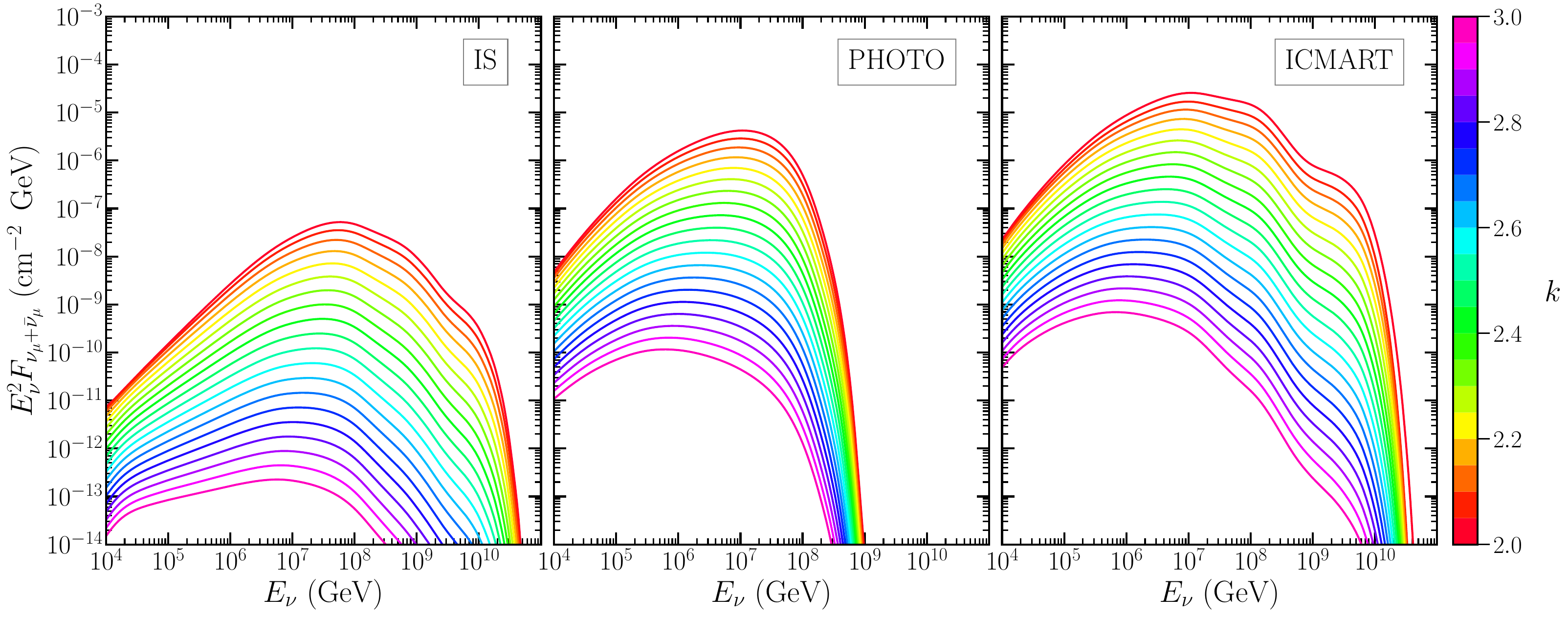} \\
        \includegraphics[width = \linewidth]{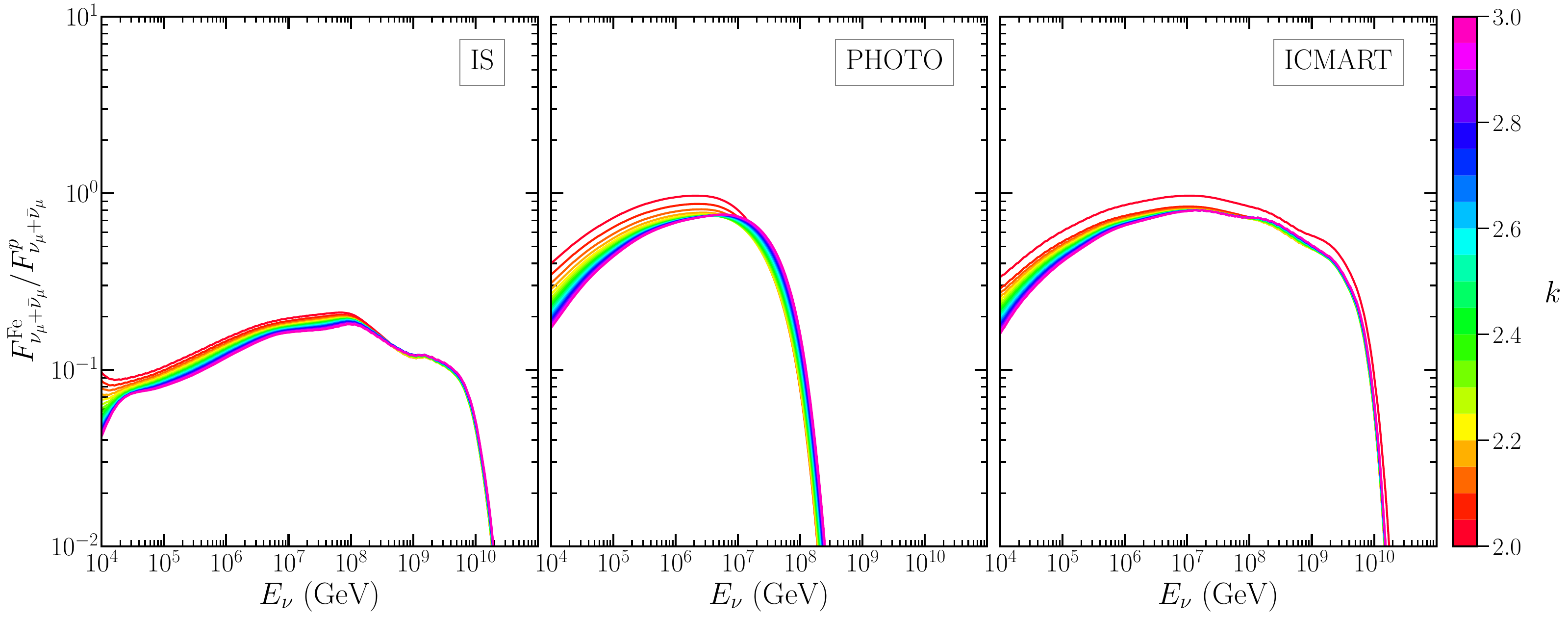}
    \caption{Dependence of the neutrino fluence on the injection spectral index of the accelerated  parent nucleus/nucleon  spectra.  \textit{Top}: $ \nu_\mu + \bar{\nu}_\mu$ neutrino fluence as a function of the neutrino energy for  $k = 2.0$--$3.0$, using our benchmark Fe-jet  (cf.~Table~\ref{tab:initial_conditions}) for the internal shock, photospheric, and ICMART jet models; the other jet parameters are defined as in Table~\ref{tab:initial_conditions}, in particular $\Gamma=300$ is fixed. \textit{Bottom}: Ratio between the Fe-jet neutrino fluence  and the  $p$-jet one as a function of the neutrino energy. A smaller $k$ allows for a larger neutrino fluence, because it favors larger  densities of  seed nuclei; such trend holds independent of the jet composition.}
    \label{fig:fluence_k_var}
\end{figure} 

The top panels of Fig.~\ref{fig:fluence_k_var} show the neutrino fluence for  varying $k$ for the internal shock, photospheric and ICMART models; all other jet parameters are kept fixed as indicated in Table~\ref{tab:initial_conditions}.    
Decreasing the power law index tends to increase the neutrino fluence, because it allows for 
 a larger fraction of the total population to be accelerated at high energies.
We can also note that decreasing $k$ slightly increases the energy $E_\nu^M$ corresponding to the  maximum neutrino fluence. 
For our benchmark jet, we also observe that  $k$  mostly scales the neutrino fluence in a similar fashion for $p$- and Fe-jets, as evident from the bottom panels of Fig.~\ref{fig:fluence_k_var}.

\begin{figure}
    \includegraphics[width = \linewidth]{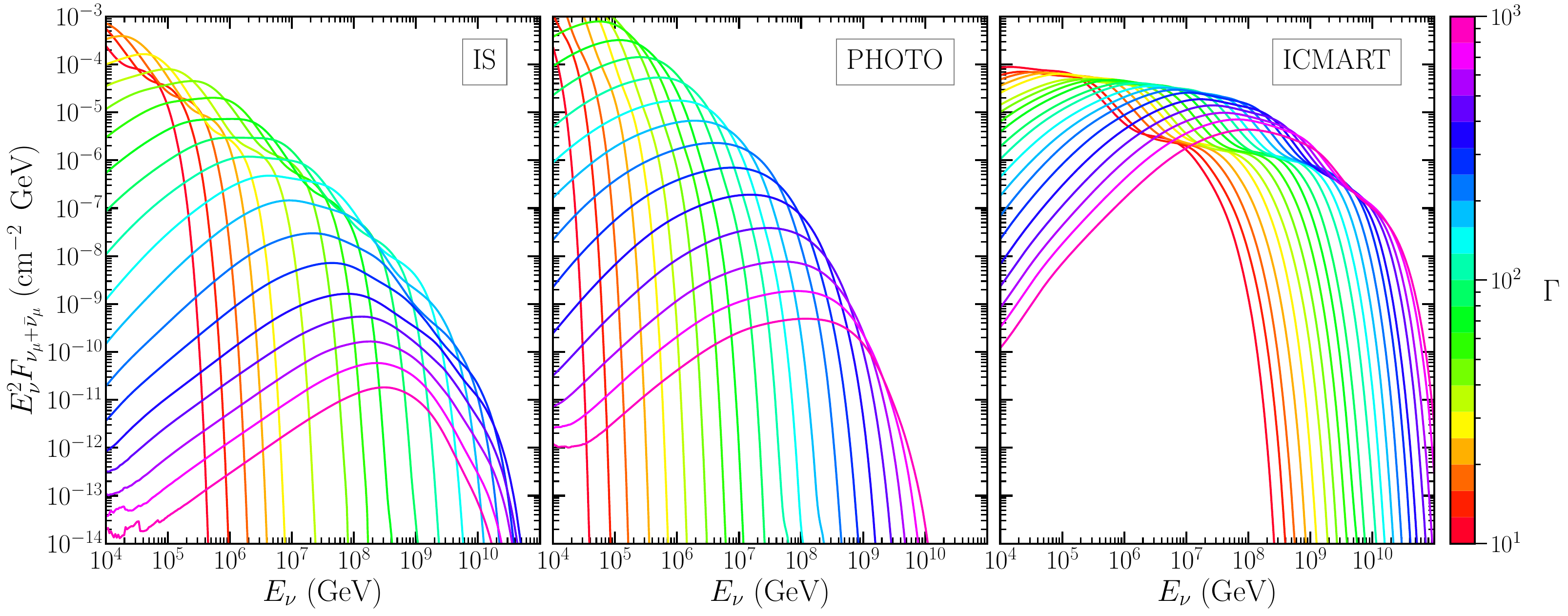} \\
    \includegraphics[width = \linewidth]{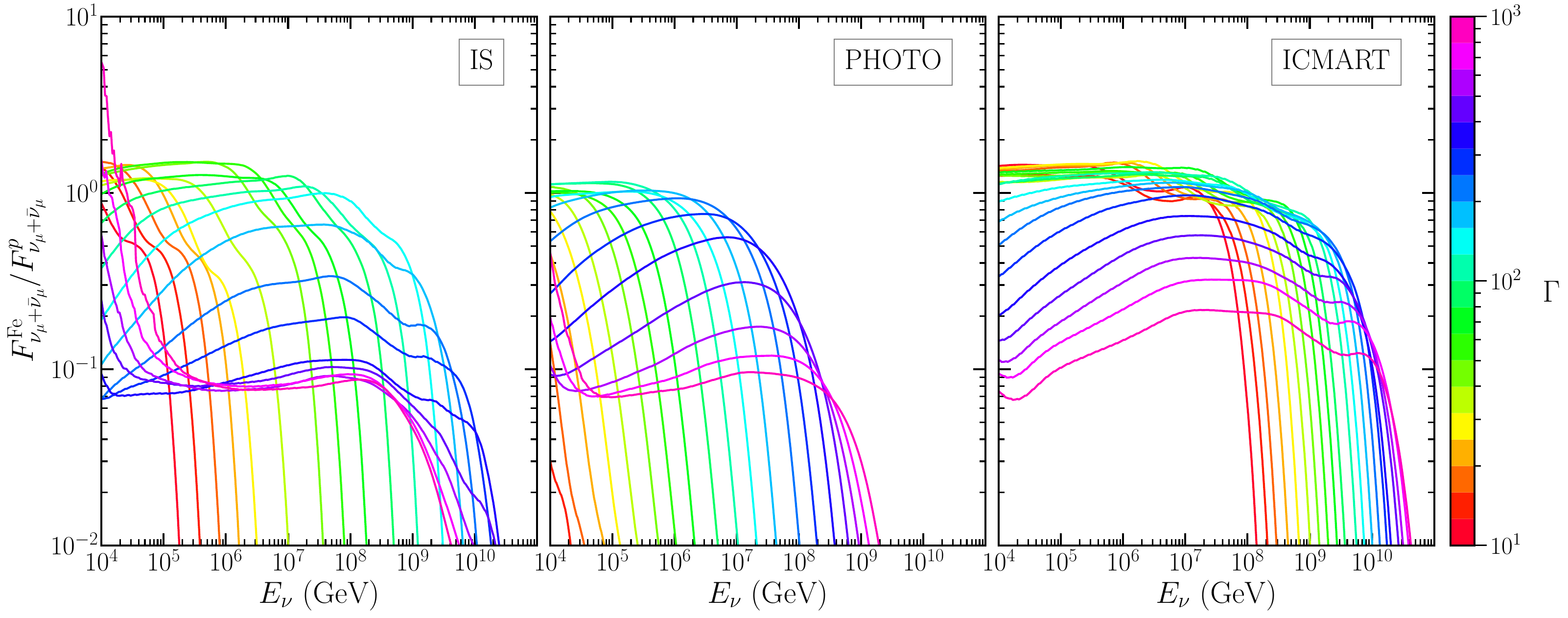}
           \caption{Same as Fig.~\ref{fig:fluence_k_var}, but for investigating the dependence of the neutrino fluence on $\Gamma$ ($k=2.2$ is fixed for both the internal shock and the photospheric models, while $k=2.0$ is used for the ICMART case). Decreasing $\Gamma$ increases the neutrino fluence, with the larger spread occurring for the internal shock and photospheric jet models. In fact, for these models, the interaction radius scales as  $\Gamma^2$, which implies that for larger  $\Gamma$,  internal shocks happen at a radius where both photon and seed nuclei have lower densities. However, this is not the case for the ICMART model, for which the radius of interaction  is independent of the bulk Lorentz factor. This explains the milder variation of the neutrino fluence on $\Gamma$ in the internal shock model. A strong variation of the neutrino fluence as a function of  $\Gamma$ occurs according to the  composition.}
           \label{fig:fluence_gamma_var}
\end{figure} 

Figure~\ref{fig:fluence_gamma_var}  explores the impact of $\Gamma$ on the neutrino emission.
For fixed composition, from the top panels of Fig.~\ref{fig:fluence_gamma_var}, we can see that 
decreasing $\Gamma$ tends to increase the neutrino fluence, in a way that depends on the jet model. 
For the internal shock and photospheric models,  changes in $\Gamma$ affect the  radius of interaction ($R_\text{IS} = 2 \Gamma^2 c \Tilde{t}_\text{var}$); when $\Gamma$ decreases,  internal shocks happen at a radius where both photon and seed nuclei have larger  densities. However, this is not the case for the ICMART model, for which the radius of interaction  is independent of the bulk Lorentz factor (internal shocks trigger magnetic reconnection events assumed to happen at a nearly constant radius). This explains the milder dependence of the neutrino fluence on $\Gamma$ in the internal shock model. 
The energy corresponding to  the maximum neutrino fluence  ($E_\nu^M$) is strongly shifted to lower energies as $\Gamma$ decreases; 
this  is because  higher photon densities are achieved as $\Gamma$ decreases, hence  the photodisintegration timescale increases and crosses the acceleration timescale at lower energies. This implies that the cutoff energy of the parent spectrum decreases, which also forces the neutrino fluence to peak at lower energies. This effect is less prominent in Poynting-flux jets than in kinetically-dominated jets since the acceleration process is more efficient in the former thanks to the larger magnetic fields.

The bottom panels of Fig.~\ref{fig:fluence_gamma_var} also highlight a strong variation of the neutrino fluence as a function of  $\Gamma$ according to the composition. The  ratio between the neutrino fluence of Fe-jet and  $p$-jet is a direct measurement of the efficiency of nuclear cascades.
For large  $\Gamma$, this ratio saturates near $10^{-1}$ because of the  less efficient production of neutrinos from $^{56}$Fe nuclei; then, there is a transition from inefficient to efficient cascades as $\Gamma$ decreases due to higher photon densities. This transition is more or less sharp depending on the model, following the dependence on the interaction radius on  $\Gamma$. For small  $\Gamma$, the ratio saturates around $1$, which is consistent with the fact that the Fe-jet with highly efficient cascades of iron into nucleons is equivalent to the $p$-jet (disintegrated $^{56}$Fe leads to $26$ protons and $30$ neutrons; photohadronic interactions are equivalent for protons and neutrons as long as no distinction between neutrino and antineutrino is considered).

\begin{figure}
        \includegraphics[width =\linewidth]{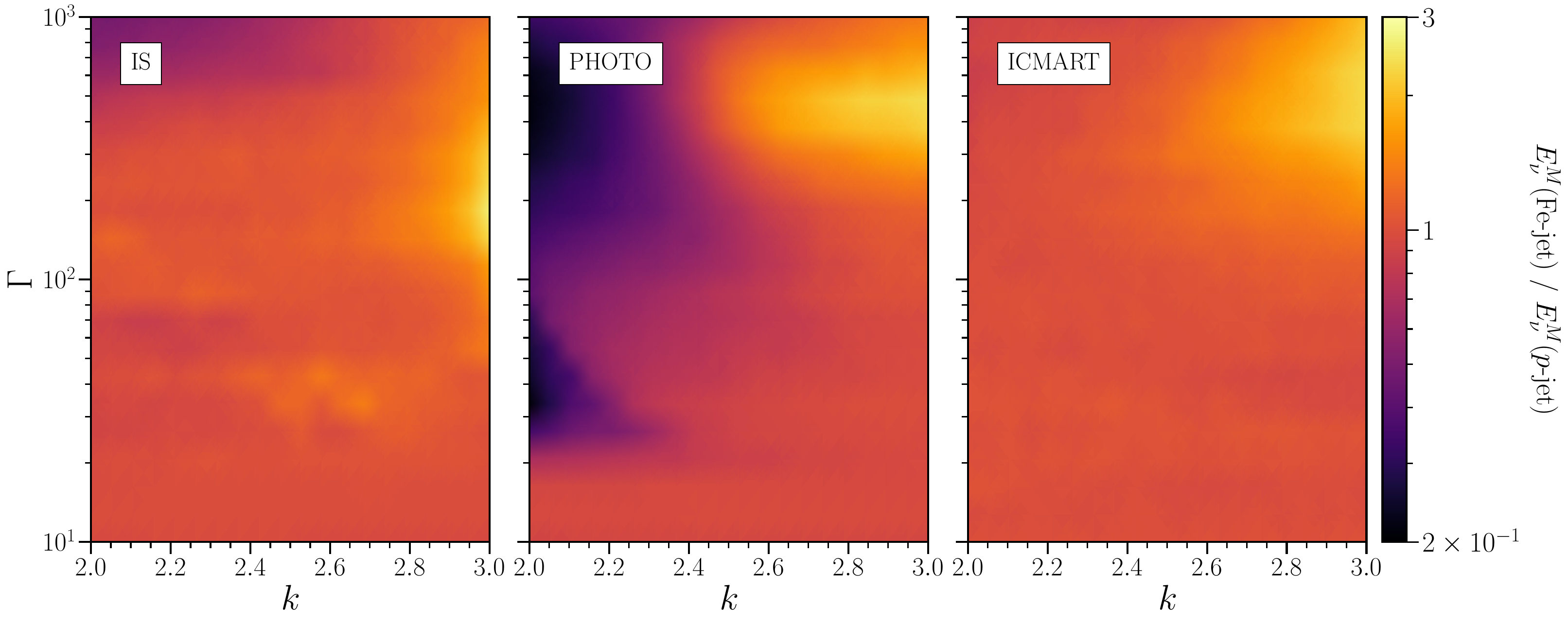}     
    \caption{Contour plots of the ratio between the neutrino energy at which the neutrino fluence peaks for the Fe-jet and the p-jet in the plane spanned by $k$ and $\Gamma$. From left to right, the contour plots for the internal shock, photospheric, and ICMART models are illustrated, respectively. The energy at which the neutrino fluence peaks does not depend on the jet composition strongly  for all considered jet models, however the normalization of the spectral energy distribution does (cf.~Figs.~\ref{fig:fluence_k_var} and \ref{fig:fluence_gamma_var}).
}
    \label{fig:E_max_nu}
\end{figure}

Figure~\ref{fig:E_max_nu}  quantifies the dependence of the neutrino energy ($E_\nu^M$) at which the fluence peaks in  Fe-jets and $p$-jets in the parameter space spanned by $k$ and $\Gamma$. 
In order to avoid degeneracies in the computation of $E_\nu^\text{M}$ for the cases  where the neutrino fluence reaches a plateau around the spectral peak, 
we compute the two energies ($E_\nu^{90\%, 1}$ and $E_\nu^{ 90\%, 2}$) where the fluence is $90 \%$ of its maximum. Then, the maximum energy is computed as
    $\log_{10}(E_\nu^M) = {1}/{2} \, [\log_{10}(E_\nu^{90\%, 1}) + \log_{10}(E_\nu^{90\%, 2})]$.
The contours of the  ratio of $E_\nu^M$ for the Fe-jet and $p$-jet in Fig.~\ref{fig:E_max_nu} highlight that $E_\nu^M$ does not change much as a function of the jet composition. 
However, $E_\nu^M$ has non-trivial  dependencies on the jet parameters according to the jet model. We note that the normalization of the neutrino fluence does strongly depend on the jet composition, as illustrated in Figs.~\ref{fig:fluence_k_var} and \ref{fig:fluence_gamma_var}. 

\section{Dependence of the diffuse neutrino flux on the jet composition}\label{sec:diffuse_flux}
In this work, we do not explore the detection prospects of neutrinos, since we are  interested in assessing the model dependence on the jet composition. 
On the other hand, in order to gauge the variation of the neutrino signal as a function of the jet composition for different jet models, we consider the diffuse emission of neutrinos from HL- LL- and sGRBs. 
The  diffuse neutrino emission is defined as follows:
\begin{equation}
    F^d_{\nu_\beta}(E_{\nu_\beta}) = \int_{z_\text{min}}^{z_\text{max}}  \frac{c}{4\pi H_0 \Gamma} \frac{1}{\sqrt{\Omega_\Lambda + (1+z)^3 \Omega_M}} \; R_\text{GRB}(z)\;  \frac{4\pi \Gamma {d_L(z)
    }^2}{(1+z)^3}F_{\nu_\beta}( E_{\nu_\beta}, \, z) \; d z\ ,
\end{equation}
where $ R_\text{GRB} (z)$ is the cosmological rate (in units of Gpc$^{-3}$ yr$^{-1}$), which is different for HL-GRBs, LL-GRBs and sGRBs. For HL-GRBs, we model the rate following Ref.~\cite{Wei:2013wza,Wanderman:2009es}, for LL-GRBs we consider Refs.~\cite{Liu:2011cua,Yuksel:2008cu}, and for sGRBs we  follow Ref.~\cite{Salafia:2023sjx}:
\begin{align}
    R_{\rm{HL-GRB}}(z) &= \rho_{0,\rm{HL-GRB}} 
    \begin{cases}
        (1+z)^{2.1} & \rm{for }\; z < z^{\rm{HL-GRB}}_\star \\
        (1+z^{\rm{HL}}_\star)^{2.1 + 0.7} \; (1+z)^{-0.7} &\rm{for } \; z > z_\star
    \end{cases}\ ,   \\
    R_{\rm{LL-GRB}}(z) &= \rho_{0,\rm{LL-GRB}} \left[ (1+z)^{-34.0} +  \left(\frac{1+z}{5000}\right)^{3.0} + \left(\frac{1+z}{9}\right)^{35.0} \right]^{-1/10}\ , \\
    R_{\rm{sGRB}}(z) &= 
    \begin{cases}
        10^{\alpha \,z^3 - \beta\,z^2 + \gamma \,z + \delta}  & \rm{for }\; z < z^{\rm{sGRB}}_\star \\
        10^{ - \eta \,z + \mu}  & \rm{for }\; z > z^{\rm{sGRB}}_\star 
    \end{cases}\; \text{Gpc$^{-3}$~yr$^{-1}$}\ , 
\end{align}
with $\rho_{0,\rm{HL-GRB}} = 200$--$2000$~Gpc$^{-3}$~yr$^{-1}$, $\rho_{0,\rm{LL-GRB}} = 0.5$--$0.8$~Gpc$^{-3}$~ yr$^{-1}$, $z^{\rm{HL-GRB}}_\star = 3.6$, and $z^{\rm{sGRB}}_\star = 6.0$. We have $\alpha = [0.049, \;0.047, \;0.048$], $\beta = [0.636,\; 0.609, \;0.586$], $\gamma = [1.9,\; 1.9,\;2.0$], $\delta = [0.01, \;0.49, \;0.98$], $\eta= [0.6,\; 0.53,\; 0.23$], and $\mu = [3.0,\; 3.6,\; 3.6$]. For each of the parameters of $R_\mathrm{sGRB}$, we indicate fit values corresponding  to the lower, average, and higher redshift rate. 
\begin{table}[t]
 \caption{Jet parameters assumed for HL-GRBs~\cite{Zitouni:2018wre, Lan:2021uuf}, LL-GRBs~\cite{Tamborra:2015qza,Rudolph:2021cvn, Liu:2011cua} and sGRBs~\cite{Tamborra:2015qza,Ito:2021asl, Wanderman:2014eza}. All other jet parameters are modeled as specified in Table~\ref{tab:initial_conditions} according to the jet model.
 }
 \centering
    \begin{tabular}{c||c|c|c|}
           & HL-GRB & LL-GRB & sGRB \\ \midrule
         $\Tilde{E}_\text{iso} $ &  $4.5\times10^{54}$~erg &  $1.5\times10^{52}$~erg  &  $1.05\times10^{53}$~erg  \\ \hline
        $\Gamma$ &  $300$ & $10$ & $300$ \\ \hline
        $t_\text{dur} $ &  $30$~s & $1000$~s & $1$~s\\ \hline
        $t_\text{var} $ &  $0.5$~s & $100$~s & $0.01$~s\\ \hline
    \end{tabular}
    \label{tab:bench_param}
\end{table}
\begin{figure}
    \includegraphics[width = \linewidth]{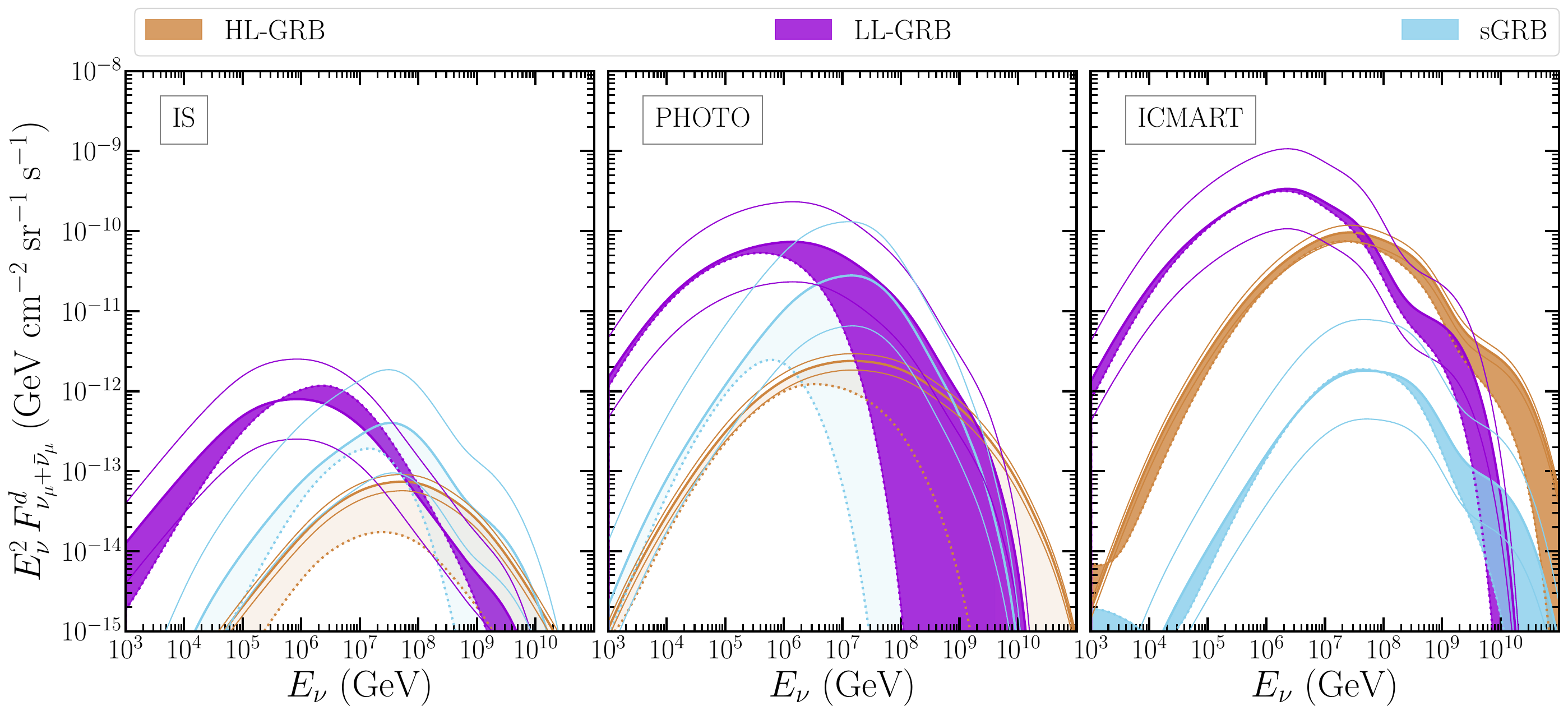}
    \caption{Diffuse $\nu_\mu + \bar{\nu}_\mu$ emission for HL-GRBs (ochre), LL-GRBs (purple) and sGRBs (light blue) for the internal shock, photospheric and ICMART models, from left to right, respectively, computed assuming  the average redshift rate. For each jet model and GRB type, the band marks the uncertainty due to the jet composition ($p$-jets are plotted as solid lines and Fe-jets as dotted lines; any mixed composition is represented by the color band). 
   For each GRB type, the color band is shadowed for the cases where nuclear contamination is expected to be disfavored.  
    The uncertainty due to the jet composition is smaller or at most comparable to  the uncertainty due to redshift rate. The latter is represented by solid lines for the $p$-jet. Independent of the GRB type, the diffuse neutrino flux is expected to be lower for the internal shock model.
   } 
    \label{fig:diffuse_flux} 
\end{figure} 

As for the jet parameters entering the neutrino fluence, we differentiate among the three GRB types as summarized in Table~\ref{tab:bench_param} and otherwise use the parameters in Table~\ref{tab:initial_conditions}.
For each GRB type, we compute the diffuse emission considering  $p$-jet and  Fe-jet. 
   
Figure~\ref{fig:diffuse_flux} shows the expected diffuse emission for  HL-GRBs, LL-GRBs and sGRBs, for the three jet models, from left to right, respectively.
For fixed jet model, for each GRB type, the  band represents the uncertainty due to the jet composition (with the lower emission corresponding to $p$-jet and the larger one to Fe-jet). 
In order to compare the uncertainty due to the jet composition with the one due to the redshift rate, we plot three solid lines for each GRB type, representing the diffuse flux of a $p$-jet assuming the lowest, average and highest GRB rate. 
Independent of the GRB type, the diffuse neutrino emission  is expected to be smaller for the  internal shock model.

In order to allow for a comparison across different jet models, we  use the same jet parameters (e.g.~isotropic energy and total duration of the burst) for the internal shock, photospheric and ICMART models. This implies that, due to the smaller microphysical parameters typical of the internal shock model, the overall gamma-ray luminosity predicted within the internal shock model is smaller than the one of the other two  models, and about two orders of magnitude below the peak of the luminosity function for each GRB family~\cite{Lan:2021uuf, Howell:2014wba,Dai:2008mw,Wanderman:2014eza}. While this choice might seem conservative, we are interested in exploring the variation of the neutrino signal across jet models and  composition relying on the same benchmark jet parameters. 

While we refrain from an exploration of the detection prospects, that would require a dedicated exploration of the GRB population and survival of nuclei according to the jet properties, it is interesting to note that the uncertainty on the GRB rate is generally larger or at most comparable to the uncertainty due to  the composition.
Magnetically dominated jets and low-luminosity ones are the most likely to guarantee the survival of heavy nuclei up to the interaction region~\cite{Horiuchi:2012by}. For this reason, we  use darker (lighter) shaded bands to distinguish between  the GRB types that are likely (unlikely) to be loaded with heavier nuclei at the interaction region.

\section{Discussion and outlook} \label{sec:discussion}

It is likely that the GRB jet is not only composed of protons, but also heavier nuclei, either synthetized  or entrained in the jet~\cite{Beloborodov:2002af,Horiuchi:2012by}. In this paper, we have explored the impact of the jet composition on the neutrino emission for kinetic dominated jets (according to the internal shock  and photospheric models) as well as for Poynting flux dominated jets (considering the ICMART model). To this purpose, we have built a Monte-Carlo algorithm that simulates nuclear cascades for any nucleus with $A\leq 56$. 

We find that the nuclear composition affects the neutrino fluence in quantitatively different ways according to the jet model, although the overall qualitative trend is similar across models. The neutrino fluence can be lower up to one order of magnitude for the Fe-jet case with respect to the $p$-jet scenario because the production of mesons from nuclei is less efficient than the one from protons. The efficiency of neutrino production is linked to the jet model;  the ICMART model, because of its fixed particle interaction radius, guarantees efficient nuclear cascades leading to  an overall larger neutrino fluence across all three jet models, even for the Fe-jet case.

\begin{figure}
        \includegraphics[width = 0.49\linewidth]{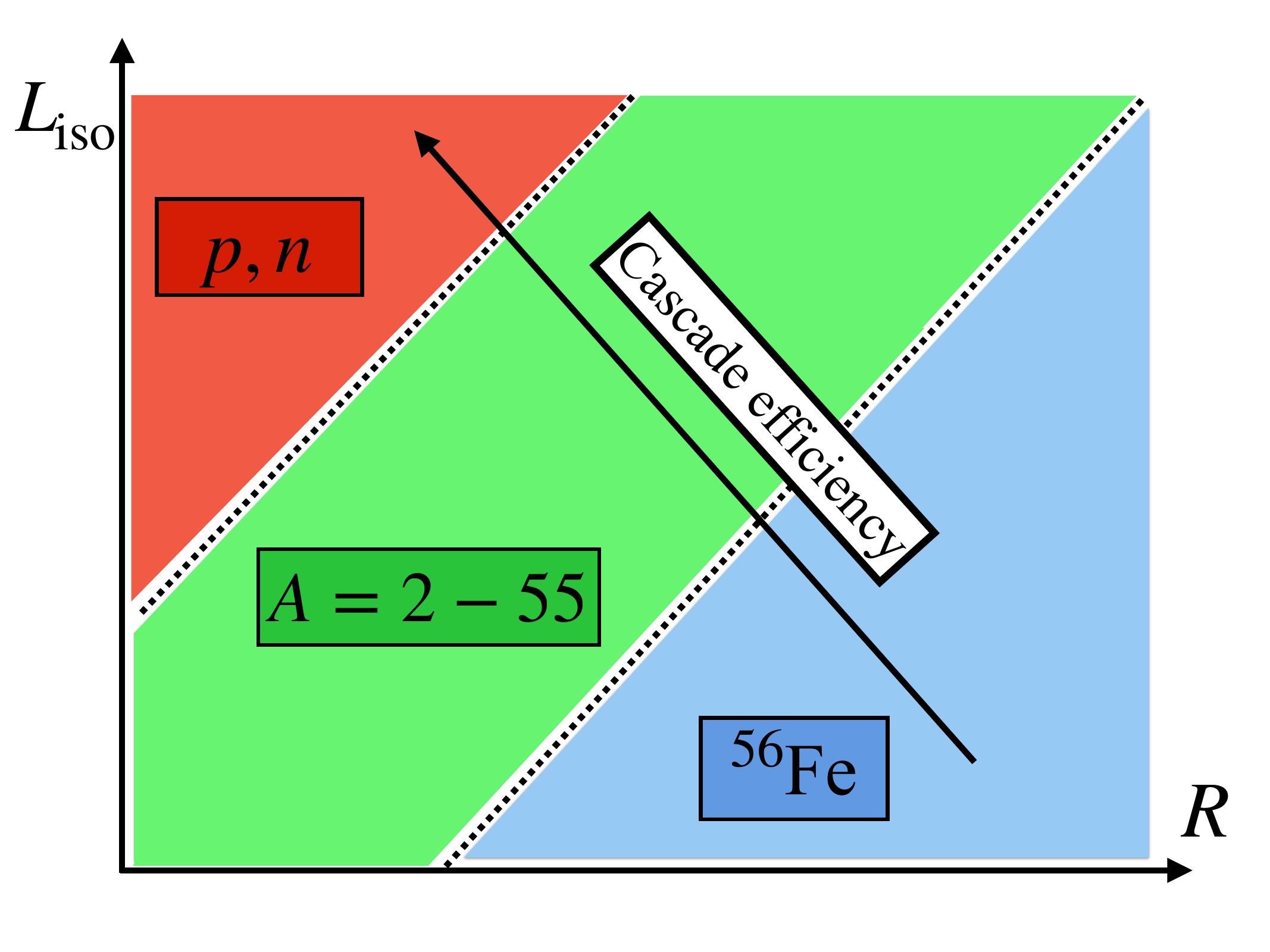}
    \includegraphics[width=0.49\linewidth]{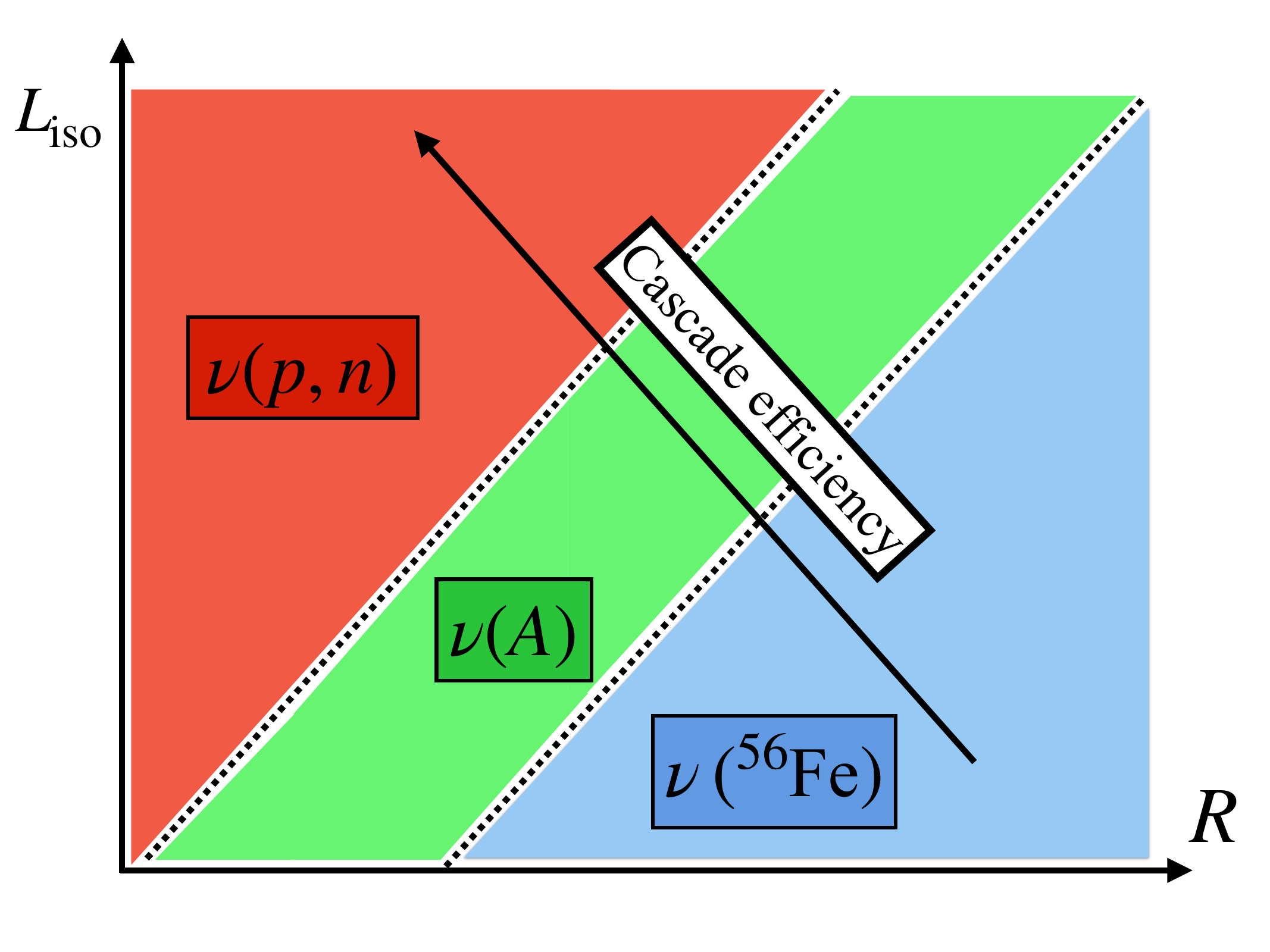}
    \caption{\textit{Left:} Sketch of the efficiency of nuclear cascades for our  one-zone jet model in the plane spanned by the radius of  neutrino production and the jet luminosity.   The nuclear species dominating the particle density are highlighted in color (protons and neutrons in red, any nucleus lighter than iron in green, and iron in light blue). 
    \textit{Right:} Same as the left panel, except for the fact that the different regions are colored depending on which isotopes contribute the most to the total neutrino fluence. The efficiency of nuclear cascades is the largest for small interaction radii and larger jet luminosity. Notably, when nuclear cascades are efficient, the neutrino production is driven by proton and neutrons in a larger region of the parameter space (cf.~green band that is smaller for the right panel). }
    \label{fig:conclusion}
\end{figure}

Independent of the jet model,  the survival of nuclei and inefficient nuclear cascades lead to a lower neutrino fluence due to the lower efficiency of nuclear photohadronic interactions. Conversely, if nuclei are disintegrated, the neutrino fluence may be similar to the one from a jet loaded  with protons. The sketch in Fig.~\ref{fig:conclusion} summarizes how the jet luminosity  and the radius of neutrino production, within a one-zone jet model, affect the nuclear cascade efficiency and therefore the neutrino production (left and right panels, respectively); see also Fig.~\ref{fig:L_R_variation}.

Exploring the jet parameter space, we showed that the power law index of the seed  spectra of nuclei mostly affects the normalization of the spectral energy distribution of neutrinos, without altering its shape. On the other hand, the bulk Lorentz factor of the jet is important to assess the efficiency of nuclear cascades and is thus highly correlated with both the composition and the GRB model. Our work focuses on neutrinos only, complementing dedicated work in this direction centered on ultra-high-energy cosmic rays, e.g.~Refs.~\cite{Biehl:2017zlw,Heinze:2020zqb}. 

We  have explored two extreme cases in terms of composition: either assuming that the jet is fully made out of protons or pure $^{56}$Fe. Any other mixed composition, involving nuclei lighter than $^{56}$Fe is expected to fall in between the two extreme scenarios considered here. However, it is important to keep in mind that, according to the jet luminosity and Lorentz boost factor, heavier nuclei may not survive in the jet~\cite{Horiuchi:2012by,Zhang:2017moz}; we do not explore the conditions for survival of nuclei in this paper, rather focus on assessing the  uncertainty linked to the jet composition across jet models. 
Further work should be dedicated to explore  the neutrino and cosmic ray emission from jets with mixed composition, guided by self-consistent simulations of the source and their ejecta composition. 
Nuclear cascades from GRBs could also contribute to the spectrum of ultra-high-energy cosmic rays~\cite{Zhang:2017moz,Murase_2008,Wang:2007xj,Biehl:2017zlw}.

Modeling the diffuse neutrino emission from HL-GRBs, LL-GRBs and sGRB, for our three jet models and varying jet composition, we find that also for Fe-jets, the diffuse neutrino flux from LL-GRBs may dominates the overall GRB diffuse emission, consistently with the findings of Refs.~\cite{Tamborra:2015qza,Murase:2006mm} for $p$-jets.  Our results  on the diffuse neutrino emission are in overall good agreement with Ref.~\cite{Murase:2010gj}, which  predicted that the neutrino production from nuclei should be up to one order of magnitude  lower in the limit of inefficient nuclear cascades. However, the uncertainty on the jet composition is generally smaller or at most comparable with the one associated to the GRB cosmological rate.

\acknowledgments
We are grateful to Annika Rudolph for involvement in the very early stages of this project and  Walter Winter for helpful comments on the manuscript. This research project has received support from  the Villum Foundation (Project No.~37358),  the Deutsche Forschungsgemeinschaft through Sonderforschungsbereich SFB 1258 ``Neutrinos and Dark Matter in Astro- and Particle Physics'' (NDM), and the {\'E}cole Normale Sup{\'e}rieure Paris-Saclay  scholarship program. V.D.L.~also thanks the Particle Astrophysics group members at the Niels Bohr Institute for their warm hospitality.

\appendix

\section{Details on the Monte Carlo simulation of nuclear cascades} \label{App:MCNC_params} 

In this appendix, we present details on the modeling of the photonuclear cross section for any nuclei with mass number $A\leq56$, as well as the half-life and radioactive decay channels of unstable isotopes. We also describe the method adopted to construct the particle spectral energy distributions and the modeling of the  nuclear photohadronic interactions.

\subsection{Modeling of the nuclei photonuclear cross section} \label{App:GDR1}

In order to model the photonuclear cross sections we distinguish among three different energy regimes:
\begin{itemize}
\item[ ]{\bf Giant Dipole Resonance 1.} 
We consider the  energy range between $0$ and $30$~MeV as being the Giant Dipole Resonance 1 (GDR1) one;  it coincides with the GDR peak. We use the fit from Ref.~\cite{2002EPJA...14..377K} to compute the total inelastic cross-section for any nucleus of mass number $A\leq 56$ in this energy region: 
\begin{equation} \label{eq:GDR1_fit}
    \displaystyle \sigma_{A\gamma}(E_{\gamma,r}, A) = \sum_{i=1,2,4,8} \frac{e^{i\,(\rho_i - z)}}{1+ e^{3i\,(\tau_i -z)}} + f_r \, (r_H + r_\Delta) + s_p\, f_p\, h_p\ , 
\end{equation}
with $z=\ln{( E_{\gamma,r} /1 \,\text{MeV})}$, and the fit parameters $\rho_i$, $\tau_i$, $f_r$, $r_H$, $r_\Delta$, $s_p$, $f_p$, and $h_p$ being defined as in Ref.~\cite{2002EPJA...14..377K}. Figure~\ref{fig:Fe_cross_section} compares this fit to the  GDR model presented in Ref.~\cite{Puget:1976nz}, displaying a satisfactory agreement in the GDR1 energy range. 

In the  GDR1 energy range, the emission of only one or two nucleons is possible.
The  fit of Ref.~\cite{2002EPJA...14..377K} does not provide this information, hence the data from Ref.~\cite{Puget:1976nz} are used to compute the branching ratio $P_\mathcal{N}$ for $\mathcal{N} = Z + N$ nucleons emission . The branching ratio $P$ is a function of the number $(Z,N)$ of protons and neutrons emitted and is computed as follows: 
\begin{equation} \label{eq:probab_br}
    P(Z,N) = P_\mathcal{N}(Z+N) \,\binom{Z+N}{Z} \, \frac{1}{2^{Z+N}}\ ,
\end{equation}
where the binomial distribution is  equivalent to the hypothesis that protons and neutrons have the same probability of being  emitted and  all emission channels are independent.
\begin{figure}
    \centering
    \includegraphics[width = 0.7\linewidth]{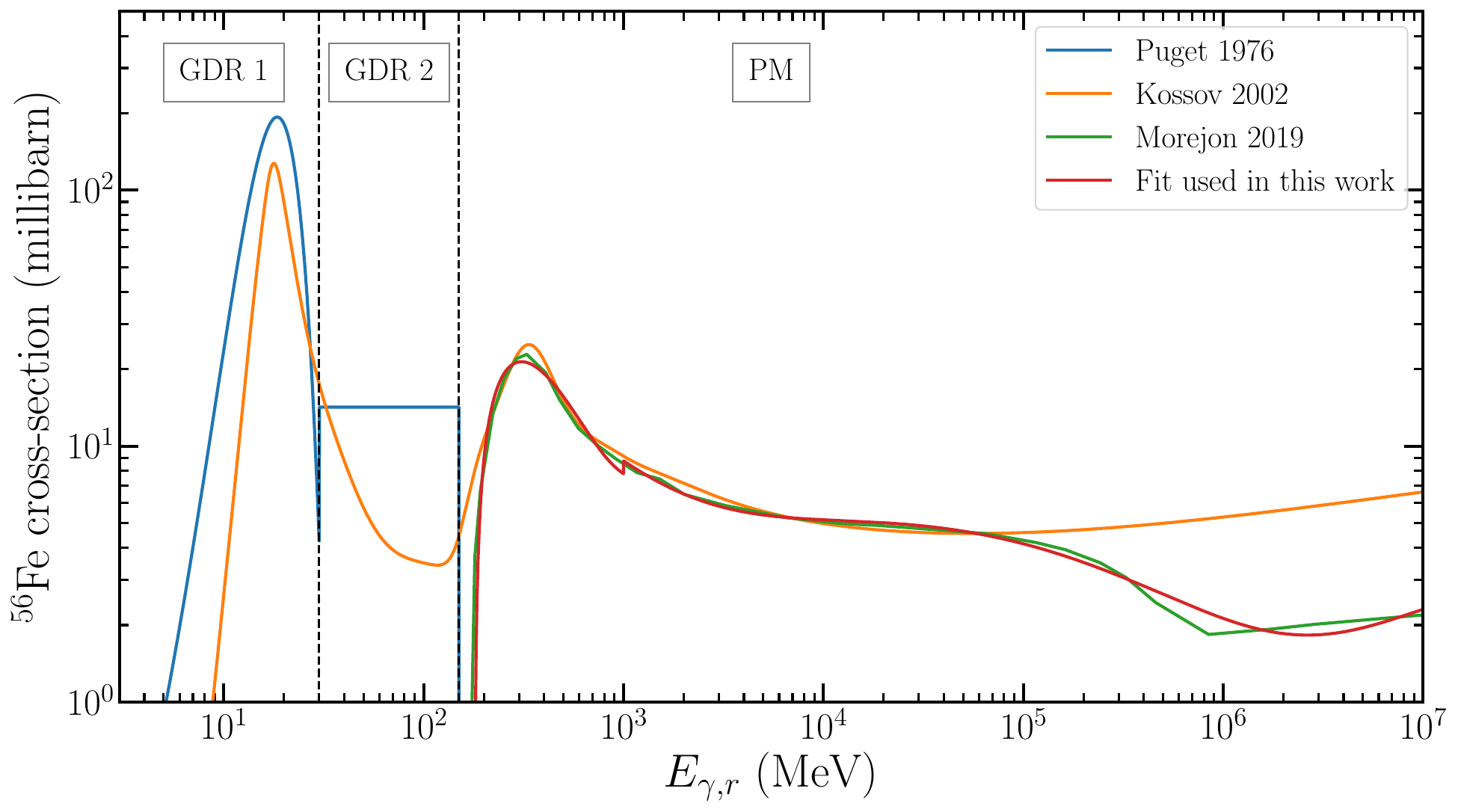}
    \caption{Comparison among different models of  the $^{56}$Fe photonuclear inelastic cross section. The blue, orange and green curves refer  to the models presented in  Refs.~\cite{Puget:1976nz}, \cite{2002EPJA...14..377K} and \cite{Morejon:2019pfu}, respectively. The dashed vertical lines highlight the different  energy ranges considered in this work (GDR1, GDR2, and PM). } 
    \label{fig:Fe_cross_section}
\end{figure}

\item[ ]{\bf Giant Dipole Resonance 2.} 
The second energy range is the Giant Dipole Resonance 2 (GDR2) one, which spans from $30$  to $150$~MeV. This range coincides with the transition between the GDR peak  and the photomeson range. The fit in Eq.~\eqref{eq:GDR1_fit} is still used to compute the cross-section in this energy range and is displayed in  Fig.~\ref{fig:Fe_cross_section}. A notable improvement on the modeling of the cross section  in this energy range  has been achieved in Ref.~\cite{2002EPJA...14..377K}, with respect to  Ref.~\cite{Puget:1976nz}. 
An estimation of the branching ratios  $P_\mathcal{N}$ can be found in Ref.~\cite{Puget:1976nz} (see Eq.~\eqref{eq:probab_br}).

\item[ ]{\bf Photomeson regime.} 
The photomeson energy range (PM) is defined for energies above $150$~MeV, where  the production of mesons is possible thanks to interactions with photons. Ref.~\cite{Morejon:2019pfu} provides  the cross-section and the estimation of the branching ratio in the PM range. We have performed a polynomial fit of the total inelastic cross-section (in microbarn): 
\begin{equation} \label{eq:PM_cs}
    \sigma_{A\gamma}^{PM}(E_{\gamma,r}, A) = A \left(\frac{A}{56}\right)^{\alpha(E_{\gamma,r}) - 1}
    \begin{cases}
    \sigma_\text{low}(E_{\gamma,r}) & \text{if } 0.18 \;\text{GeV} \leq E_{\gamma,r} \leq 1 \;\text{GeV}\\
    \sigma_\text{mid}(E_{\gamma,r}) & \text{if } 1 \;\text{GeV} \leq E_{\gamma,r} \leq 10^4 \;\text{GeV}\ .\\
    \sigma_\text{high}(E_{\gamma,r}) & \text{if } E_{\gamma,r} \geq 10^4 \;\text{GeV}
    \end{cases}
\end{equation}
Using $z= \log_{10}(E_{\gamma,r} / 1 \text{GeV})$, the expressions for the cross section above in the three energy ranges are respectively:
\begin{eqnarray}
    \sigma_\text{low}({E_{\gamma,r}}) &=& 2720.9 z^5 -1979.5 z^4 -1139.4 z^3 +  902.4 z^2 -165.1 z+ 139.1 \\
    \sigma_\text{mid}({E_{\gamma,r}}) &=& -2.02 z^5 + 25.32 z^4 - 112.23 z^3 +  212.91 z^2 -188.22 z+ 156.26\ ; \\
    \sigma_\text{high}({E_{\gamma,r}}) &=& 39.1
\end{eqnarray}
while the function $\alpha(E_{\gamma,r})$ is defined as follows
\begin{equation} \label{alpha_cs}
    \alpha(E_{\gamma,r}) = 
    \begin{cases}
    1 & \text{if }  E_{\gamma,r} \leq 0.2 \;\text{GeV}\\
    0.004 z^5 - 0.023 z^4 + 0.019 z^3 + 0.018 z^2 - 0.043 z+ 0.97 & \text{if } 0.2 \;\text{GeV} \leq E_{\gamma,r} \leq 10^3 \;\text{GeV}\\
    2/3 & \text{if } E_{\gamma,r} \geq 10^3 \;\text{GeV}
    \end{cases}
\end{equation}
This PM parametrization of the cross section is displayed in Fig.~\ref{fig:Fe_cross_section}.

The cross-sections of each process averaged over the  PM energy range are used to compute the branching ratios for the five types of photonuclear reactions: direct proton production, direct neutron production, multi neutron production, spallation and pion production. The branching ratio of a process is estimated as the ratio of its cross-section over the total cross-section. 
\end{itemize}

\subsection{Radioactive decay of unstable isotopes} \label{App:Radioactivity}

\begin{figure}
    \centering    
    \includegraphics[width = 0.49\linewidth]{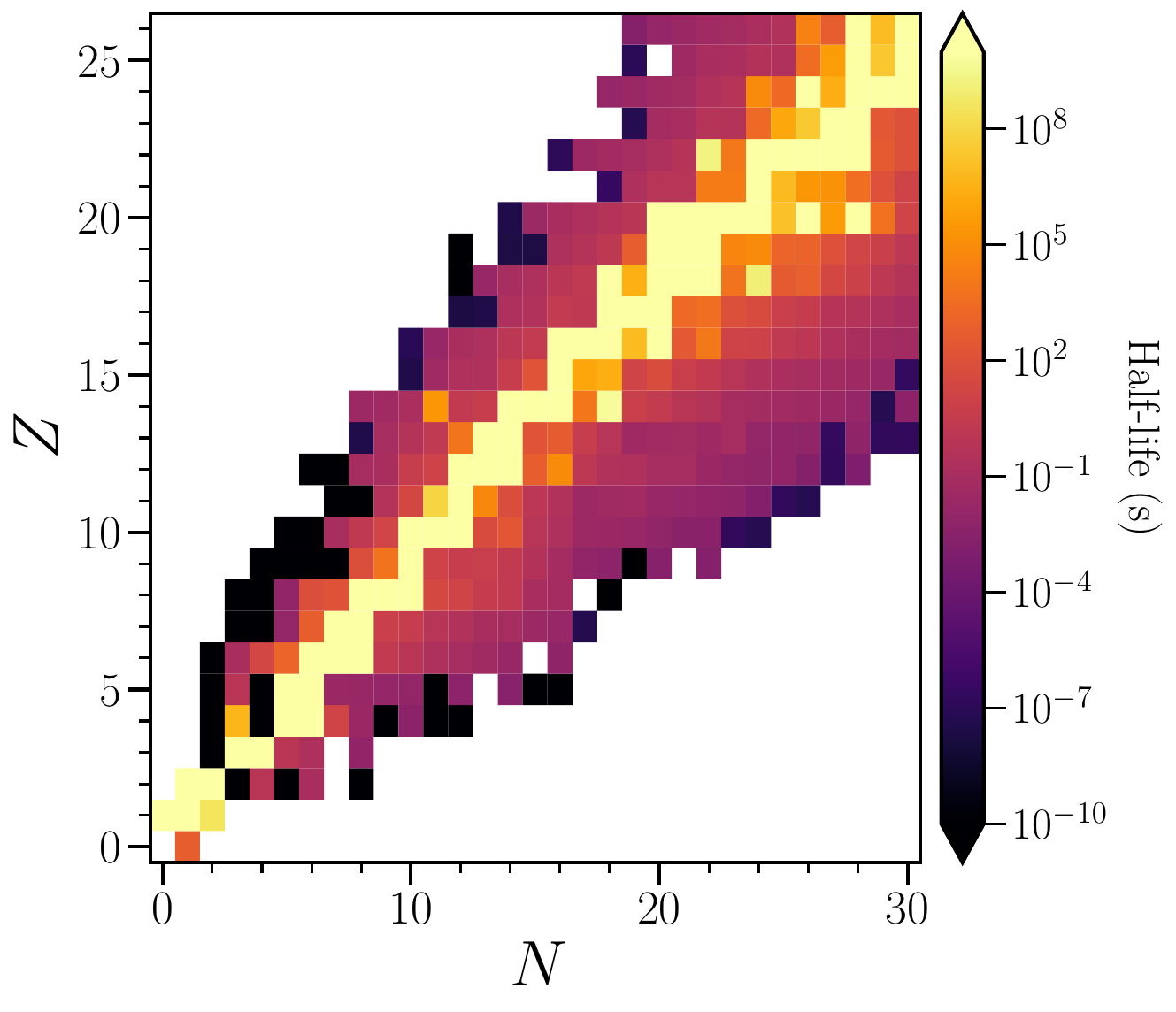}
    \includegraphics[width = 0.5\linewidth]{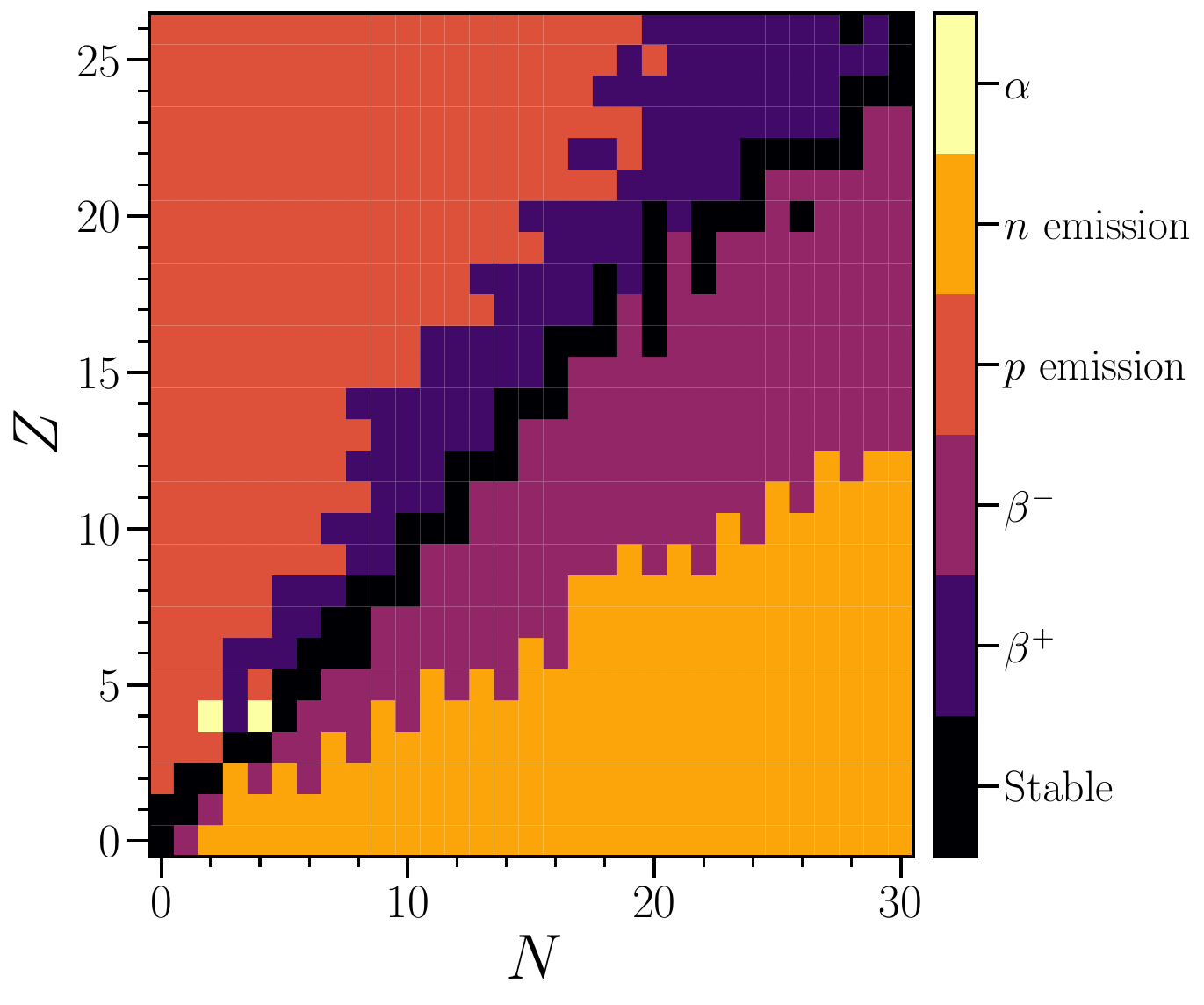}
    \caption{Contour plots of the half-life (left panel) and radioactive decay channels (right panel) for  nuclei of mass number $A\leq 56$ \cite{IAEA} in the plane spanned by $N$ and $Z$. The  lifetime of different isotopes can largely vary.  The isotope stability  decreases exponentially as the isotope  strays from the valley of stability ($ Z = N = A / 2 $; corresponding to the black region on the right panel); in this case, the isotope can undergo  five different  types of radioactive decays: $\beta^+$ decay, $\beta^-$ decay, $\alpha$ decay, single proton emission and single neutron emission. }
    \label{fig:nuclear_data}
\end{figure}

For each isotope, we extract the  mean lifetime (or half-life) and decay channel from the International Atomic Energy Agency database~\cite{IAEA}. Figure~\ref{fig:nuclear_data} shows  the half-life (left panel) and decay type (right panel) for isotopes of mass number $A\leq 56$. 
We can see that the  lifetimes can vary by orders of magnitudes. 
Moreover, the isotope stability decreases exponentially as the isotope position in the plane spanned by $N$ and $Z$ strays from the valley of stability (which corresponds  to $ Z = N = A / 2 $). An unstable isotope can undergo  five  types of radioactive decays: $\beta^+$ decay, $\beta^-$ decay, $\alpha$ decay, single proton emission and single neutron emission. Single proton and neutron emission are unusual decay types that describe the behavior of the most proton-rich/neutron-rich isotopes. Such isotopes  are so unstable that they emit  a proton/neutron almost instantaneously. No precise measurement of their lifetime is available; this explains why they are not highlighted in color in the left panel of Fig.~\ref{fig:nuclear_data}. Consequently, if a nucleus arrives in the proton/neutron emission region, the emission of single nucleons can be integrated out until the isotope reaches a $\beta^+$, $\beta^- $, $\alpha$ or stable state (as indicated in the contour scale of the right panel).

\subsection{Spectral energy distributions} \label{App:comp_spectra}
In order to generate the spectral energy distribution for each nucleon, we first select the seed isotope energy randomly out of a  logarithmic distribution:
\begin{equation}
    \log_{10}(E'_0) = \log_{10}(E'_{\text{min}}) + (\log_{10}(E'_{\text{max}}) - \log_{10}(E'_{\text{min}}) )\; U(0,1)\ ,
\end{equation}
with $U(0,1)$ being the uniform probability density function  between $0$ and $1$, $E'_{\text{min}}$ and $E'_{\text{max}}$ the minimum and maximum energies.

The spectral density $n'_A(E'_A)$ is a set of $N_s$ points located at energies $E'_{s,i}$ with $0 \leq i \leq N_s -1$. Hence, to each nucleus  with  initial energy $E'_0$, we can associate  a number density:
\begin{equation} \label{eq:d_0}
    d_0(E'_0) =  \int_{E'_{s,i_0}}^{E'_{s,i_0+1}} n'_A(E') dE' \approx n'_A(E'_{s,i_0}) \;  ( E'_{s,i_0+1} - E'_{s,i_0} )\ ,
\end{equation}
with $i_0$ such that $E'_{s,i_0} \leq E'_0 \leq E'_{s,i_0+1}$. 
If a test nucleus undergoes a photonuclear reaction that generates $(N_p, N_n)$ free protons and neutrons, this  creates the nucleon density $d_\mathcal{N} = N_\mathcal{N} \; d_0(E'_0)$, with $\mathcal{N} = (p,n)$ representing a proton or a neutron. Note that these particle densities are  independent of the energy at which the test nucleus interacts with a photon and the energy of the produced nucleons. 

We then add these nucleons together to populate the spectral energy distribution for each species. Nucleons have an energy $E'_f$ at the end of the Monte-Carlo run. Their density is computed as

\begin{equation} \label{eq:d_x}
     d_\mathcal{N}(E'_f) =  \int_{E'_{s,i_f}}^{E'_{s,i_f+1}} n'_\mathcal{N}(E') dE' \approx n'_\mathcal{N}(E'_{s,i_f}) \;  ( E'_{s,i_f+1} - E'_{s,i_f} )\ ,
\end{equation}
with $i_f$ such that $E'_{s,i_f} \leq E'_f \leq 'E_{s,i_f+1}$. 
Using $d_\mathcal{N} = N_\mathcal{N} \; d(E'_0) $ and Eqs.~\eqref{eq:d_0} and \eqref{eq:d_x}, we get 
\begin{equation}
    n'_\mathcal{N}(E'_{s,i_f}) = N_\mathcal{N} \; n'_A(E'_{s,i_0}) \; \frac{E'_{s,i_0+1} - E'_{s,i_0}}{E'_{s,i_f+1} - E'_{s,f_0}}\ .
\end{equation}
We note that these calculations presented in this section can be easily generalized to compute the spectrum of the associated photodisintegrated nuclei.

\section{Nuclear photohadronic model}
\label{App:photomeson}
In order to estimate the meson production, the function $f$ (cf.~Eq.~\eqref{eq:Q_formula}) should be identified for each energy interval $\Delta E = [y_{\min},\; y_{\max}]$:
\begin{equation}
    f^{\Delta E}(y) =   \frac{1}{2y^2}\; \int_{E^{\mathrm{th}, \Delta E}_{\gamma, r} }^{2y} \;E_{\gamma,r} \;\sigma_{A\gamma\rightarrow \mathcal{M}}(E_{\gamma,r}) \; d E_{\gamma,r}  
\end{equation}
To this purpose, we compute the total  function ($f^\mathrm{tot}$) using the total cross-section fit from Eq.~\eqref{eq:PM_cs}.

The function $f^\mathrm{tot}$  does not only describe the meson production, but also all the other processes  in this energy range. That is why it needs to be scaled by the branching ratios of meson production. Using data for the 
$\pi^0$ production cross-section~\cite{Morejon:2019pfu}, we compute an approximation of the product of the multiplicity and the branching ratio and find 
$M_{\pi^0} \; {\sigma_{A\gamma \rightarrow \pi^0 }^\mathrm{PM}}/{\sigma_{A\gamma}^\mathrm{PM}} \approx 0.18$.
Then, using data for $(p,n)$--$\gamma$ interactions from Ref.~\cite{Hummer:2010vx} and assuming that they are still valid for nuclei, we compute the ratios $f_{\pi^+} / f_{\pi^0}$ and $f_{\pi^-} / f_{\pi^0}$ as approximations of the multiplicity ratio $M_{\pi^+} / M_{\pi^0}$ and $M_{\pi^-} / M_{\pi^0}$.
\begin{figure}
\centering
    \includegraphics[width = 0.48\linewidth]{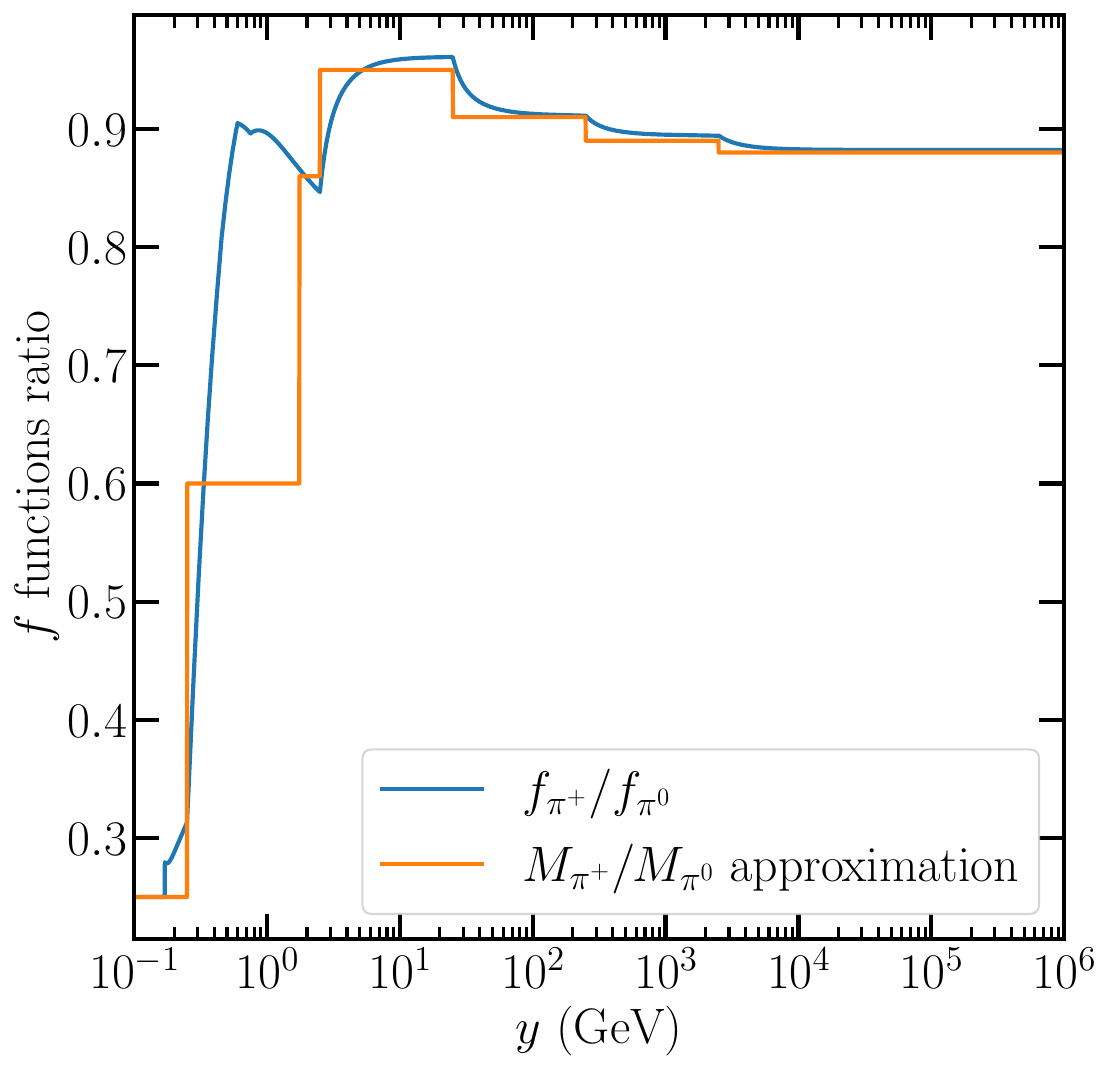}
   \caption{ 
   Ratio of  the function $f_{\pi^+}/f_{\pi^0}$  and  multiplicity ratio $M_{\pi^+} / M_{\pi^0} $  in the nuclear photohadronic interaction model as functions of $y$.}
    \label{fig:photo_model_details}
\end{figure} 
This is shown in  Fig.~\ref{fig:photo_model_details}. We highlight  that $M_{\pi^+} / M_{\pi^0} = M_{\pi^-} / M_{\pi^0}$ because  $50 \%$ of interactions are expected to happen with a proton and the rest with  a neutron of the nucleus. Indeed, the parameters describing $p\gamma$ and $n\gamma$ photohadronic interactions are the same, except that produced mesons have opposite charge~\cite{Hummer:2010vx}. As a consequence, we have symmetry between the production of  $\pi^+$ and the one of $\pi^-$. Following  Ref.~\cite{Hummer:2010vx}, we also compute approximations of the fraction of energy transferred to the produced meson $\chi_\mathcal{M}$. 
Both $M_\pi$ and $\chi_\mathcal{M}$ need to be constant as  functions of the energy for a given energy range.
 Thus, we consider seven energy intervals $\Delta E$, each associated to a range where the multiplicity and the energy fraction is approximately constant, as summarized in  Table \ref{tab:mult_photomeson}. 
\begin{table}
    \centering
        \caption{Multiplicities and fraction of the energy transferred to pions for the seven considered energy ranges.}
    \begin{tabular}{c||c|c|c|c}
       Intervals $\Delta E$  &  $y_{\min} = E^\mathrm{th}_{\gamma, r} / 2$ & $y_{\max} = E^{\max}_{\gamma, r} / 2$ & $M_{\pi^+} / M_{\pi^0} = M_{\pi^-} / M_{\pi^0}$ & $\chi_\pi \times A$\\ \hline
        1 & 0.1 GeV & 0.25 GeV & 0.25 & 0.22\\ \hline
        2 & 0.25 GeV & 1.75 GeV & 0.6 & 0.2\\ \hline
        3 & 1.75 GeV & 2.5 GeV & 0.86 & 0.25\\ \hline
        4 & 2.5 GeV & 25 GeV & 0.95 & 0.18\\ \hline
        5 & 25 GeV & 250 GeV & 0.91 & 0.25\\ \hline
        6 & 250 GeV & 2500 GeV & 0.89 & 0.25\\ \hline
        7 & 2500 GeV & $+ \infty$ & 0.88 & 0.3\\ \hline
    \end{tabular}
    \label{tab:mult_photomeson}
\end{table} 

The product of the multiplicity and the function $f^{\Delta E}$, taking into account both branching ratio and multiplicity, is:

\begin{equation}
    M_{\pi^+}\, f^{\Delta E} (y) \approx  \left(\frac{M_{\pi^+}}{M_{\pi^0}}\right)^{\Delta E} \left( M_{\pi^0} \; \frac{\sigma_{A\gamma \rightarrow \pi^0 }^\mathrm{PM}}{\sigma_{A\gamma}^\mathrm{PM}} \right) \left( f^\mathrm{tot}(y) - f^\mathrm{tot}\left(y^{\Delta E}_\mathrm{th}\right) \left(\frac{y^{\Delta E}_\mathrm{th}}{y}\right)^2\right) . 
\end{equation}
On the right hand side of the equation above, the second term is subtracted from $f^\mathrm{tot}(y)$ since the latter is an integral with lower limit equal to $E^{\mathrm{th}, \mathrm{PM}}_{\gamma, r} = 0.2$~GeV (and not from $E^{\mathrm{th}, \Delta E}_{\gamma, r}$): 
\begin{eqnarray}
    f^\mathrm{tot}(y) &=& \frac{1}{2y^2}\; \int_{E^{\mathrm{th},\mathrm{PM}}_{\gamma, r} }^{2y} \;E_{\gamma,r} \;\sigma_{A\gamma}(E_{\gamma,r}) \; d E_{\gamma,r} 
     = f^\mathrm{tot}\left(y^{\Delta E}_\mathrm{th}\right) \left(\frac{y^{\Delta E}_\mathrm{th}}{y}\right)^2 + \\\nonumber  
     &+&\frac{1}{2y^2}\;  \int_{E^{\mathrm{th}, \Delta E}_{\gamma, r} }^{2y} \;E_{\gamma,r} \;\sigma_{A\gamma}(E_{\gamma,r}) \; d E_{\gamma,r}\ .
\end{eqnarray}
Finally, knowing the parameters $M_\pi$, $\chi_\pi$ and $f^{\Delta E}$, we use Eq.~\eqref{eq:Q_formula} to compute the pion production.

\section{Comparison with existing work}
\label{App:Comparison_Biehl}
This appendix provides an assessment of the impact on the neutrino fluence of our NCMC approach in comparison with the one adopted in Ref.~\cite{Biehl:2017zlw}. 
Reference~\cite{Biehl:2017zlw} solved the nuclear cascade problem through a system of partial differential equations, while we rely on the NCMC method to randomly generate photodisintegration events. 
In order to allow for a fair comparison between our results, in this appendix, we set the initial parameters as in Ref.~\cite{Biehl:2017zlw}: 
$\Gamma = 300$, $\tilde{t}_\mathrm{var} = 0.01$ s, $\tilde{t}_\mathrm{dur} = 10$ s and $z=2$. The photon spectrum is  a Band function with $\alpha = -1.0$ and $\beta = -2.0$, and we compute the  spectral energy density of nuclei  for three different isotropic luminosities: $\tilde{L}_\mathrm{iso} = 10^{49}$ erg/s, $\tilde{L}_\mathrm{iso} = 10^{51}$ erg/s, and $\tilde{L}_\mathrm{iso} = 10^{53}$ erg/s. 

\begin{figure}
    \centering
    \includegraphics[width = \linewidth]{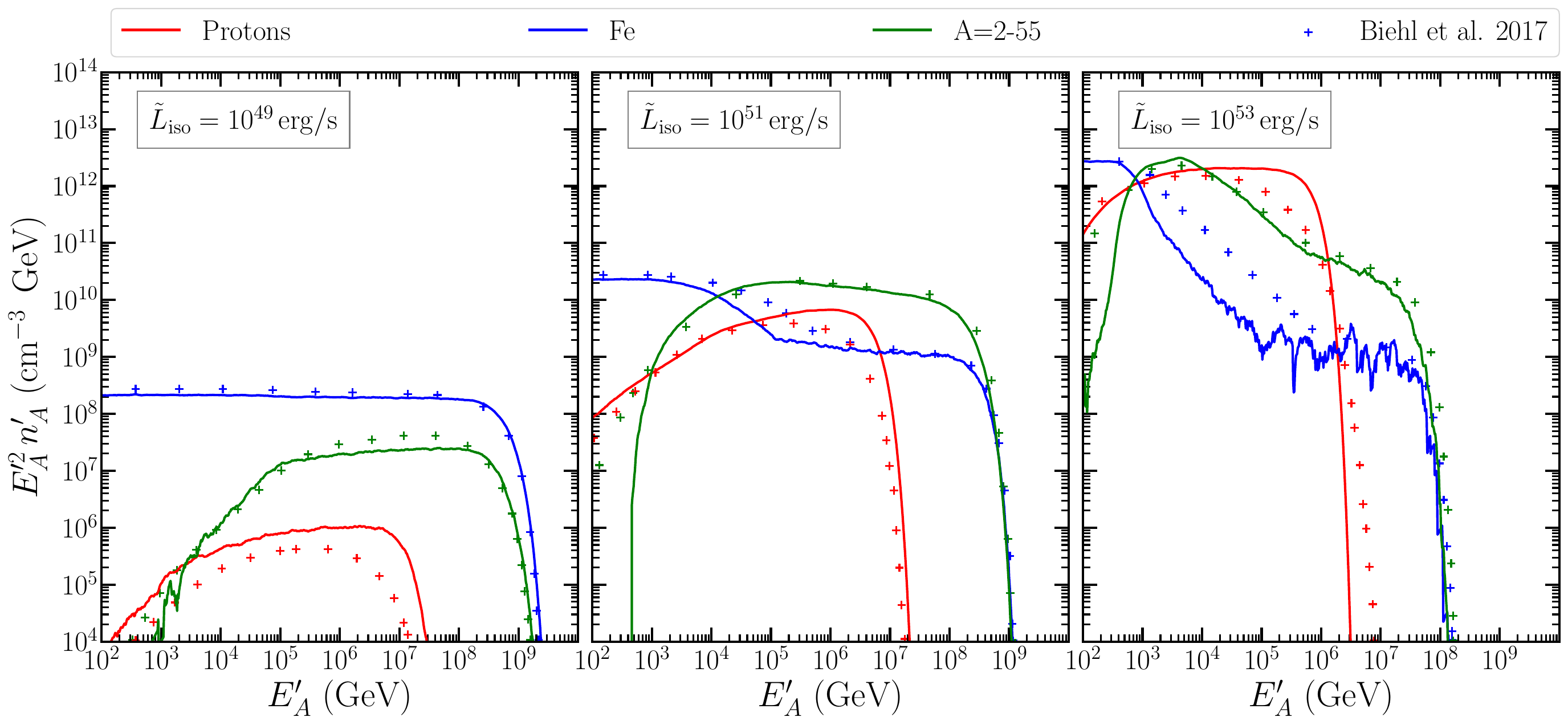}
    \caption{Spectral energy density of nuclei as a function of the particle energy for three different values of $\tilde{L}_\mathrm{iso}$, from left to right respectively. The crosses indicate the results obtained by Ref.~\cite{Biehl:2017zlw} with comparable initial conditions. 
    The results obtained through our NCMC  algorithm and the ones from Ref.~\cite{Biehl:2017zlw} are in overall good agreement. We  note that our approach  tends to overestimate the proton density in the acceleration region because we assume that  all secondary products are emitted as nucleons. Conversely, the density of nuclei tends to be underestimated. 
}
    \label{fig:compar_biehl_density}
\end{figure}

Figure~\ref{fig:compar_biehl_density} displays the spectral energy density of nuclei obtained through our NCMC approach and the  results of Ref.~\cite{Biehl:2017zlw}  obtained for three different values of $\tilde{L}_\mathrm{iso}$. We can see that there is a general good agreement between the spectra computed relying on our NCMC algorithm and the method based on the solution of  partial differential equations of Ref.~\cite{Biehl:2017zlw}.
In order to limit the computational time, we assume that only nucleons, and no intermediate mass nuclei, can be produced during a photodisintegration event.  Because of this, we overestimate the density of nucleons in the jet. 
This simplification mostly affects  the spectral density of nucleons and intermediate nuclei for scenarios where nuclear cascades are not efficient (cf.~the red and green lines in the left panel of Fig.~\ref{fig:compar_biehl_density}). In the opposite scenario of efficient cascades, any additional intermediate nuclei would be rapidly photodisintegrated into nucleons by the dense photon field, leading to similar results.
Overall, we conclude that the agreement between our model and the one of Ref.~\cite{Biehl:2017zlw}  is satisfactory  to inspect the trends in the  neutrino fluence across the jet parameter space. 

In order to compute the neutrino fluence from the spectral energy densities of nuclei and nucleons, Ref.~\cite{Biehl:2017zlw} adopts the single-particle model (SPM), namely  nuclei are approximated by the sum of independent nucleons to compute their photohadronic cross-section. On the other hand,  we rely on the empirical model (EM) cross-section of Ref.~\cite{Morejon:2019pfu}. 
A comparison of the neutrino fluence obtained adopting these two approaches is shown in Fig.~\ref{fig:compar_biehl_neutrino}.  Note that there is no difference between the neutrino fluences from nucleons assuming either the SPM or the EM since we do not extrapolate the EM for the case $A=1$ and prefer to use the photohadronic model from Ref.~\cite{Hummer:2010vx} {(note that red solid and dashed lines are on top of each other in all panels of Fig.~\ref{fig:compar_biehl_neutrino}). We can also note that there is a systematic larger neutrino production from nucleons at low energies in Ref.~\cite{Biehl:2017zlw}; this is due to the fact that they do not take into account the cooling of free neutrons, which leads to an overestimation of the number of neutrinos from $\beta^-$ decay. Using the SPM, we obtain the  neutrino fluence for each species as the one presented in Ref.~\cite{Biehl:2017zlw}.
However, there is a  discrepancy when we rely on the  EM. Relying on the latter, Ref.~\cite{Morejon:2019pfu} found that the photohadronic cross section for meson production is smaller than the SPM one. 
Using inputs from the EM instead of the SPM, we find a difference of approximately one order of magnitude when computing the neutrino fluence from nuclei. 
Therefore, for cases of inefficient cascades when the neutrino production is dominated by $A\gamma$ interactions, we find a  total neutrino fluence up to one order of magnitude smaller than then one obtained with the SPM 
(see left panel and middle panels of Fig.~\ref{fig:compar_biehl_neutrino}). Conversely, the total neutrino production agrees well for the efficient cascade scenarios, since the contribution from nuclei is negligible with respect to the one from nucleons (see right panel of Fig.~\ref{fig:compar_biehl_neutrino}).

In conclusion, when nuclear cascades are inefficient and the total neutrino fluence is dominated by nuclei, we report an overall smaller neutrino fluence than Ref.~~\cite{Biehl:2017zlw}. 
However, the employment of the EM and the SPM has no impact when considering efficient cascades since the neutrino production is dominated by protons and neutrons. 

\begin{figure}
    \includegraphics[width = \linewidth]{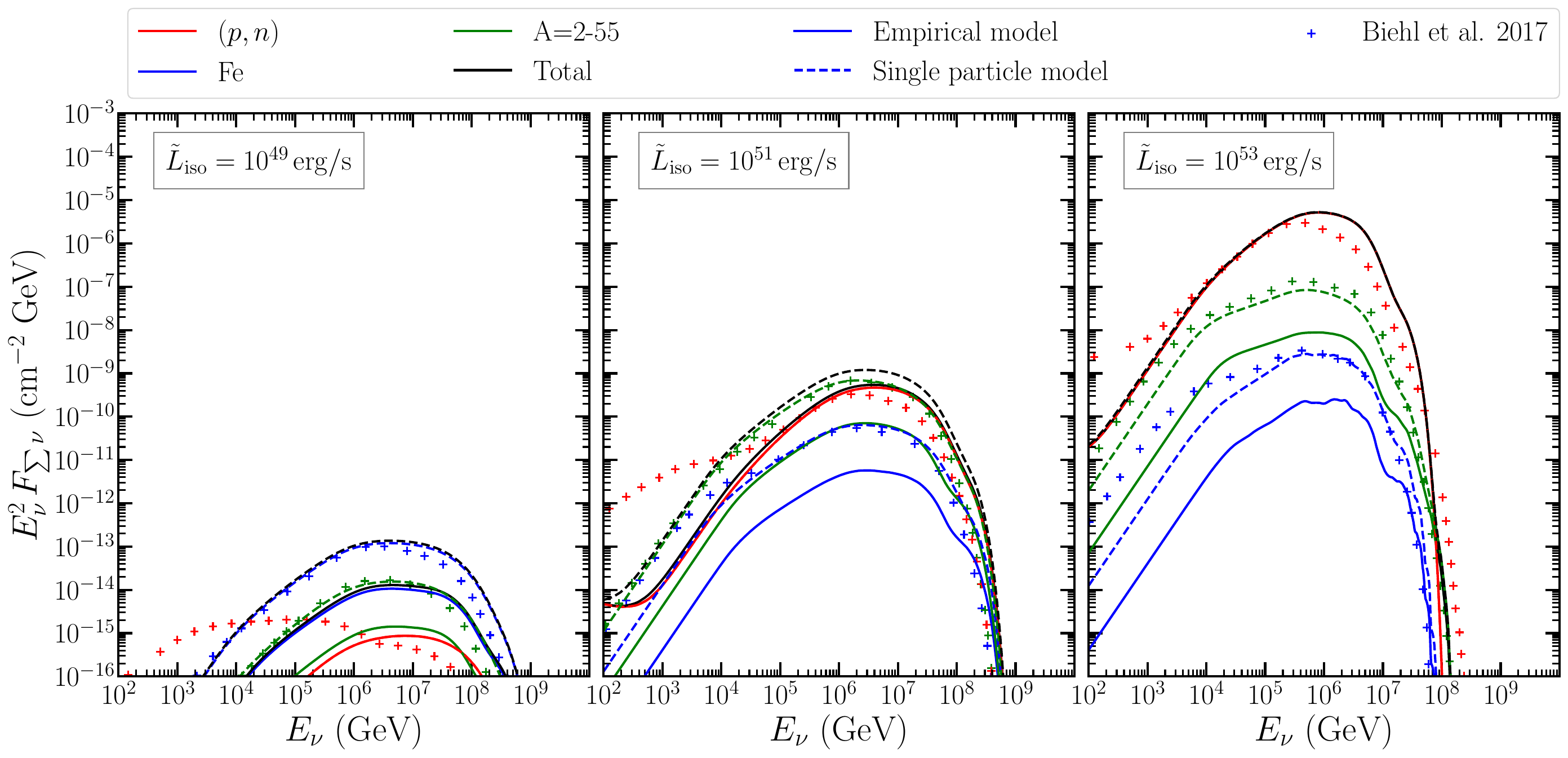} \\
           \caption{All flavor neutrino fluence as a function of the neutrino energy for three different isotropic luminosities, from left to right, respectively. The neutrino fluence computed relying on the EM (SPM)  is plotted with solid (dashed) lines. The different colors mark  neutrino production from nucleons (red), iron (blue), and intermediate nuclei (green), and the total one (black). 
           Crosses indicate the results obtained by Ref.~\cite{Biehl:2017zlw} for comparable initial conditions. The neutrino production from nuclei is one order of magnitude lower when  the EM cross section is used. The total neutrino fluence is also one order of magnitude smaller when nuclear cascades are inefficient. The neutrino production from nucleons, however, does not depend on the   cross-section model (SPM vs.~EM, cf.~red lines, the solid  and dashed lines are exactly on top of each other and therefore not distinguishable).
           }
           \label{fig:compar_biehl_neutrino}
\end{figure}

\bibliographystyle{JHEP}
\bibliography{ref.bib}

\providecommand{\href}[2]{#2}\begingroup\raggedright\begin{thebibliography}{100}

\bibitem{Frederiks:2023bxg}
D.~Frederiks et~al., \emph{{Properties of the Extremely Energetic GRB 221009A
  from Konus-WIND and SRG/ART-XC Observations}},
  \href{https://doi.org/10.3847/2041-8213/acd1eb}{\emph{Astrophys. J. Lett.}
  {\bfseries 949} (2023) L7},
  [\href{https://arxiv.org/abs/2302.13383}{{\ttfamily 2302.13383}}].

\bibitem{Eichler:1989ve}
D.~Eichler, M.~Livio, T.~Piran and D.~N. Schramm, \emph{{Nucleosynthesis,
  Neutrino Bursts and Gamma-Rays from Coalescing Neutron Stars}},
  \href{https://doi.org/10.1038/340126a0}{\emph{Nature} {\bfseries 340} (1989)
  126--128}.

\bibitem{Paczynski:1991aq}
B.~Paczynski, \emph{{Cosmological gamma-ray bursts}}, {\emph{Acta Astron.}
  {\bfseries 41} (1991) 257--267}.

\bibitem{Nakar:2007yr}
E.~Nakar, \emph{{Short-Hard Gamma-Ray Bursts}},
  \href{https://doi.org/10.1016/j.physrep.2007.02.005}{\emph{Phys. Rept.}
  {\bfseries 442} (2007) 166--236},
  [\href{https://arxiv.org/abs/astro-ph/0701748}{{\ttfamily
  astro-ph/0701748}}].

\bibitem{LIGOScientific:2017zic}
{\scshape LIGO Scientific, Virgo, Fermi-GBM, INTEGRAL} collaboration, B.~P.
  Abbott et~al., \emph{{Gravitational Waves and Gamma-rays from a Binary
  Neutron Star Merger: GW170817 and GRB 170817A}},
  \href{https://doi.org/10.3847/2041-8213/aa920c}{\emph{Astrophys. J. Lett.}
  {\bfseries 848} (2017) L13},
  [\href{https://arxiv.org/abs/1710.05834}{{\ttfamily 1710.05834}}].

\bibitem{LIGOScientific:2017ync}
{\scshape LIGO Scientific, Virgo, Fermi GBM, INTEGRAL, IceCube, AstroSat
  Cadmium Zinc Telluride Imager Team, IPN, Insight-Hxmt, ANTARES, Swift, AGILE
  Team, 1M2H Team, Dark Energy Camera GW-EM, DES, DLT40, GRAWITA, Fermi-LAT,
  ATCA, ASKAP, Las Cumbres Observatory Group, OzGrav, DWF (Deeper Wider Faster
  Program), AST3, CAASTRO, VINROUGE, MASTER, J-GEM, GROWTH, JAGWAR,
  CaltechNRAO, TTU-NRAO, NuSTAR, Pan-STARRS, MAXI Team, TZAC Consortium, KU,
  Nordic Optical Telescope, ePESSTO, GROND, Texas Tech University, SALT Group,
  TOROS, BOOTES, MWA, CALET, IKI-GW Follow-up, H.E.S.S., LOFAR, LWA, HAWC,
  Pierre Auger, ALMA, Euro VLBI Team, Pi of Sky, Chandra Team at McGill
  University, DFN, ATLAS Telescopes, High Time Resolution Universe Survey,
  RIMAS, RATIR, SKA South Africa/MeerKAT} collaboration, B.~P. Abbott et~al.,
  \emph{{Multi-messenger Observations of a Binary Neutron Star Merger}},
  \href{https://doi.org/10.3847/2041-8213/aa91c9}{\emph{Astrophys. J. Lett.}
  {\bfseries 848} (2017) L12},
  [\href{https://arxiv.org/abs/1710.05833}{{\ttfamily 1710.05833}}].

\bibitem{Mooley:2017enz}
K.~P. Mooley et~al., \emph{{A mildly relativistic wide-angle outflow in the
  neutron star merger GW170817}},
  \href{https://doi.org/10.1038/nature25452}{\emph{Nature} {\bfseries 554}
  (2018) 207}, [\href{https://arxiv.org/abs/1711.11573}{{\ttfamily
  1711.11573}}].

\bibitem{Goldstein:2017mmi}
A.~Goldstein et~al., \emph{{An Ordinary Short Gamma-Ray Burst with
  Extraordinary Implications: Fermi-GBM Detection of GRB 170817A}},
  \href{https://doi.org/10.3847/2041-8213/aa8f41}{\emph{Astrophys. J. Lett.}
  {\bfseries 848} (2017) L14},
  [\href{https://arxiv.org/abs/1710.05446}{{\ttfamily 1710.05446}}].

\bibitem{Woosley:2006fn}
S.~E. Woosley and J.~S. Bloom, \emph{{The Supernova Gamma-Ray Burst
  Connection}},
  \href{https://doi.org/10.1146/annurev.astro.43.072103.150558}{\emph{Ann. Rev.
  Astron. Astrophys.} {\bfseries 44} (2006) 507--556},
  [\href{https://arxiv.org/abs/astro-ph/0609142}{{\ttfamily
  astro-ph/0609142}}].

\bibitem{Modjaz:2015cca}
M.~Modjaz, Y.~Q. Liu, F.~B. Bianco and O.~Graur, \emph{{The Spectral SN-GRB
  Connection: Systematic Spectral Comparisons between Type Ic Supernovae, and
  broad-lined Type Ic Supernovae with and without Gamma-Ray Bursts}},
  \href{https://doi.org/10.3847/0004-637X/832/2/108}{\emph{Astrophys. J.}
  {\bfseries 832} (2016) 108},
  [\href{https://arxiv.org/abs/1509.07124}{{\ttfamily 1509.07124}}].

\bibitem{Yang:2023mqt}
Y.-H. Yang et~al., \emph{{A lanthanide-rich kilonova in the aftermath of a long
  gamma-ray burst}},
  \href{https://doi.org/10.1038/s41586-023-06979-5}{\emph{Nature} {\bfseries
  626} (Feb., 2024) 742--745},
  [\href{https://arxiv.org/abs/2308.00638}{{\ttfamily 2308.00638}}].

\bibitem{Troja:2022yya}
E.~Troja et~al., \emph{{A nearby long gamma-ray burst from a merger of compact
  objects}}, \href{https://doi.org/10.1038/s41586-022-05327-3}{\emph{Nature}
  {\bfseries 612} (2022) 228--231},
  [\href{https://arxiv.org/abs/2209.03363}{{\ttfamily 2209.03363}}].

\bibitem{Gillanders:2023zys}
J.~H. Gillanders et~al., \emph{{Heavy element nucleosynthesis associated with a
  gamma-ray burst}},  \href{https://arxiv.org/abs/2308.00633}{{\ttfamily
  2308.00633}}.

\bibitem{Sun:2015bda}
H.~Sun, B.~Zhang and Z.~Li, \emph{{Extragalactic High-energy Transients: Event
  Rate Densities and Luminosity Functions}},
  \href{https://doi.org/10.1088/0004-637X/812/1/33}{\emph{Astrophys. J.}
  {\bfseries 812} (2015) 33},
  [\href{https://arxiv.org/abs/1509.01592}{{\ttfamily 1509.01592}}].

\bibitem{Virgili:2008gp}
F.~Virgili, E.~Liang and B.~Zhang, \emph{{Low-Luminosity Gamma-Ray Bursts as a
  Distinct GRB Population:A Monte Carlo Analysis}},
  \href{https://doi.org/10.1111/j.1365-2966.2008.14063.x}{\emph{Mon. Not. Roy.
  Astron. Soc.} {\bfseries 392} (2009) 91},
  [\href{https://arxiv.org/abs/0801.4751}{{\ttfamily 0801.4751}}].

\bibitem{Paczynski:1986px}
B.~Paczynski, \emph{{Gamma-ray bursters at cosmological distances}},
  \href{https://doi.org/10.1086/184740}{\emph{Astrophys. J. Lett.} {\bfseries
  308} (1986) L43--L46}.

\bibitem{Drenkhahn:2002ug}
G.~Drenkhahn and H.~C. Spruit, \emph{{Efficient acceleration and radiation in
  Poynting flux powered GRB outflows}},
  \href{https://doi.org/10.1051/0004-6361:20020839}{\emph{Astron. Astrophys.}
  {\bfseries 391} (2002) 1141},
  [\href{https://arxiv.org/abs/astro-ph/0202387}{{\ttfamily
  astro-ph/0202387}}].

\bibitem{Rees:1994nw}
M.~J. Rees and P.~M{\'e}sz{\'a}ros, \emph{{Unsteady outflow models for
  cosmological gamma-ray bursts}},
  \href{https://doi.org/10.1086/187446}{\emph{Astrophys. J. Lett.} {\bfseries
  430} (1994) L93--L96},
  [\href{https://arxiv.org/abs/astro-ph/9404038}{{\ttfamily
  astro-ph/9404038}}].

\bibitem{Spitkovsky:2008fi}
A.~Spitkovsky, \emph{{Particle acceleration in relativistic collisionless
  shocks: Fermi process at last?}},
  \href{https://doi.org/10.1086/590248}{\emph{Astrophys. J. Lett.} {\bfseries
  682} (2008) L5}, [\href{https://arxiv.org/abs/0802.3216}{{\ttfamily
  0802.3216}}].

\bibitem{Beloborodov:2017use}
A.~M. Beloborodov and P.~M\'esz\'aros, \emph{{Photospheric Emission of
  Gamma-Ray Bursts}},
  \href{https://doi.org/10.1007/s11214-017-0348-6}{\emph{Space Sci. Rev.}
  {\bfseries 207} (2017) 87--110},
  [\href{https://arxiv.org/abs/1701.04523}{{\ttfamily 1701.04523}}].

\bibitem{Peer:2016mqn}
A.~Pe'er and F.~Ryde, \emph{{Photospheric Emission in Gamma-Ray Bursts}},
  \href{https://doi.org/10.1142/S021827181730018X}{\emph{Int. J. Mod. Phys. D}
  {\bfseries 26} (2017) 1730018},
  [\href{https://arxiv.org/abs/1603.05058}{{\ttfamily 1603.05058}}].

\bibitem{Spruit:2000zm}
H.~C. Spruit, F.~Daigne and G.~Drenkhahn, \emph{{Large scale magnetic fields
  and their dissipation in grb fireballs}},
  \href{https://doi.org/10.1051/0004-6361:20010131}{\emph{Astron. Astrophys.}
  {\bfseries 369} (2001) 694},
  [\href{https://arxiv.org/abs/astro-ph/0004274}{{\ttfamily
  astro-ph/0004274}}].

\bibitem{Rudolph:2023auv}
A.~Rudolph, I.~Tamborra and O.~Gottlieb, \emph{{Subphotospheric Emission from
  Short Gamma-Ray Bursts: Protons Mold the Multimessenger Signals}},
  \href{https://doi.org/10.3847/2041-8213/ad1525}{\emph{Astrophys. J. Lett.}
  {\bfseries 961} (2024) L7},
  [\href{https://arxiv.org/abs/2309.08667}{{\ttfamily 2309.08667}}].

\bibitem{Waxman:2003vh}
E.~Waxman, \emph{{Gamma-ray bursts: The Underlying model}},
  \href{https://doi.org/10.1007/3-540-45863-8_19}{\emph{Lect. Notes Phys.}
  {\bfseries 598} (2003) 393},
  [\href{https://arxiv.org/abs/astro-ph/0303517}{{\ttfamily
  astro-ph/0303517}}].

\bibitem{Meszaros:2017fcs}
P.~M{\'e}sz{\'a}ros, \emph{{Astrophysical Sources of High Energy Neutrinos in
  the IceCube Era}},
  \href{https://doi.org/10.1146/annurev-nucl-101916-123304}{\emph{Ann. Rev.
  Nucl. Part. Sci.} {\bfseries 67} (2017) 45--67},
  [\href{https://arxiv.org/abs/1708.03577}{{\ttfamily 1708.03577}}].

\bibitem{Piran:2004ba}
T.~Piran, \emph{{The physics of gamma-ray bursts}},
  \href{https://doi.org/10.1103/RevModPhys.76.1143}{\emph{Rev. Mod. Phys.}
  {\bfseries 76} (2004) 1143--1210},
  [\href{https://arxiv.org/abs/astro-ph/0405503}{{\ttfamily
  astro-ph/0405503}}].

\bibitem{Moore:2023sgo}
E.~Moore, B.~Gendre, N.~B. Orange and F.~H. Panther, \emph{{Constraints on the
  ultra-high energy cosmic ray output of gamma-ray bursts}},
  \href{https://doi.org/10.1093/mnras/stae873}{\emph{Mon. Not. Roy. Astron.
  Soc.} {\bfseries 530} (2024) 555--559},
  [\href{https://arxiv.org/abs/2303.13781}{{\ttfamily 2303.13781}}].

\bibitem{Guarini:2023rnd}
E.~Guarini, I.~Tamborra, R.~Margutti and E.~Ramirez-Ruiz, \emph{{Transients
  stemming from collapsing massive stars: The missing pieces to advance joint
  observations of photons and high-energy neutrinos}},
  \href{https://doi.org/10.1103/PhysRevD.108.083035}{\emph{Phys. Rev. D}
  {\bfseries 108} (2023) 083035},
  [\href{https://arxiv.org/abs/2308.03840}{{\ttfamily 2308.03840}}].

\bibitem{Boncioli:2018lrv}
D.~Boncioli, D.~Biehl and W.~Winter, \emph{{On the common origin of cosmic rays
  across the ankle and diffuse neutrinos at the highest energies from
  low-luminosity Gamma-Ray Bursts}},
  \href{https://doi.org/10.3847/1538-4357/aafda7}{\emph{Astrophys. J.}
  {\bfseries 872} (2019) 110},
  [\href{https://arxiv.org/abs/1808.07481}{{\ttfamily 1808.07481}}].

\bibitem{Beloborodov:2002af}
A.~M. Beloborodov, \emph{{Nuclear composition of gamma-ray burst fireballs}},
  \href{https://doi.org/10.1086/374217}{\emph{Astrophys. J.} {\bfseries 588}
  (2003) 931--944}, [\href{https://arxiv.org/abs/astro-ph/0210522}{{\ttfamily
  astro-ph/0210522}}].

\bibitem{Horiuchi:2012by}
S.~Horiuchi, K.~Murase, K.~Ioka and P.~M{\'e}sz{\'a}ros, \emph{{The survival of
  nuclei in jets associated with core-collapse supernovae and gamma-ray
  bursts}}, \href{https://doi.org/10.1088/0004-637X/753/1/69}{\emph{Astrophys.
  J.} {\bfseries 753} (2012) 69},
  [\href{https://arxiv.org/abs/1203.0296}{{\ttfamily 1203.0296}}].

\bibitem{Biehl:2017zlw}
D.~Biehl, D.~Boncioli, A.~Fedynitch and W.~Winter, \emph{{Cosmic-Ray and
  Neutrino Emission from Gamma-Ray Bursts with a Nuclear Cascade}},
  \href{https://doi.org/10.1051/0004-6361/201731337}{\emph{Astron. Astrophys.}
  {\bfseries 611} (2018) A101},
  [\href{https://arxiv.org/abs/1705.08909}{{\ttfamily 1705.08909}}].

\bibitem{Pitik:2021xhb}
T.~Pitik, I.~Tamborra and M.~Petropoulou, \emph{{Neutrino signal dependence on
  gamma-ray burst emission mechanism}},
  \href{https://doi.org/10.1088/1475-7516/2021/05/034}{\emph{JCAP} {\bfseries
  05} (2021) 034}, [\href{https://arxiv.org/abs/2102.02223}{{\ttfamily
  2102.02223}}].

\bibitem{Heinze:2020zqb}
J.~Heinze, D.~Biehl, A.~Fedynitch, D.~Boncioli, A.~Rudolph and W.~Winter,
  \emph{{Systematic parameter space study for the UHECR origin from GRBs in
  models with multiple internal shocks}},
  \href{https://doi.org/10.1093/mnras/staa2751}{\emph{Mon. Not. Roy. Astron.
  Soc.} {\bfseries 498} (2020) 5990--6004},
  [\href{https://arxiv.org/abs/2006.14301}{{\ttfamily 2006.14301}}].

\bibitem{Rudolph:2022ppp}
A.~Rudolph, M.~Petropoulou, {\v{Z}}.~Bo\v{s}njak and W.~Winter,
  \emph{{Multicollision Internal Shock Lepto-hadronic Models for Energetic
  Gamma-Ray Bursts (GRBs)}},
  \href{https://doi.org/10.3847/1538-4357/acc861}{\emph{Astrophys. J.}
  {\bfseries 950} (2023) 28},
  [\href{https://arxiv.org/abs/2212.00765}{{\ttfamily 2212.00765}}].

\bibitem{Murase:2008sp}
K.~Murase, \emph{{Prompt High-Energy Neutrinos from Gamma-Ray Bursts in the
  Photospheric and Synchrotron Self-Compton Scenarios}},
  \href{https://doi.org/10.1103/PhysRevD.78.101302}{\emph{Phys. Rev. D}
  {\bfseries 78} (2008) 101302},
  [\href{https://arxiv.org/abs/0807.0919}{{\ttfamily 0807.0919}}].

\bibitem{Wang:2007xj}
X.-Y. Wang, S.~Razzaque and P.~M{\'e}sz{\'a}ros, \emph{{On the Origin and
  Survival of UHE Cosmic-Ray Nuclei in GRBs and Hypernovae}},
  \href{https://doi.org/10.1086/529018}{\emph{Astrophys. J.} {\bfseries 677}
  (2008) 432--440}, [\href{https://arxiv.org/abs/0711.2065}{{\ttfamily
  0711.2065}}].

\bibitem{Murase:2010gj}
K.~Murase and J.~F. Beacom, \emph{{Neutrino Background Flux from Sources of
  Ultrahigh-Energy Cosmic-Ray Nuclei}},
  \href{https://doi.org/10.1103/PhysRevD.81.123001}{\emph{Phys. Rev. D}
  {\bfseries 81} (2010) 123001},
  [\href{https://arxiv.org/abs/1003.4959}{{\ttfamily 1003.4959}}].

\bibitem{Toma:2010xw}
K.~Toma, X.-F. Wu and P.~M{\'e}sz{\'a}ros, \emph{{A Photosphere-Internal Shock
  Model of Gamma-Ray Bursts: Case Studies of Fermi/LAT Bursts}},
  \href{https://doi.org/10.1111/j.1365-2966.2011.18807.x}{\emph{Mon. Not. Roy.
  Astron. Soc.} {\bfseries 415} (2011) 1663--1680},
  [\href{https://arxiv.org/abs/1002.2634}{{\ttfamily 1002.2634}}].

\bibitem{Zhang:2010jt}
B.~Zhang and H.~Yan, \emph{{The Internal-Collision-Induced Magnetic
  Reconnection and Turbulence (ICMART) Model of Gamma-Ray Bursts}},
  \href{https://doi.org/10.1088/0004-637X/726/2/90}{\emph{Astrophys. J.}
  {\bfseries 726} (2011) 90},
  [\href{https://arxiv.org/abs/1011.1197}{{\ttfamily 1011.1197}}].

\bibitem{Bromberg:2011fg}
O.~Bromberg, E.~Nakar, T.~Piran and R.~Sari, \emph{{The propagation of
  relativistic jets in external media}},
  \href{https://doi.org/10.1088/0004-637X/740/2/100}{\emph{Astrophys. J.}
  {\bfseries 740} (2011) 100},
  [\href{https://arxiv.org/abs/1107.1326}{{\ttfamily 1107.1326}}].

\bibitem{Kobayashi:1997jk}
S.~Kobayashi, T.~Piran and R.~Sari, \emph{{Can internal shocks produce the
  variability in GRBs?}},
  \href{https://doi.org/10.1086/512791}{\emph{Astrophys. J.} {\bfseries 490}
  (1997) 92--98}, [\href{https://arxiv.org/abs/astro-ph/9705013}{{\ttfamily
  astro-ph/9705013}}].

\bibitem{Daigne:1998xc}
F.~Daigne and R.~Mochkovitch, \emph{{Gamma-ray bursts from internal shocks in a
  relativistic wind: temporal and spectral properties}},
  \href{https://doi.org/10.1046/j.1365-8711.1998.01305.x}{\emph{Mon. Not. Roy.
  Astron. Soc.} {\bfseries 296} (1998) 275},
  [\href{https://arxiv.org/abs/astro-ph/9801245}{{\ttfamily
  astro-ph/9801245}}].

\bibitem{Guetta:2000ye}
D.~Guetta, M.~Spada and E.~Waxman, \emph{{Efficiency and spectrum of internal
  gamma-ray burst shocks}},
  \href{https://doi.org/10.1086/321543}{\emph{Astrophys. J.} {\bfseries 557}
  (2001) 399}, [\href{https://arxiv.org/abs/astro-ph/0011170}{{\ttfamily
  astro-ph/0011170}}].

\bibitem{Bustamante:2016wpu}
M.~Bustamante, K.~Murase, W.~Winter and J.~Heinze, \emph{{Multi-messenger light
  curves from gamma-ray bursts in the internal shock model}},
  \href{https://doi.org/10.3847/1538-4357/837/1/33}{\emph{Astrophys. J.}
  {\bfseries 837} (2017) 33},
  [\href{https://arxiv.org/abs/1606.02325}{{\ttfamily 1606.02325}}].

\bibitem{Rudolph:2019ccl}
A.~Rudolph, J.~Heinze, A.~Fedynitch and W.~Winter, \emph{{Impact of the
  Collision Model on the Multi-messenger Emission from Gamma-Ray Burst Internal
  Shocks}}, \href{https://doi.org/10.3847/1538-4357/ab7ea7}{\emph{Astrophys.
  J.} {\bfseries 893} (2020) 72},
  [\href{https://arxiv.org/abs/1907.10633}{{\ttfamily 1907.10633}}].

\bibitem{Globus:2014fka}
N.~Globus, D.~Allard, R.~Mochkovitch and E.~Parizot, \emph{{UHECR acceleration
  at GRB internal shocks}},
  \href{https://doi.org/10.1093/mnras/stv893}{\emph{Mon. Not. Roy. Astron.
  Soc.} {\bfseries 451} (2015) 751--790},
  [\href{https://arxiv.org/abs/1409.1271}{{\ttfamily 1409.1271}}].

\bibitem{Sironi:2010rb}
L.~Sironi and A.~Spitkovsky, \emph{{Particle Acceleration in Relativistic
  Magnetized Collisionless Electron-Ion Shocks}},
  \href{https://doi.org/10.1088/0004-637X/726/2/75}{\emph{Astrophys. J.}
  {\bfseries 726} (2011) 75},
  [\href{https://arxiv.org/abs/1009.0024}{{\ttfamily 1009.0024}}].

\bibitem{Crumley:2018kvf}
P.~Crumley, D.~Caprioli, S.~Markoff and A.~Spitkovsky, \emph{{Kinetic
  simulations of mildly relativistic shocks \textendash{} I. Particle
  acceleration in high Mach number shocks}},
  \href{https://doi.org/10.1093/mnras/stz232}{\emph{Mon. Not. Roy. Astron.
  Soc.} {\bfseries 485} (2019) 5105--5119},
  [\href{https://arxiv.org/abs/1809.10809}{{\ttfamily 1809.10809}}].

\bibitem{Lipari:2007su}
P.~Lipari, M.~Lusignoli and D.~Meloni, \emph{{Flavor Composition and Energy
  Spectrum of Astrophysical Neutrinos}},
  \href{https://doi.org/10.1103/PhysRevD.75.123005}{\emph{Phys. Rev. D}
  {\bfseries 75} (2007) 123005},
  [\href{https://arxiv.org/abs/0704.0718}{{\ttfamily 0704.0718}}].

\bibitem{Wang:2015vpa}
X.-G. Wang et~al., \emph{{How bad or Good are the External Forward Shock
  Afterglow Models of Gamma-ray Bursts?}},
  \href{https://doi.org/10.1088/0067-0049/219/1/9}{\emph{Astrophys. J. Suppl.}
  {\bfseries 219} (2015) 9},
  [\href{https://arxiv.org/abs/1503.03193}{{\ttfamily 1503.03193}}].

\bibitem{IceCube:2017amx}
{\scshape IceCube} collaboration, M.~G. Aartsen et~al., \emph{{Extending the
  search for muon neutrinos coincident with gamma-ray bursts in IceCube data}},
  \href{https://doi.org/10.3847/1538-4357/aa7569}{\emph{Astrophys. J.}
  {\bfseries 843} (2017) 112},
  [\href{https://arxiv.org/abs/1702.06868}{{\ttfamily 1702.06868}}].

\bibitem{Ghirlanda:2017opl}
G.~Ghirlanda, F.~Nappo, G.~Ghisellini, A.~Melandri, G.~Marcarini, L.~Nava
  et~al., \emph{{Bulk Lorentz factors of Gamma-Ray Bursts}},
  \href{https://doi.org/10.1051/0004-6361/201731598}{\emph{Astron. Astrophys.}
  {\bfseries 609} (2018) A112},
  [\href{https://arxiv.org/abs/1711.06257}{{\ttfamily 1711.06257}}].

\bibitem{Zitouni:2018wre}
H.~Zitouni, N.~Guessoum, K.~M. ALQassimi and O.~Alaryani, \emph{{Distributions
  of Pseudo-Redshifts and Durations (Observed and Intrinsic) of Fermi GRBs}},
  \href{https://doi.org/10.1007/s10509-018-3449-0}{\emph{Astrophys. Space Sci.}
  {\bfseries 363} (2018) 223},
  [\href{https://arxiv.org/abs/1810.04124}{{\ttfamily 1810.04124}}].

\bibitem{ANTARES:2020vzs}
{\scshape ANTARES} collaboration, A.~Albert et~al., \emph{{Constraining the
  contribution of Gamma-Ray Bursts to the high-energy diffuse neutrino flux
  with 10 yr of ANTARES data}},
  \href{https://doi.org/10.1093/mnras/staa3503}{\emph{Mon. Not. Roy. Astron.
  Soc.} {\bfseries 500} (2020) 5614--5628},
  [\href{https://arxiv.org/abs/2008.02127}{{\ttfamily 2008.02127}}].

\bibitem{Deng:2015xea}
W.~Deng, H.~Li, B.~Zhang and S.~Li, \emph{{Relativistic MHD simulations of
  collision-induced magnetic dissipation in Poynting-flux-dominated
  jets/outflows}},
  \href{https://doi.org/10.1088/0004-637X/805/2/163}{\emph{Astrophys. J.}
  {\bfseries 805} (2015) 163},
  [\href{https://arxiv.org/abs/1501.07595}{{\ttfamily 1501.07595}}].

\bibitem{Sironi:2015eoa}
L.~Sironi, M.~Petropoulou and D.~Giannios, \emph{{Relativistic Jets Shine
  through Shocks or Magnetic Reconnection?}},
  \href{https://doi.org/10.1093/mnras/stv641}{\emph{Mon. Not. Roy. Astron.
  Soc.} {\bfseries 450} (2015) 183--191},
  [\href{https://arxiv.org/abs/1502.01021}{{\ttfamily 1502.01021}}].

\bibitem{Groselj:2024dnv}
D.~Groselj, L.~Sironi and A.~Spitkovsky, \emph{{Long-term Evolution of
  Relativistic Unmagnetized Collisionless Shocks}},
  \href{https://doi.org/10.3847/2041-8213/ad2c8c}{\emph{Astrophys. J. Lett.}
  {\bfseries 963} (2024) L44},
  [\href{https://arxiv.org/abs/2401.02392}{{\ttfamily 2401.02392}}].

\bibitem{Sironi:2013ri}
L.~Sironi, A.~Spitkovsky and J.~Arons, \emph{{The Maximum Energy of Accelerated
  Particles in Relativistic Collisionless Shocks}},
  \href{https://doi.org/10.1088/0004-637X/771/1/54}{\emph{Astrophys. J.}
  {\bfseries 771} (2013) 54},
  [\href{https://arxiv.org/abs/1301.5333}{{\ttfamily 1301.5333}}].

\bibitem{Zhang:2023lvw}
H.~Zhang, L.~Sironi, D.~Giannios and M.~Petropoulou, \emph{{The Origin of
  Power-law Spectra in Relativistic Magnetic Reconnection}},
  \href{https://doi.org/10.3847/2041-8213/acfe7c}{\emph{Astrophys. J. Lett.}
  {\bfseries 956} (2023) L36},
  [\href{https://arxiv.org/abs/2302.12269}{{\ttfamily 2302.12269}}].

\bibitem{Sironi:2014jfa}
L.~Sironi and A.~Spitkovsky, \emph{{Relativistic Reconnection: an Efficient
  Source of Non-Thermal Particles}},
  \href{https://doi.org/10.1088/2041-8205/783/1/L21}{\emph{Astrophys. J. Lett.}
  {\bfseries 783} (2014) L21},
  [\href{https://arxiv.org/abs/1401.5471}{{\ttfamily 1401.5471}}].

\bibitem{Band:1993eg}
D.~Band et~al., \emph{{BATSE observations of gamma-ray burst spectra. 1.
  Spectral diversity.}}, \href{https://doi.org/10.1086/172995}{\emph{Astrophys.
  J.} {\bfseries 413} (1993) 281--292}.

\bibitem{Gruber:2014iza}
D.~Gruber et~al., \emph{{The Fermi GBM Gamma-Ray Burst Spectral Catalog: Four
  Years Of Data}},
  \href{https://doi.org/10.1088/0067-0049/211/1/12}{\emph{Astrophys. J. Suppl.}
  {\bfseries 211} (2014) 12},
  [\href{https://arxiv.org/abs/1401.5069}{{\ttfamily 1401.5069}}].

\bibitem{Amati:2006ky}
L.~Amati, \emph{{The E(p,i) - E(iso) correlation in grbs: updated observational
  status, re-analysis and main implications}},
  \href{https://doi.org/10.1111/j.1365-2966.2006.10840.x}{\emph{Mon. Not. Roy.
  Astron. Soc.} {\bfseries 372} (2006) 233--245},
  [\href{https://arxiv.org/abs/astro-ph/0601553}{{\ttfamily
  astro-ph/0601553}}].

\bibitem{Bahcall:2000sa}
J.~N. Bahcall and P.~M{\'e}sz{\'a}ros, \emph{{5-GeV to 10-GeV neutrinos from
  gamma-ray burst fireballs}},
  \href{https://doi.org/10.1103/PhysRevLett.85.1362}{\emph{Phys. Rev. Lett.}
  {\bfseries 85} (2000) 1362--1365},
  [\href{https://arxiv.org/abs/hep-ph/0004019}{{\ttfamily hep-ph/0004019}}].

\bibitem{Murase:2013hh}
K.~Murase, K.~Kashiyama and P.~M\'esz\'aros, \emph{{Subphotospheric Neutrinos
  from Gamma-Ray Bursts: The Role of Neutrons}},
  \href{https://doi.org/10.1103/PhysRevLett.111.131102}{\emph{Phys. Rev. Lett.}
  {\bfseries 111} (2013) 131102},
  [\href{https://arxiv.org/abs/1301.4236}{{\ttfamily 1301.4236}}].

\bibitem{Kashiyama:2013ata}
K.~Kashiyama, K.~Murase and P.~M\'esz\'aros, \emph{{Neutron-Proton-Converter
  Acceleration Mechanism at Subphotospheres of Relativistic Outflows}},
  \href{https://doi.org/10.1103/PhysRevLett.111.131103}{\emph{Phys. Rev. Lett.}
  {\bfseries 111} (2013) 131103},
  [\href{https://arxiv.org/abs/1304.1945}{{\ttfamily 1304.1945}}].

\bibitem{Wang:2008zm}
X.-Y. Wang and Z.-G. Dai, \emph{{Prompt TeV neutrinos from dissipative
  photospheres of gamma-ray bursts}},
  \href{https://doi.org/10.1088/0004-637X/691/2/L67}{\emph{Astrophys. J. Lett.}
  {\bfseries 691} (2009) L67--L71},
  [\href{https://arxiv.org/abs/0807.0290}{{\ttfamily 0807.0290}}].

\bibitem{Guarini:2022hry}
E.~Guarini, I.~Tamborra and O.~Gottlieb, \emph{{State-of-the-art collapsar jet
  simulations imply undetectable subphotospheric neutrinos}},
  \href{https://doi.org/10.1103/PhysRevD.107.023001}{\emph{Phys. Rev. D}
  {\bfseries 107} (2023) 023001},
  [\href{https://arxiv.org/abs/2210.03757}{{\ttfamily 2210.03757}}].

\bibitem{2018pgrb.book.....Z}
B.~{Zhang}, \emph{{The Physics of Gamma-Ray Bursts}}.
\newblock Cambridge University Press, 2018,
  \href{https://doi.org/10.1017/9781139226530}{10.1017/9781139226530}.

\bibitem{Gottlieb:2022old}
O.~Gottlieb, A.~Tchekhovskoy and R.~Margutti, \emph{{Shocked jets in CCSNe can
  power the zoo of fast blue optical transients}},
  \href{https://doi.org/10.1093/mnras/stac910}{\emph{Mon. Not. Roy. Astron.
  Soc.} {\bfseries 513} (2022) 3810--3817},
  [\href{https://arxiv.org/abs/2201.04636}{{\ttfamily 2201.04636}}].

\bibitem{Werner:2016fxe}
G.~R. Werner, D.~A. Uzdensky, M.~C. Begelman, B.~Cerutti and K.~Nalewajko,
  \emph{{Non-thermal particle acceleration in collisionless relativistic
  electron\textendash{}proton reconnection}},
  \href{https://doi.org/10.1093/mnras/stx2530}{\emph{Mon. Not. Roy. Astron.
  Soc.} {\bfseries 473} (2018) 4840--4861},
  [\href{https://arxiv.org/abs/1612.04493}{{\ttfamily 1612.04493}}].

\bibitem{Guo:2015ydj}
F.~Guo, X.~Li, H.~Li, W.~Daughton, B.~Zhang, N.~Lloyd-Ronning et~al.,
  \emph{{Efficient Production of High-energy Nonthermal Particles During
  Magnetic Reconnection in a Magnetically Dominated Ion\textendash{}electron
  Plasma}}, \href{https://doi.org/10.3847/2041-8205/818/1/L9}{\emph{Astrophys.
  J. Lett.} {\bfseries 818} (2016) L9},
  [\href{https://arxiv.org/abs/1511.01434}{{\ttfamily 1511.01434}}].

\bibitem{Gao:2012ay}
S.~Gao, K.~Asano and P.~M{\'e}sz{\'a}ros, \emph{{High Energy Neutrinos from
  Dissipative Photospheric Models of Gamma Ray Bursts}},
  \href{https://doi.org/10.1088/1475-7516/2012/11/058}{\emph{JCAP} {\bfseries
  11} (2012) 058}, [\href{https://arxiv.org/abs/1210.1186}{{\ttfamily
  1210.1186}}].

\bibitem{2009herb.book.....D}
C.~D. {Dermer} and G.~{Menon}, \emph{{High Energy Radiation from Black Holes:
  Gamma Rays, Cosmic Rays, and Neutrinos}}.
\newblock Princeton University Press, 2009.

\bibitem{PhysRev.137.B1306}
F.~C. Jones, \emph{Inverse compton scattering of cosmic-ray electrons},
  \href{https://doi.org/10.1103/PhysRev.137.B1306}{\emph{Phys. Rev.} {\bfseries
  137} (Mar, 1965) B1306--B1311}.

\bibitem{Kafexhiu:2014cua}
E.~Kafexhiu, F.~Aharonian, A.~M. Taylor and G.~S. Vila, \emph{{Parametrization
  of gamma-ray production cross-sections for pp interactions in a broad proton
  energy range from the kinematic threshold to PeV energies}},
  \href{https://doi.org/10.1103/PhysRevD.90.123014}{\emph{Phys. Rev. D}
  {\bfseries 90} (2014) 123014},
  [\href{https://arxiv.org/abs/1406.7369}{{\ttfamily 1406.7369}}].

\bibitem{Lebedev_1964}
A.~M. Lebedev, S.~A. Slavatinskii and B.~V. Tolkachev, \emph{Interaction cross
  section and energy conserved by a nucleon colliding with complex nuclei},
  {\emph{JETP} {\bfseries 46} (Jun, 1964) 2151}.

\bibitem{Morejon:2019pfu}
L.~Morejon, A.~Fedynitch, D.~Boncioli, D.~Biehl and W.~Winter, \emph{{Improved
  photomeson model for interactions of cosmic ray nuclei}},
  \href{https://doi.org/10.1088/1475-7516/2019/11/007}{\emph{JCAP} {\bfseries
  11} (2019) 007}, [\href{https://arxiv.org/abs/1904.07999}{{\ttfamily
  1904.07999}}].

\bibitem{Hummer:2010vx}
S.~Hummer, M.~Ruger, F.~Spanier and W.~Winter, \emph{{Simplified models for
  photohadronic interactions in cosmic accelerators}},
  \href{https://doi.org/10.1088/0004-637X/721/1/630}{\emph{Astrophys. J.}
  {\bfseries 721} (2010) 630--652},
  [\href{https://arxiv.org/abs/1002.1310}{{\ttfamily 1002.1310}}].

\bibitem{Anchordoqui:2013dnh}
L.~A. Anchordoqui et~al., \emph{{Cosmic Neutrino Pevatrons: A Brand New Pathway
  to Astronomy, Astrophysics, and Particle Physics}},
  \href{https://doi.org/10.1016/j.jheap.2014.01.001}{\emph{JHEAp} {\bfseries
  1-2} (2014) 1--30}, [\href{https://arxiv.org/abs/1312.6587}{{\ttfamily
  1312.6587}}].

\bibitem{Esteban:2020cvm}
I.~Esteban, M.~C. Gonzalez-Garcia, M.~Maltoni, T.~Schwetz and A.~Zhou,
  \emph{{The fate of hints: updated global analysis of three-flavor neutrino
  oscillations}}, \href{https://doi.org/10.1007/JHEP09(2020)178}{\emph{JHEP}
  {\bfseries 09} (2020) 178},
  [\href{https://arxiv.org/abs/2007.14792}{{\ttfamily 2007.14792}}].

\bibitem{Planck:2018vyg}
{\scshape Planck} collaboration, N.~Aghanim et~al., \emph{{Planck 2018 results.
  VI. Cosmological parameters}},
  \href{https://doi.org/10.1051/0004-6361/201833910}{\emph{Astron. Astrophys.}
  {\bfseries 641} (2020) A6},
  [\href{https://arxiv.org/abs/1807.06209}{{\ttfamily 1807.06209}}].

\bibitem{Khan:2004nd}
E.~Khan, S.~Goriely, D.~Allard, E.~Parizot, T.~Suomijarvi, A.~J. Koning et~al.,
  \emph{{Photodisintegration of ultra-high-energy cosmic rays revisited}},
  \href{https://doi.org/10.1016/j.astropartphys.2004.12.007}{\emph{Astropart.
  Phys.} {\bfseries 23} (2005) 191--201},
  [\href{https://arxiv.org/abs/astro-ph/0412109}{{\ttfamily
  astro-ph/0412109}}].

\bibitem{Rachen:1996zeh}
J.~P. Rachen, \emph{{Interaction Processes and Statistical Properties of the
  Propagation of Cosmic Rays in Photon Backgrounds}}, Ph.D. thesis, University
  of Bonn / Max-Planck-Institute for Radioastronomy, 8, 1996.
\newblock 10.5281/zenodo.3242300.

\bibitem{Stecker:1998ib}
F.~W. Stecker and M.~H. Salamon, \emph{{Photodisintegration of ultrahigh-energy
  cosmic rays: A New determination}},
  \href{https://doi.org/10.1086/306816}{\emph{Astrophys. J.} {\bfseries 512}
  (1999) 521--526}, [\href{https://arxiv.org/abs/astro-ph/9808110}{{\ttfamily
  astro-ph/9808110}}].

\bibitem{GEANT4:2002zbu}
{\scshape GEANT4} collaboration, S.~Agostinelli et~al., \emph{{GEANT4--a
  simulation toolkit}},
  \href{https://doi.org/10.1016/S0168-9002(03)01368-8}{\emph{Nucl. Instrum.
  Meth. A} {\bfseries 506} (2003) 250--303}.

\bibitem{2002EPJA...14..377K}
M.~V. {Kossov}, \emph{{Approximation of photonuclear interaction
  cross-sections}}, {\emph{European Physical Journal A} {\bfseries 14} (Jan.,
  2002) 377--392}.

\bibitem{Puget:1976nz}
J.~L. Puget, F.~W. Stecker and J.~H. Bredekamp, \emph{{Photonuclear
  Interactions of Ultrahigh-Energy Cosmic Rays and their Astrophysical
  Consequences}}, \href{https://doi.org/10.1086/154321}{\emph{Astrophys. J.}
  {\bfseries 205} (1976) 638--654}.

\bibitem{IAEA}
IAEA, \emph{Table of nuclides - nuclear strcuture and decay data},  2024.

\bibitem{Achterberg:2001rx}
A.~Achterberg, Y.~A. Gallant, J.~G. Kirk and A.~W. Guthmann, \emph{{Particle
  acceleration by ultrarelativistic shocks: Theory and simulations}},
  \href{https://doi.org/10.1046/j.1365-8711.2001.04851.x}{\emph{Mon. Not. Roy.
  Astron. Soc.} {\bfseries 328} (2001) 393},
  [\href{https://arxiv.org/abs/astro-ph/0107530}{{\ttfamily
  astro-ph/0107530}}].

\bibitem{Wei:2013wza}
J.-J. Wei, X.-F. Wu, F.~Melia, D.-M. Wei and L.-L. Feng, \emph{{Cosmological
  tests using gamma-ray bursts, the star formation rate and possible abundance
  evolution}}, \href{https://doi.org/10.1093/mnras/stu166}{\emph{Mon. Not. Roy.
  Astron. Soc.} {\bfseries 439} (2014) 3329--3341},
  [\href{https://arxiv.org/abs/1306.4415}{{\ttfamily 1306.4415}}].

\bibitem{Wanderman:2009es}
D.~Wanderman and T.~Piran, \emph{{The luminosity function and the rate of
  Swift's Gamma Ray Bursts}},
  \href{https://doi.org/10.1111/j.1365-2966.2010.16787.x}{\emph{Mon. Not. Roy.
  Astron. Soc.} {\bfseries 406} (2010) 1944--1958},
  [\href{https://arxiv.org/abs/0912.0709}{{\ttfamily 0912.0709}}].

\bibitem{Liu:2011cua}
R.-Y. Liu, X.-Y. Wang and Z.-G. Dai, \emph{{Nearby low-luminosity GRBs as the
  sources of ultra-high energy cosmic rays revisited}},
  \href{https://doi.org/10.1111/j.1365-2966.2011.19590.x}{\emph{Mon. Not. Roy.
  Astron. Soc.} {\bfseries 418} (2011) 1382},
  [\href{https://arxiv.org/abs/1108.1551}{{\ttfamily 1108.1551}}].

\bibitem{Yuksel:2008cu}
H.~Yuksel, M.~D. Kistler, J.~F. Beacom and A.~M. Hopkins, \emph{{Revealing the
  High-Redshift Star Formation Rate with Gamma-Ray Bursts}},
  \href{https://doi.org/10.1086/591449}{\emph{Astrophys. J. Lett.} {\bfseries
  683} (2008) L5--L8}, [\href{https://arxiv.org/abs/0804.4008}{{\ttfamily
  0804.4008}}].

\bibitem{Salafia:2023sjx}
O.~S. Salafia, M.~E. Ravasio, G.~Ghirlanda and I.~Mandel, \emph{{The short
  gamma-ray burst population in a quasi-universal jet scenario}},
  \href{https://doi.org/10.1051/0004-6361/202347298}{\emph{Astron. Astrophys.}
  {\bfseries 680} (2023) A45},
  [\href{https://arxiv.org/abs/2306.15488}{{\ttfamily 2306.15488}}].

\bibitem{Lan:2021uuf}
G.-X. Lan, J.-J. Wei, H.-D. Zeng, Y.~Li and X.-F. Wu, \emph{{Revisiting the
  luminosity and redshift distributions of long gamma-ray bursts}},
  \href{https://doi.org/10.1093/mnras/stab2508}{\emph{Mon. Not. Roy. Astron.
  Soc.} {\bfseries 508} (2021) 52--68},
  [\href{https://arxiv.org/abs/2109.00766}{{\ttfamily 2109.00766}}].

\bibitem{Tamborra:2015qza}
I.~Tamborra and S.~Ando, \emph{{Diffuse emission of high-energy neutrinos from
  gamma-ray burst fireballs}},
  \href{https://doi.org/10.1088/1475-7516/2015/9/036}{\emph{JCAP} {\bfseries
  09} (2015) 036}, [\href{https://arxiv.org/abs/1504.00107}{{\ttfamily
  1504.00107}}].

\bibitem{Rudolph:2021cvn}
A.~Rudolph, {\v{Z}}.~Bo\v{s}njak, A.~Palladino, I.~Sadeh and W.~Winter,
  \emph{{Multiwavelength radiation models for low-luminosity GRBs and the
  implications for UHECRs}},
  \href{https://doi.org/10.1093/mnras/stac433}{\emph{Mon. Not. Roy. Astron.
  Soc.} {\bfseries 511} (2022) 5823--5842},
  [\href{https://arxiv.org/abs/2107.04612}{{\ttfamily 2107.04612}}].

\bibitem{Ito:2021asl}
H.~Ito, O.~Just, Y.~Takei and S.~Nagataki, \emph{{A Global Numerical Model of
  the Prompt Emission in Short Gamma-ray Bursts}},
  \href{https://doi.org/10.3847/1538-4357/ac0cf9}{\emph{Astrophys. J.}
  {\bfseries 918} (2021) 59},
  [\href{https://arxiv.org/abs/2105.09323}{{\ttfamily 2105.09323}}].

\bibitem{Wanderman:2014eza}
D.~Wanderman and T.~Piran, \emph{{The rate, luminosity function and time delay
  of non-Collapsar short GRBs}},
  \href{https://doi.org/10.1093/mnras/stv123}{\emph{Mon. Not. Roy. Astron.
  Soc.} {\bfseries 448} (2015) 3026--3037},
  [\href{https://arxiv.org/abs/1405.5878}{{\ttfamily 1405.5878}}].

\bibitem{Howell:2014wba}
E.~J. Howell, D.~M. Coward, G.~Stratta, B.~Gendre and H.~Zhou,
  \emph{{Constraining the rate and luminosity function of Swift gamma-ray
  bursts}}, \href{https://doi.org/10.1093/mnras/stu1403}{\emph{Mon. Not. Roy.
  Astron. Soc.} {\bfseries 444} (2014) 15--28},
  [\href{https://arxiv.org/abs/1407.2333}{{\ttfamily 1407.2333}}].

\bibitem{Dai:2008mw}
X.~Dai, \emph{{Intensity Distribution and Luminosity Function of the Swift
  Gamma-Ray Bursts}},
  \href{https://doi.org/10.1088/0004-637X/697/1/L68}{\emph{Astrophys. J. Lett.}
  {\bfseries 697} (2009) L68--L71},
  [\href{https://arxiv.org/abs/0812.4466}{{\ttfamily 0812.4466}}].

\bibitem{Zhang:2017moz}
B.~T. Zhang, K.~Murase, S.~S. Kimura, S.~Horiuchi and P.~M\'esz\'aros,
  \emph{{Low-luminosity gamma-ray bursts as the sources of ultrahigh-energy
  cosmic ray nuclei}},
  \href{https://doi.org/10.1103/PhysRevD.97.083010}{\emph{Phys. Rev. D}
  {\bfseries 97} (2018) 083010},
  [\href{https://arxiv.org/abs/1712.09984}{{\ttfamily 1712.09984}}].

\bibitem{Murase_2008}
K.~Murase, K.~Ioka, S.~Nagataki and T.~Nakamura, \emph{High-energy cosmic-ray
  nuclei from high- and low-luminosity gamma-ray bursts and implications for
  multimessenger astronomy},
  \href{https://doi.org/10.1103/physrevd.78.023005}{\emph{Physical Review D}
  {\bfseries 78} (July, 2008) }.

\bibitem{Murase:2006mm}
K.~Murase, K.~Ioka, S.~Nagataki and T.~Nakamura, \emph{{High Energy Neutrinos
  and Cosmic-Rays from Low-Luminosity Gamma-Ray Bursts?}},
  \href{https://doi.org/10.1086/509323}{\emph{Astrophys. J. Lett.} {\bfseries
  651} (2006) L5--L8},
  [\href{https://arxiv.org/abs/astro-ph/0607104}{{\ttfamily
  astro-ph/0607104}}].

\end{thebibliography}\endgroup

\end{document}